\documentclass{article}

\usepackage[english]{babel}
\usepackage[
    backend=biber,
    style=vancouver,
  ]{biblatex}
\usepackage[english]{babel}

\usepackage[letterpaper,top=2cm,bottom=2cm,left=3cm,right=3cm,marginparwidth=1.75cm]{geometry}

\usepackage{authblk}
\usepackage{mdframed}
\usepackage{amsmath}
\usepackage{amssymb}
\usepackage{tcolorbox}
\tcbuselibrary{theorems}
\usepackage{caption}
\usepackage{subcaption}
\usepackage{float}
\usepackage{enumitem}
\usepackage{varwidth}
\usepackage{multirow}
\usepackage{booktabs}
\usepackage{eurosym}
\usepackage{subcaption}
\usepackage{hyperref}

\allowdisplaybreaks

\newcommand\independent{\protect\mathpalette{\protect\independenT}{\perp}}
\def\independenT#1#2{\mathrel{\rlap{$#1#2$}\mkern2mu{#1#2}}}

\title{Decision-analytical models as causal models}

\author[1,*]{Maurice N. Korf}
\author[1,2,3]{M.G. Myriam Hunink}
\author[1,4,$\dagger$]{Richard A.J. Post}
\author[1,$\dagger$]{Jeremy A. Labrecque}

\affil[1]{Department of Epidemiology, Erasmus MC University Medical Center, Rotterdam, The Netherlands}
\affil[2]{Department of Radiology and Nuclear Medicine, Erasmus MC University Medical Center, Rotterdam, The Netherlands}
\affil[3]{Centre for Health Decision Science, Harvard T.H. Chan School of Public Health, Boston, USA}
\affil[4]{Department of Biostatistics, Erasmus MC University Medical Center, Rotterdam, The Netherlands}
\affil[*]{Corresponding author: m.korf@erasmusmc.nl}
\affil[ ]{$\dagger$ Joint last authorship}

\date{}

\begin{document}
\maketitle

\begin{abstract}

Health economic evaluations are fundamentally concerned with answering causal questions by targeting estimands that contrast the costs and health consequences that would be observed under at least two different interventions. This requires the joint distribution of potential outcomes under each specific level of the intervention, which, with appropriate causal assumptions, can in principle be identified from the joint distribution of observed health outcomes. Such data, however, are rarely, if ever, available from a single source. This limitation has motivated the use of decision-analytical models to approximate the joint distribution of outcomes under each intervention directly, informed by causal parameters drawn and synthesized from multiple sources, so that the potential outcomes of interest can be approximated as an expectation over the model-implied outcome trajectories. The validity of this approach, however, rests on the extent to which the underlying (causal) assumptions are deemed plausible. In this work, we formalize this procedure as a task of causal inference, defining and decomposing decision-analytical model bias into components arising from the model structure itself (\textit{model bias}) and the causal input parameters it requires (\textit{target bias}). Because decision-analytical models often rely on unconventional target parameters lacking straightforward observable analogues, and because bias in these parameters can propagate through the model, target bias may arise even in simple settings, a point of central focus in this work. More broadly, this work provides a unifying foundation for medical decision-analytical modelling and causal inference, making explicit the potential for decision-analytical model bias and the role of causal assumptions contributing to it. Framing this as a task of causal inference facilitates precise articulation of the quantities of interest and the parameters required, encouraging analysts to report, justify, and critically reflect on the causal assumptions relied upon. Because these causal assumptions are almost always violated to some degree in practice, this naturally invites a next step: formal assessment through causal bias analyses. After all, the resulting clinical decision is only as credible as the assumptions underlying it.
\end{abstract}
\vspace{0.5cm}

\section{Introduction}
Decision-analytical models are a cornerstone of health economic evaluations, commonly used to inform choices between competing healthcare interventions \cite{siebert2003should}. Approaches such as cost-effectiveness analysis (CEA), cost-benefit analysis, and cost-utility analysis combine clinical evidence, patient preferences, and economic considerations to compare alternative strategies in terms of both their expected costs and health consequences, typically measured in quality-adjusted life years (QALYs) or utilities. At their core, these evaluations aim to answer a simple but profound question: what if we choose intervention $X_1$ instead of $X_2$? This is, fundamentally, a causal question \cite{hernan2020causal}. Whether the model is implemented analytically or through simulation, the underlying logic involves the joint distribution of potential outcomes (i.e. the costs and  health consequences) for each level of the intervention. These health economic evaluations are informative only to the extent that the assumptions embedded in the decision-analytical model, including those about disease progression, treatment effects, risks, and preferences, are deemed plausible \cite{kuhne2022causal}. In this sense, decision-analytical models rely on a causal structure, whether explicit or implicit, to derive meaningful comparisons between interventions. In this work, we formalize this causal perspective on decision-analytical modelling and use a cost-effectiveness setting as a running example to illustrate the estimands and assumptions involved. By doing so, we highlight both the connections between decision-analytical modelling and causal inference, and the practical implications for the evaluation of healthcare interventions.

\section{Causal target estimands in decision-analytical modelling}

Health economic evaluations target causal estimands as they involve, in one way or another, a contrast between costs and health consequences of the same group of people under different interventions. For instance, the incremental cost-effectiveness ratio (ICER), a widely used estimand, is the ratio of the difference in average costs to the difference in average effectiveness between two interventions within the specific target context of interest, characterized by dimensions including population, time, and setting. Conventionally, the ICER is defined as follows:

\begin{equation} \label{EQ:ICER}
     \frac{\displaystyle\mathop{\mathbb{E}} \bigl[ \text{Costs}_1 \bigr] \  - \  \mathop{\mathbb{E}} \bigl[\text{Costs}_0 \bigr]}{\displaystyle\mathop{\mathbb{E}} \bigl[ \text{Effectiveness}_1 \bigr] \  - \  \mathop{\mathbb{E}} \bigl[ \text{Effectiveness}_0 \bigr]}
\end{equation}
\vspace{0.1cm}

\noindent
where $\displaystyle\mathop{\mathbb{E}}[\text{Costs}_1]$ and $\displaystyle\mathop{\mathbb{E}}[\text{Effectiveness}_1]$ denote, respectively, the average costs and effectiveness resulting from intervention 1, and are analogously defined for intervention 0, with effectiveness representing a target health outcome, such as QALYs \cite{hunink2014decision}. To emphasize its causal nature, the ICER can be rewritten using the potential (counterfactual) outcomes framework \cite{hernan2020causal}, as:

\begin{equation} \label{EQ:ICER_C}
     \frac{\displaystyle\mathop{\mathbb{E}} \bigl[ \text{Costs}^{a=1} \bigr] \  - \  \mathop{\mathbb{E}} \bigl[\text{Costs}^{a=0} \bigr]}{\displaystyle\mathop{\mathbb{E}} \bigl[ \text{Effectiveness}^{a=1} \bigr] \  - \  \mathop{\mathbb{E}} \bigl[ \text{Effectiveness}^{a=0} \bigr]} 
\end{equation}
\vspace{0.1cm}

\noindent
where $a$ is an indicator for treatment and $\textrm{Costs}^{a}$ is the value the costs would take when setting the population to intervention $a$, and similarly for $\textrm{Effectiveness}^{a}$. It follows that the ICER represents the average incremental cost associated with a one-unit difference in the average measure of effectiveness when the entire target population would be set intervention $a=1$, as compared to being set $a=0$. Importantly, costs and effectiveness are aggregate quantities, each composed of multiple health outcome variables or events that together make up total costs and/or effectiveness. Interventions exert their impact on costs and effectiveness by altering the underlying distribution of all relevant health outcomes and events, whereby each outcome or event has a specific cost and/or effectiveness value. Each cost and effectiveness term is accordingly modeled as a function of health outcomes (or events) under a specific intervention, where each outcome must be affected by the intervention and, in turn, influences costs, effectiveness, or both. Let the random variable $\mathbf{V}^a$ denote the vector of all relevant health outcomes under intervention $a$, each contributing to costs and/or effectiveness (i.e. the random vector of outcomes in the universe where all individuals are assigned to intervention $a$). The actual but unknown costs and effectiveness equal $f(\boldsymbol{V}^a)$ and $g(\boldsymbol{V}^a)$, respectively, for some functions $f(\cdot)$ and $g(\cdot)$ that map the potential health outcomes to the corresponding total cost and effectiveness. To capture possible upfront costs that arise directly from the decision to intervene through $a$, we define a total cost function $f_{\lambda}(a,\mathbf{V}^a) = \lambda(a) + f(\mathbf{V}^a)$. Here, $\lambda(a)= \lambda_1 \mathbb{I}(a=1) + \lambda_0 \mathbb{I}(a=0)$ represents the fixed costs corresponding to setting the intervention to $a$, where $\mathbb{I}()$ denotes the indicator function. Unlike costs, there are no fixed upfront effectiveness values, as these are, by definition, defined with respect to potential health outcomes and thus fully captured by $g(\boldsymbol{V}^a)$. For example, side effects resulting from an intervention procedure are represented through corresponding potential health outcomes incorporated into $\boldsymbol{V}^a$. Importantly, each $f(\boldsymbol{V}^a)$ and $g(\boldsymbol{V}^a)$ should evaluate, in principle, the same set of health outcomes $\boldsymbol{V}$ under each intervention level $a$ to ensure well-defined causal contrasts, while the specific outcome sets may differ between $f(\cdot)$ and $g(\cdot)$.

Under this notation, each target outcome term in Equation (\ref{EQ:ICER_C}) can be expressed in terms of its respective total cost or effectiveness function, at the specified level of the intervention, as illustrated in Equation (\ref{EQ:ICER_EV}).

\begin{equation} \label{EQ:ICER_EV}
    \frac{\displaystyle\mathop{\mathbb{E}}  \Bigl[ \lambda_1 + f(\mathbf{V}^{a=1}) \Bigr] \  - \  \displaystyle\mathop{\mathbb{E}} \Bigl[ \lambda_0 + f(\mathbf{V}^{a=0}) \Bigr] }{\displaystyle\mathop{\mathbb{E}} \Bigl[ g(\mathbf{V}^{a=1}) \bigr] \  - \ \displaystyle\mathop{\mathbb{E}} \bigl[ g(\mathbf{V}^{a=0}) \Bigr]}
\end{equation}
\vspace{0.1cm}

\subsection{Missing data as motivation for decision-analytical modelling}

While the target estimand precisely defines the quantity of interest, two considerations arise: why focus is typically restricted to population-level quantities, and why decision-analytical models are used to target them. Both are related to missing data, albeit operating at different levels. We first revisit the fundamental problem of causal inference in the context of decision-analytical modelling, an aspect that is often not made explicit in practice, and then discuss how the absence of a single, complete data source precludes direct identification of marginal, population-level quantities, thereby motivating the use of decision-analytical models.

\subsubsection{The fundamental problem of causal inference}
In full generality, the target estimand of interest may depend on the pair of potential outcomes for each individual, such as the $i$-CER \cite{basu2009individualization} or the individual-level net monetary benefit \cite{van2012role}. As an illustrative example using the introduced notation, consider an alternative hypothetical estimand, defined as the ratio of the expected cost ratio to the expected effectiveness ratio, as shown in Equation (\ref{EQ:ICER_R}).

\begin{equation} \label{EQ:ICER_R}
\frac{\displaystyle
\mathbb{E}\left[ \frac{\lambda_1 + f(\mathbf{V}^{a=1})}{\;\lambda_0 + f(\mathbf{V}^{a=0})} \right]
}{\displaystyle
\mathbb{E}\left[ \frac{g(\mathbf{V}^{a=1})}{\;  g(\mathbf{V}^{a=0})} \right]
}
\end{equation}

\noindent
Since the expectation is taken over each respective ratio, this requires having the individual-level cost and effectiveness ratios. Specifically, for each individual $i$, we would need to have access to information set $\mathcal{I}_i$, where $\boldsymbol{V}_i^a$ denotes the vector of potential outcomes for individual $i$ under intervention $a$. Note that the cost and effectiveness values have been omitted from $\mathcal{I}_i$, as they are derived from the potential outcomes via analyst-specified cost and effectiveness functions. While the specification of the cost and effectiveness functions is an important aspect of decision-analytical modelling, their inclusion in $\mathcal{I}_i$ offers no additional insight for the purpose of present discussion.

\[\mathcal{I}_i=\{\boldsymbol{V}_i^{a=1}, \boldsymbol{V}_i^{a=0} \} \]

\noindent
If the complete information set $\mathcal{I}_i$ were available for all individuals $i$, then, given the specified cost and effectiveness functions, the target estimand in Equation (\ref{EQ:ICER_R}) could be directly estimated. In practice, however, we only observe the vector of potential outcomes corresponding to the intervention actually received by individual $i$, denoted by $A_i$. Let $\boldsymbol{V}_i$ represent the corresponding vector of observed outcomes; together with $A_i$, these define the observed information set $\mathcal{I}^{obs}_{i,A_i}$,

\[ \mathcal{I}^{obs}_{i,A_i}=\{A_i, \boldsymbol{V}_i\}. \]

\noindent
Under the stable unit treatment value assumption (SUTVA) \cite{hernan2020causal, imbens2015causal}, the observed outcomes equal the potential outcomes under the received intervention; that is, $\boldsymbol{V}_i=\boldsymbol{V}^{a}_i$ if $A_i=a$. Consequently, the potential outcomes information set that would be available for each individual is given by

\[\mathcal{I}_{i}=\{\boldsymbol{V}_i^{A_i}\}. \]

\noindent
This highlights the fundamental problem of causal inference: only, at most, one of the potential outcome vectors can be observed for each individual. This is, at its core, a missing data problem \cite{holland1986statistics}. As a result, individual causal effects, such as those required for Equation (\ref{EQ:ICER_R}), are in principle not identifiable without imposing strong assumptions \cite{post2025beyond}. For this reason, we instead target estimands that depend on marginal causal quantities, such as the counterfactual ICER in Equation (\ref{EQ:ICER_EV}), which can nonetheless be identified without necessitating access to the pair of potential outcomes at the individual level. As an illustration, consider a perfect marginally randomized controlled trial in which individuals are randomized to either $A=1$ or $A=0$. For each individual, we observe $\mathcal{I}^{obs}_{i,A_i}$ under the assigned intervention $A$, contributing either to the aggregated information set $\mathcal{I}^{obs}_{A=1}$ or $\mathcal{I}^{obs}_{A=0}$. Under the causal assumptions of exchangeability, positivity, and SUTVA \cite{hernan2020causal}, together with the specified cost and effectiveness functions, the expected potential costs and effectiveness under each intervention $a$ are identifiable from the observed aggregated information sets. That is, for costs, $\displaystyle\mathbb{E}[f(\boldsymbol{V}^a)] = \displaystyle\mathbb{E}[f(\boldsymbol{V} \mid A=a)]$, and similarly for effectiveness, $\displaystyle\mathbb{E}[g(\boldsymbol{V}^a)] = \displaystyle\mathbb{E}[g(\boldsymbol{V} \mid A=a)]$.

\subsubsection{Absence of a complete single data source}

While identification, by definition, entails linking potential outcomes to the observed data distribution under specified assumptions, the estimands targeted in health economic evaluations often rely on outcome data, such as $\boldsymbol{V}$, that are not all observed within any single data source. For example, we may be interested in $\boldsymbol{V}=\{V_1,V_2,V_3\}$ under intervention $A$, but no single data source contains all of these outcomes. Instead, one source may have measured $\{V_1,V_2\}$, while another measured $\{V_2,V_3\}$, assuming both considered the same intervention.

Fundamentally, the absence of a single data source informing the joint distribution of interest gives rise to an additional missing data problem: neither the individual-level information sets $\mathcal{I}^{\text{obs}}_{i,A_i}$ nor the corresponding aggregated sets $\mathcal{I}^{\text{obs}}_{A}$ for each $A$ are fully observed. As a result, without further assumptions, the potential outcomes $\boldsymbol{V}^a$ under each intervention $a$ cannot be readily identified, leaving even marginal causal quantities unidentifiable. This challenge has, in turn, motivated the use of decision-analytical models to approximate the underlying data-generating process. A statistical decision-analytical model $\mathcal{M}$ is specified with input probability parameters $\mathcal{P_M}$, whose nature governs both the interpretation and the scope of the resulting inference. Unlike standard identification, $\mathcal{M}$ is specified a priori via $\mathcal{P_M}$ to approximate either the joint distribution of observed health outcomes (when $\mathcal{P_M}$ is non-causal) or the joint distribution of potential health outcomes (when $\mathcal{P_M}$ is causal). In the latter case, $\mathcal{M}$ is referred to as a causal decision-analytical analytical model, as it encodes the model-implied joint distribution of potential health outcomes $\boldsymbol{V}^a$. The causal assumptions for a specific intervention embedded within $\mathcal{M}$ concern the identification of the causal parameters in $\mathcal{P_M}$; however, due to the multifaceted nature of these assumptions, a comprehensive discussion is postponed to section 4.2. Simulating units $j$ through $\mathcal{M}$, when causal, generates potential outcomes under intervention $a$, where each $j$ denotes a sub-population defined by shared attributes. For each unit $j$ under intervention $a$, this defines the information set $\mathcal{I}_{j,a}^\mathcal{M}$, and by aggregating over all $j$ yields the corresponding population-level set $\mathcal{I}_{a}^\mathcal{M}$, which serves as the basis for estimating the expected potential costs and effectiveness.

\section{Bias in decision-analytical models}

Although using a decision-analytical model $\mathcal{M}$ to simulate (potential) outcome data is a practical workaround for unavailable observations, its validity relies on the credibility of the underlying assumptions. Notably, modelling imposes structural assumptions that may not reflect the true data-generating process, whether $\mathcal{M}$ is employed for causal or non-causal purposes, potentially introducing \textit{model bias}. When used for causal purposes, each parameter in $\mathcal{P_M}$ must individually satisfy the necessary causal assumptions. Violation of any assumption for any input parameter may introduce \textit{target bias}. We proceed to formalize the notions of model bias and target bias and to define their combined contribution as decision-analytical model bias.

Let $\mathcal{T}$ denote the target estimand of interest, defined throughout the remainder of this paper as a functional of the expected potential costs and expected potential effectiveness under each level of the intervention under the true underlying data generating process $\mathcal{P}$. Using the previously defined representations of each respective target outcome, the target estimand $\mathcal{T}$ can be expressed as follows:

\begin{equation} \label{eq:T_definition}
\mathcal{T} = h\Big(
    \mathbb{E}_{\mathcal{P}}\big[\lambda_1 + f(\boldsymbol{V}^{a=1}) \big],
    \mathbb{E}_{\mathcal{P}}\big[\lambda_0 + f(\boldsymbol{V}^{a=0})\big],
    \mathbb{E}_{\mathcal{P}}\big[g(\boldsymbol{V}^{a=1}) \big],
    \mathbb{E}_{\mathcal{P}}\big[g(\boldsymbol{V}^{a=0}) \big]
\Big)
\end{equation}

\noindent
Depending on the choice of $h(\cdot)$, $\mathcal{T}$ may represent  different quantities of interest, including the counterfactual ICER. The best attainable approximation of $\mathcal{T}$ under model $\mathcal{M}$ with parametrization $\mathcal{P_M}$ is denoted $\mathcal{T}_{\mathcal{M}}$. We refer to $\mathcal{M}$, parametrized by $\mathcal{P_M}$, as the \textit{target decision-analytical model}\textemdash an idealized model that fully specifies the set of parameter estimands $\mathcal{P_M}$, required to characterize the decision problem under the structure $\mathcal{M}$, with all parameters taking their true values. Accordingly, $\mathcal{T}_{\mathcal{M}}$ is given by:

\begin{equation} \label{eq:T_M_definition}
\mathcal{T}_{\mathcal{M}}
= h\Big(
    \mathbb{E}_{\mathcal{P_M}}[\lambda_1 + f(\boldsymbol{V}^{a=1})],\ 
    \mathbb{E}_{\mathcal{P_M}}[\lambda_0 + f(\boldsymbol{V}^{a=0})],\ 
    \mathbb{E}_{\mathcal{P_M}}[g(\boldsymbol{V}^{a=1})],\ 
    \mathbb{E}_{\mathcal{P_M}}[g(\boldsymbol{V}^{a=0})]
\Big)
\end{equation}

\noindent
where $\mathop{\mathbb{E}}_{\mathcal{P_M}}[\cdot]$ denotes the expectation taken with respect to the distribution implied by inputs $\mathcal{P_M}$ under model structure $\mathcal{M}$. Any discrepancy between $\mathcal{T}$ and $\mathcal{T_M}$ is defined as \textit{model bias}, resulting from the assumptions involved in approximating the true data-generating process using a model. For instance, model bias would be introduced when attempting to reconstruct the unobserved joint distribution $P(A,V_1,V_2, V_3)$ using a model that combines inputs $P(V_1,V_2 \mid A)$ and $P(V_1,V_3 \mid A)$, each obtained from a different source, when in reality $V_2 \not \independent V_3 \mid A, V_1$.

Bias may also arise from the estimation of $\mathcal{P}_{\mathcal{M}}$, as this is constrained by the available data and evidence. In particular, the decision-analytical model that can be constructed in practice using estimated parameters is referred to as the \textit{empirical decision-analytical model}. Although this model is intended to emulate the target decision-analytical model \cite{hernanemulation}, achieving this in practice may be challenging, particularly because the estimated parameters are typically drawn from heterogeneous sources. To explicitly represent this source heterogeneity, we introduce the necessary notation, which subsequently permits a formal characterization of target bias. Let $\mathcal{S}=\{s_1,\ldots,s_m\}$ denote a collection of relevant data sources, where each $s_d \in \mathcal{S}$ for $d=1,\ldots,m$ corresponds to either an experimental or observational data source. For each target parameter $P_k \in \mathcal{P_M}=\{P_k :k=1,\ldots,N \}$, the analyst selects a subset of relevant data sources to estimate it, denoted by $\mathcal{S}_k \subseteq \mathcal{S}$. Accordingly, let $\hat{P}_{k \mid \mathcal{S}_k}$ represent the estimate for parameter $P_k$ based on data (i.e. the observed outcomes) in $\mathcal{S}_k$. At the model level, source heterogeneity in the empirical decision model’s evidence base is represented by the source mixture vector $\boldsymbol{S}=(S_1,\ldots,S_m)$, where each $S_p \in \boldsymbol{S}$ for $p=1,\ldots,m$ denotes a binary indicator, with $S_p\in\{0,1\}$. If $S_p=1$, this indicates that source $s_p \in \mathcal{S}$ contributed, implicitly through the input parameters, to the quantity produced under the empirical decision-analytical model. Consequently, we define the set of estimated parameters as $\hat{\mathcal{P}}_{\mathcal{M} \mid \boldsymbol{S}}=\{\hat{P}_{k\mid \mathcal{S}_k} : k=1,\ldots,N\}$, which reflects that the parameters entering the empirical decision-analytical model are sourced from multiple, distinct data sources. With this notation, we define the estimate $\hat{\mathcal{T}}_{\mathcal{M}}$ of $\mathcal{T_M}$, obtained under the empirical decision-analytical model, as follows:

\begin{align} \label{eq:T_M_hat_definition}
\hat{\mathcal{T}}_{\mathcal{M}} = 
h\Big(
&\mathbb{E}\Big[\mathbb{E}_{\hat{\mathcal{P}}_{\mathcal{M} \mid \boldsymbol{S}}}[\lambda_1 + f(\boldsymbol{V} \mid A=1, \boldsymbol{S})]\Big],\ 
\mathbb{E}\Big[\mathbb{E}_{\hat{\mathcal{P}}_{\mathcal{M} \mid \boldsymbol{S}}}[\lambda_0 + f(\boldsymbol{V} \mid A=0, \boldsymbol{S})]\Big],\ 
\mathbb{E}\Big[\mathbb{E}_{\hat{\mathcal{P}}_{\mathcal{M} \mid \boldsymbol{S}}}[g(\boldsymbol{V} \mid A=1, \boldsymbol{S})]\Big],\notag \\ 
&\mathbb{E}\Big[\mathbb{E}_{\hat{\mathcal{P}}_{\mathcal{M} \mid \boldsymbol{S}}}[g(\boldsymbol{V} \mid A=0, \boldsymbol{S})]\Big]
\Big)
\end{align}
 
\noindent
Here, the outer expectation, $\mathbb{E}[\cdot]$, accounts for the fact that the parameter estimates themselves are uncertain. When $\mathcal{T_M} \neq \hat{\mathcal{T}}_{\mathcal{M}}$, the resulting discrepancy is defined as \textit{target bias}, which may arise when one or more estimated parameters in the empirical model are internally and/or externally biased relative to the target parameters in the target decision-analytical model. For instance, consider again reconstructing the unobserved joint distribution $P(A, V_1, V_2, V_3)$ by using a model that combines both conditional distributions 
$P(V_1, V_2 \mid A, S_1=1)$ and $P(V_1, V_3 \mid A, S_2=1)$, now with each source explicitly represented by the binary indicators $S_1$ and $S_2$. Target bias may arise if either source involves non-randomized assignment of $A$, potentially due to confounding (i.e. internal validity bias), or possibly because differences in the distribution of effect modifiers between the source and target result in differences in the expected outcomes under $A$ (i.e. external validity bias). Numerical examples illustrating these two sources of target bias are presented in Appendix A.

Having defined both model bias and target bias, we can define total decision-analytical model bias as the overall difference between the target estimand, $\mathcal{T}$, and the expected estimate from the empirical decision-analytical model, $\hat{\mathcal{T}}_{\mathcal{M}}$ (Figure 1).

\begin{figure}[h!]
    \centering
    \includegraphics[width=0.9\textwidth]{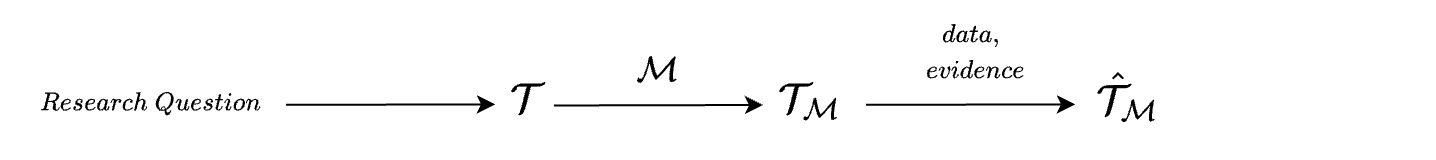}
    \caption{Flow of reasoning, from research question to estimate}
    \label{fig:flow_reasoning}
\end{figure}
\vspace{0.1cm}

To further illustrate the concept of decision-analytical model bias, we consider the counterfactual ICER in Equation (\ref{EQ:ICER_EV}) as target estimand, $\mathcal{T}$, with total decision-analytical model bias (bias$_\text{DM}$) expressed as the difference between $\mathcal{T}$ and the ICER estimated under model $\mathcal{M}$ with parameters $\hat{\mathcal{P}}_{\mathcal{M}}$, as shown in Equation (\ref{EQ:ICER_bias}). This total bias can equivalently be represented as a function $z(\cdot)$ of the target outcome–specific biases, denoted $\text{bias}_{f,a}$ for costs and $\text{bias}_{g,a}$ for effectiveness under intervention $a$, where $z(\cdot)$ is a non-additive function (details in Appendix B).

\begin{align} \label{EQ:ICER_bias}
\text{Bias}_{\text{DM}} &=  \mathcal{T} - \hat{\mathcal{T}}_{\mathcal{M} }\notag \\[0.3cm]
&= \frac{
    \displaystyle\mathbb{E}_{\mathcal{P}} \Bigl[ \lambda_1 + f(\mathbf{V}^{a=1}) \Bigr] - \displaystyle\mathbb{E}_{\mathcal{P}} \Bigl[\lambda_0 + f(\mathbf{V}^{a=0}) \Bigr]
}{
    \displaystyle\mathbb{E}_{\mathcal{P}} \Bigl[ g(\mathbf{V}^{a=1}) \Bigr] - \displaystyle\mathbb{E}_{\mathcal{P}} \Bigl[ g(\mathbf{V}^{a=0}) \Bigr]
} \notag \\
&\quad - 
\frac{
    \displaystyle\mathbb{E}\biggl[\mathbb{E}_{\hat{\mathcal{P}}_{\mathcal{M} \mid \boldsymbol{S}}} \bigl[ \lambda_1 + f(\mathbf{V} \mid A=1, \boldsymbol{S}) \bigr]\biggr] - \displaystyle\mathbb{E}\biggl[\mathbb{E}_{\hat{\mathcal{P}}_{\mathcal{M} \mid \boldsymbol{S}}} \bigl[ \lambda_0 + f(\mathbf{V} \mid A=0, \boldsymbol{S}) \bigr]\biggr]
}{
    \displaystyle\mathbb{E}\biggl[\mathbb{E}_{\hat{\mathcal{P}}_{\mathcal{M} \mid \boldsymbol{S}}} \bigl[ g(\mathbf{V} \mid A=1, \boldsymbol{S}) \bigr]\biggr] - \displaystyle\mathbb{E}\biggl[\mathbb{E}_{\hat{\mathcal{P}}_{\mathcal{M} \mid \boldsymbol{S}}} \bigl[ g(\mathbf{V} \mid A=0, \boldsymbol{S}) \bigr]\biggr]
}  \\[0.3cm]
&= z\Big(\text{bias}_{f,a=1}, \ \text{bias}_{f,a=0},\ \text{bias}_{g,a=1},\ \text{bias}_{g,a=0}\Big) \notag
\end{align}

\noindent
More specifically, target outcome-specific bias is defined as the difference between the target outcome (e.g. costs$^a$) and its expected estimate, as illustrated for costs ($\text{bias}_{f,a}$) in Equation (\ref{EQ:bias_decomp}). We can then apply the bias decomposition introduced above, here illustrated for costs, allowing target bias to be further decomposed into internal and external validity bias \cite{westreich2019target}, as shown in the last line of Equation (\ref{EQ:bias_decomp}). For simplicity, we have assumed that $\boldsymbol{V}$ is discrete, otherwise, the sum can be replaced by an integral. We further note that the fixed costs cancel out.

\begin{align} \label{EQ:bias_decomp}
    \text{bias}_{f,a} 
    &= \mathbb{E_{\mathcal{P}}}[f(\boldsymbol{V}^a)] - \mathbb{E}\bigl[\mathbb{E}_{\hat{\mathcal{P}}_{\mathcal{M} \mid \boldsymbol{S}}}[f(\boldsymbol{V} \mid A=a, \boldsymbol{S})]\bigr] \notag \\[0.3cm]
    &= \Bigl(\mathbb{E}_{\mathcal{P}}[f(\boldsymbol{V}^a)] - \mathbb{E}_{\mathcal{P}_{\mathcal{M}}}[f(\boldsymbol{V}^a)]\Bigr) + \Bigl(\mathbb{E}_{\mathcal{P}_{\mathcal{M}}}[f(\boldsymbol{V}^a)] - \mathbb{E}\bigl[\mathbb{E}_{\hat{\mathcal{P}}_{\mathcal{M} \mid \boldsymbol{S}}}[f(\boldsymbol{V} \mid A=a, \boldsymbol{S})]\bigr]\Bigr) \notag \\[0.3cm]
    &= \Bigl(\mathbb{E}_{\mathcal{P}}[f(\boldsymbol{V}^a)] - \mathbb{E}_{\mathcal{P}_{\mathcal{M}}}[f(\boldsymbol{V}^a)]\Bigr) + \Bigl(\mathbb{E}_{\mathcal{P}_{\mathcal{M}}}[f(\boldsymbol{V}^a)] - \mathbb{E}_{\mathcal{P}_{\mathcal{M} \mid \boldsymbol{S}}}[f(\boldsymbol{V}^a \mid \boldsymbol{S})]\Bigr) + \notag \\[0.1cm]
    &\quad \  \Bigl(\mathbb{E}_{\mathcal{P}_{\mathcal{M} \mid \boldsymbol{S}}}[f(\boldsymbol{V}^a \mid \boldsymbol{S})] - \mathbb{E}\bigl[\mathbb{E}_{\hat{\mathcal{P}}_{\mathcal{M} \mid \boldsymbol{S}}}[f(\boldsymbol{V} \mid A=a, \boldsymbol{S})] \Bigr)  \\[0.3cm]
    &= \underbrace{\left( \sum_{\boldsymbol{v}^a} f(\boldsymbol{v}^a) P(\boldsymbol{v}^a) - \sum_{\boldsymbol{v}^a} f(\boldsymbol{v}^a) P_{\mathcal{M}}(\boldsymbol{v}^a) \right)}_{\text{model bias}} + \underbrace{\left( \sum_{\boldsymbol{v}^a} f(\boldsymbol{v}^a) P_{\mathcal{M}}(\boldsymbol{v}^a) - \sum_{\boldsymbol{v}^a} f(\boldsymbol{v}^a \mid \boldsymbol{S}) P_{\mathcal{M} \mid \boldsymbol{S}}(\boldsymbol{v}^a \mid \boldsymbol{S}) \right)}_{\text{external validity bias}} \notag \\
    &\quad + \underbrace{\left( \sum_{\boldsymbol{v}^a} f(\boldsymbol{v}^a \mid \boldsymbol{S}) P_{\mathcal{M} \mid \boldsymbol{S}}(\boldsymbol{v}^a \mid \boldsymbol{S}) - \sum_{\boldsymbol{v}} f(\boldsymbol{v} \mid A = a, \boldsymbol{S}) \hat{P}_{\mathcal{M} \mid \boldsymbol{S}}(\boldsymbol{v} \mid A=a, \boldsymbol{S}) \right)}_{\text{internal validity bias}} \notag
\end{align}

\noindent
The bias naturally decomposes into three components. The model component need only be assessed once for the entire model structure, whereas external and internal biases should be assessed for each parameter individually. Especially because parameters are typically drawn from different sources, the conditions required to ensure external and internal validity are parameter-specific, influenced by factors such as the nature of the available data, the data-generating mechanism, the sample population, and the identification strategy employed. These considerations further depend on whether a parameter is estimated from a single source or multiple sources (e.g. causal meta-analysis context).

At a higher level, internal validity concerns the validity of the estimate of a given parameter within the context of a specific source or set of sources. Given the source data, it is necessary to assess whether the estimate is valid by examining factors such as confounding, the definition and assignment of the intervention, measurement bias, and statistical estimation procedures, among others. Crucially, in a causal context, internal validity depends on the credibility of identification assumptions specific to each parameter given the data source.  Even if an estimate is deemed internally valid, this alone is not sufficient. We must also ensure that the estimated parameter is applicable to the target context in which the decision-analytical model is defined. Here, \textit{context} is understood broadly, but only those dimensions believed to differ meaningfully between the source and target, such that it causes meaningful changes in the estimate, are explicitly considered. These dimensions may include, but are not limited to, population, intervention, outcome, and time \cite{findley2021external, degtiar2023review}.

\section{From target to estimate}

In the following sections, we discuss the bias components in greater detail using a hypothetical cost-effectiveness example, with particular emphasis on internal and external validity. We begin by specifying the model component, using the example to illustrate the relationship between the causal target estimand and the target decision-analytical model, and how this relationship is shaped by the assumed underlying causal structure. Throughout, we treat the model as given, since a complete survey of model bias is beyond the scope of this work (see e.g. \cite{brennan2006taxonomy}). With the target decision-analytical model specified, we extend the discussion to the empirical decision-analytical model, making explicit the role of the causal assumptions required to ensure internal and external validity, which in turn permit the empirical and target decision-analytical models to be linked, that is, to link $\hat{\mathcal{P}}_{\mathcal{M}}$ to $\mathcal{P}_{\mathcal{M}}$.

\subsection{The model component}
\begin{figure}[h!]
    \centering
    \includegraphics[width=0.6\textwidth]{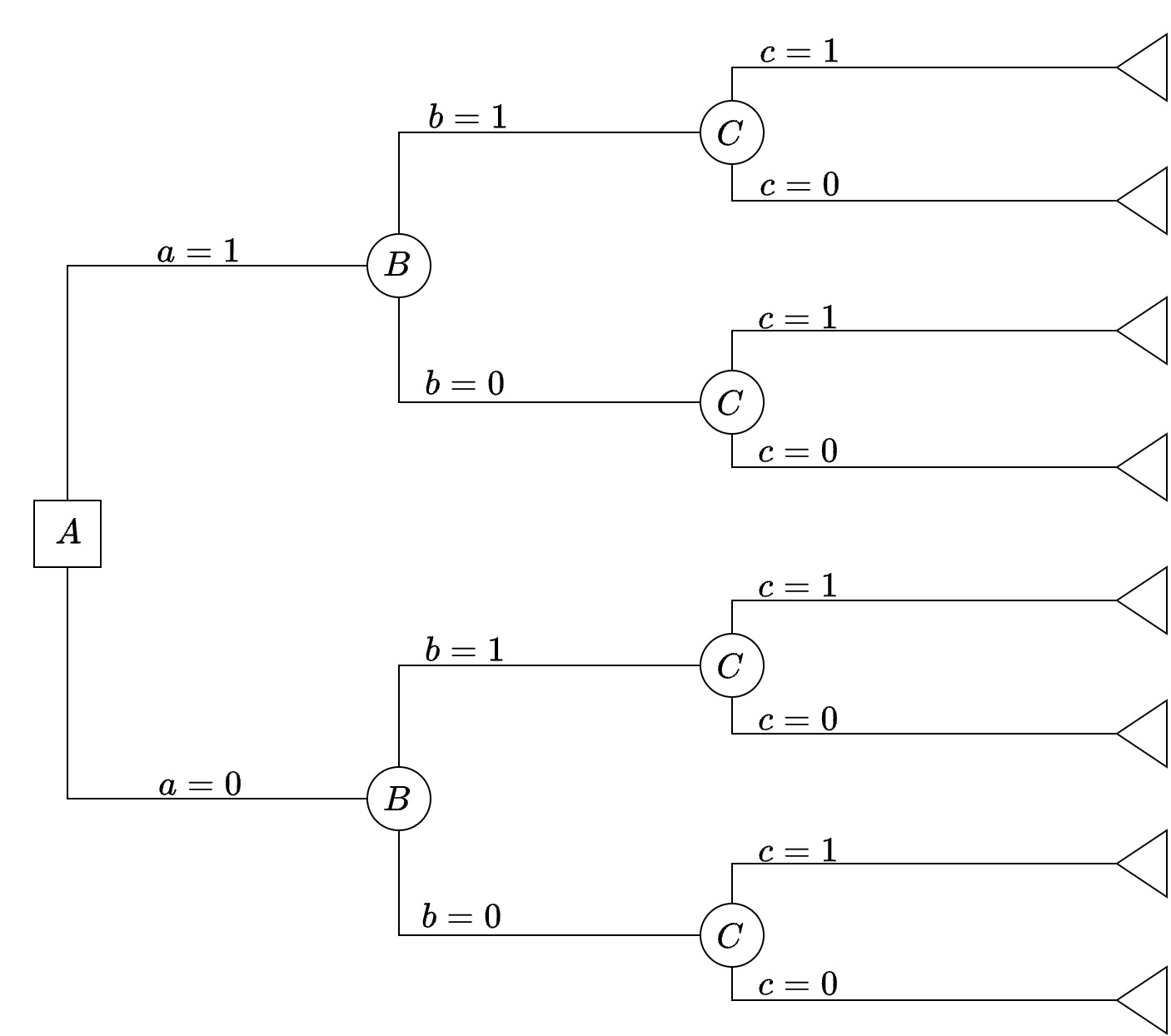}
    \caption{A conventional visualised decision tree, with the squares representing a decision node ($A$), circles representing chance nodes ($B$ and $C$), and triangles denoting terminal nodes.}
    \label{fig1}
\end{figure}

In our simple example, we are interested in the ICER of intervention $a=1$ versus $a=0$, where effectiveness is measured in Quality-Adjusted Life Years (QALYs). We consider three dichotomous variables: $A,B$ and $C$, where $A$ denotes the intervention, and $B$ and $C$ different health outcomes affected by $A$, which in turn affect costs and QALYs. Here, decision-analytical model $\mathcal{M}$ is specified as a rollback decision tree \cite{brennan2006taxonomy,hunink2014decision}, which would conventionally be visualised as in Figure (\ref{fig1}), where lowercase letters represent realizations of the corresponding random variables.

\newpage
\noindent
Ultimately, we are interested in the counterfactual ICER as defined in Equation (\ref{EQ:ICER_example}), representing target estimand $\mathcal{T}$, with  $\mathbf{V}^a=(B^a,C^a)$.

\begin{equation} \label{EQ:ICER_example}
\frac{\displaystyle\mathbb{E}_{\mathcal{P}} \Bigl[ \lambda_1 + f(B^{a=1},C^{a=1}) \Bigr] \  - \  \displaystyle\mathbb{E}_{\mathcal{P}} \Bigl[ \lambda_0 + f(B^{a=0},C^{a=0}) \Bigr] }{\displaystyle\mathbb{E}_{\mathcal{P}}\Bigl[  g(B^{a=1},C^{a=1}) \bigr] \  - \ \displaystyle\mathbb{E}_{\mathcal{P}}\bigl[  g(B^{a=0},C^{a=0}) \Bigr]}
\end{equation}
\vspace{0.1cm}

\noindent
However, following the introduced setup, the best attainable approximation of $\mathcal{T}$ is assumed to be under a rollback-decision tree model, $\mathcal{M}$. Accordingly, we aim to target the counterfactual ICER under model $\mathcal{M}$ with parameters $\mathcal{P_M}$, as illustrated in Equation (\ref{EQ:ICER_example_M}).

\begin{equation} \label{EQ:ICER_example_M}
\frac{\displaystyle\mathbb{E}_{\mathcal{P_M}} \Bigl[ \lambda_1 + f(B^{a=1},C^{a=1}) \Bigr] \  - \  \displaystyle\mathbb{E}_{\mathcal{P_M}} \Bigl[ \lambda_0 + f(B^{a=0},C^{a=0}) \Bigr] }{\displaystyle\mathbb{E}_{\mathcal{P_M}}\Bigl[ g(B^{a=1},C^{a=1}) \bigr] \  - \ \displaystyle\mathbb{E}_{\mathcal{P_M}}\bigl[  g(B^{a=0},C^{a=0}) \Bigr]}
\end{equation}
\vspace{0.1cm}

\noindent
To estimate this quantity, we must first identify the necessary causal input probabilities. When the expectation is expressed explicitly, as shown in Equation (\ref{EQ:ICER_example2}), it becomes evident that the joint distribution of potential health outcomes under intervention $a$, $P(B^a,C^a)$, is required. 

\begin{equation} \label{EQ:ICER_example2}
     \frac{ \displaystyle \left[ \lambda_1 + \mathop{\sum_{b,c}}  f(B^{a=1},C^{a=1})  P_{\mathcal{M}}(B^{a=1},C^{a=1}) \right] \  - \  \left[ \lambda_0 + \mathop{\sum_{b,c}}  f(B^{a=0},C^{a=0}) P_{\mathcal{M}}(B^{a=0},C^{a=0}) \right]}
     { \displaystyle \left[ \mathop{\sum_{b,c}}  g(B^{a=1},C^{a=1}) P_{\mathcal{M}}(B^{a=1},C^{a=1}) \right] \  - \  \left[  \mathop{\sum_{b,c}}  g(B^{a=0},C^{a=0}) P_{\mathcal{M}}(B^{a=0},C^{a=0}) \right]}
\end{equation}
\vspace{0.1cm}

\noindent
However, its joint observed analogue, $P(A,B,C)$, is rarely, if ever, directly available. Consequently, we instead factorize the joint distribution of potential health outcomes according to the assumed underlying causal structure, which decomposes it into smaller components. By doing so, we aim to identify each component separately, leveraging different data sources. The following section presents this causal factorization in detail and highlights the added value of using causal graphs to support this.

\subsubsection{The target decision-analytical model}
The joint distribution of potential health outcomes, in our example $P_{\mathcal{M}}(A,B^a,C^a)$, is factorized in practice to break down estimation into multiple lower-dimensional estimation challenges as well as to account for the fact that full data on the underlying observed joint distribution are rarely available. In causal inference, the joint distribution of potential outcomes is factorized according to the assumed causal structure, which is often described using a causal graph.  Causal graphs are commonly represented using tools such as causal directed acyclic graphs (DAGs) \cite{greenland1999causal} and single world intervention graphs (SWIGs) \cite{richardson2013single}, both of which effectively communicate assumptions about the causal structure. In this work, we adopt the SWIGs framework as it integrates potential outcomes notation, thereby offering an intuitive connection between the joint distribution of potential health outcomes under intervention $a$ from equation (\ref{EQ:ICER_example2}) and the presumed causal structure.

\begin{figure}[h]
    \centering
    \begin{subfigure}[t]{0.6\textwidth}
        \centering
        \includegraphics[width=\textwidth]{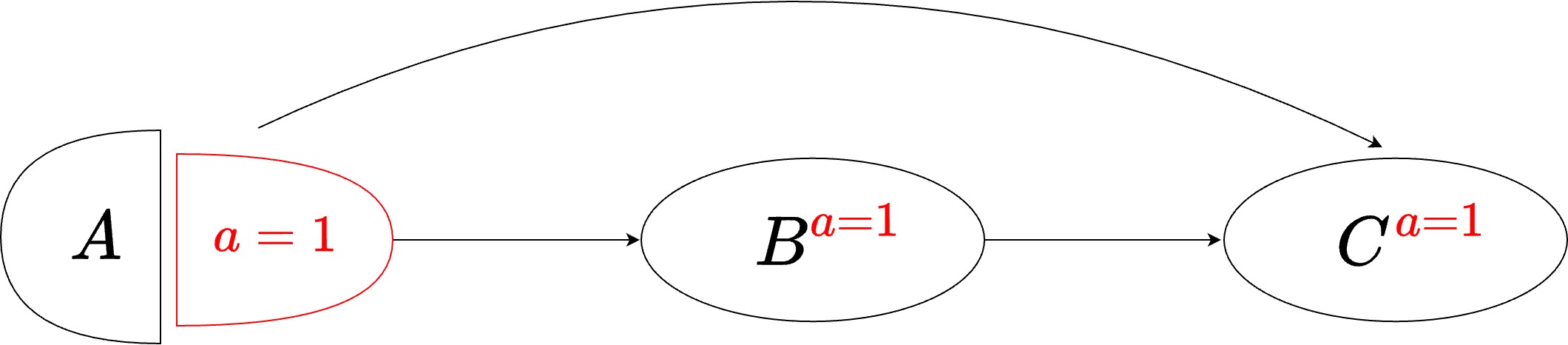}
        \caption{Single world intervention graph $\mathcal{G}(a=1)$}
        \label{fig:SWIGa1}
    \end{subfigure}
    \\[0.5cm]
    \begin{subfigure}[t]{0.6\textwidth}
        \vspace{0.5cm}
        \centering 
        \includegraphics[width=\textwidth]{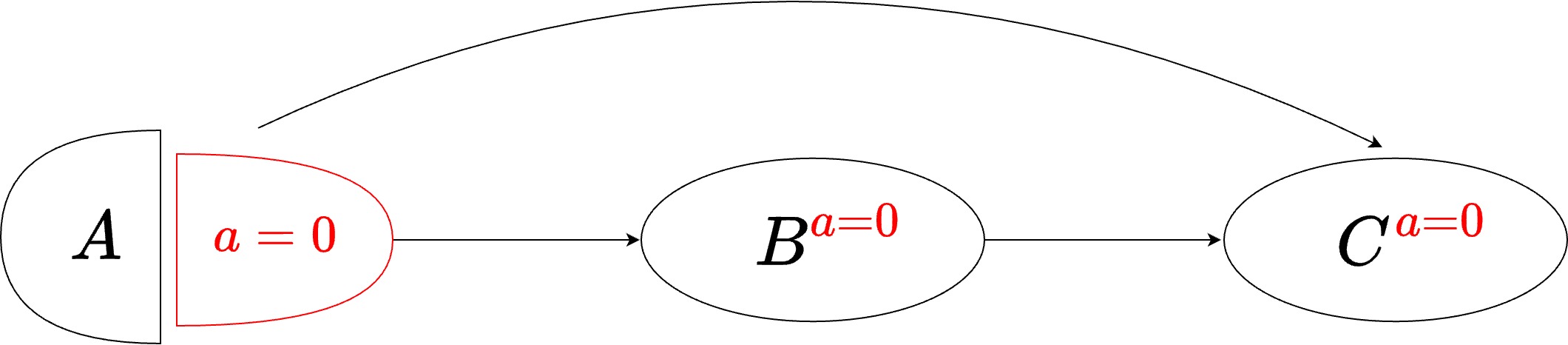} 
        \caption{Single world intervention graph $\mathcal{G}(a=0)$}
        \label{fig:SWIGa0}
    \end{subfigure}
    \\[0.2cm]
    \caption{Single world interventions graphs}
    \label{fig:swigs_combined}
\end{figure}

Suppose we assume, or know from prior studies, that both interventions $a=1$ and $a=0$ have a direct effect on $C$, as well as an indirect effect on $C$ through $B$. We can visualize this causal structure using the SWIGs illustrated in Figure (\ref{fig:swigs_combined}). Note that, for the purpose of defining the target parameter estimands, we focus exclusively on causal paths \cite{dijk2025directed}, which we define as directed paths from the intervention to the outcome of interest (i.e. $A \to \cdots \to C$). While the causal pathways in this example could have been represented by a single world intervention template (SWIT), we generally prefer to denote the SWIG per level of the intervention as this may provide more detailed information specific to each intervention (i.e. \textit{world}).

According to the assumed causal structures shown in Figure (\ref{fig:swigs_combined}), we factorize the joint distributions of the potential health outcomes, $P_{\mathcal{M}}(A, B^{a=1}, C^{a=1})$ and  $P_{\mathcal{M}}(A, B^{a=0}, C^{a=0})$, as detailed in Equations (\ref{EQ8.1}) and (\ref{EQ9.1}), respectively. Under the principle of truncated factorization, the term $P(A)$ can be omitted from both sides of each expression, since $(B^a,C^a \independent A)$ \cite{pearl2009causal,richardson2013single}. As a result, the joint distribution of the potential health outcomes under intervention $a$, $P_{\mathcal{M}}(B^a, C^a)$ in Equation (\ref{EQ:ICER_example2}), can be expressed as the product \( P_{\mathcal{M}}(C^a \mid B^a)\, P_{\mathcal{M}}(B^a) \), following the causal factorization implied by the assumed causal structure.

\begin{align} 
     \mathcal{G}(a=1): \quad P_{\mathcal{M}}(A,B^{a=1},C^{a=1}) \  &= \  P_{\mathcal{M}}(C^{a=1} \mid B^{a=1})P_{\mathcal{M}}(B^{a=1}) P_{\mathcal{M}}(A) \label{EQ8.1}\\
     \mathcal{G}(a=0): \quad P_{\mathcal{M}}(A,B^{a=0},C^{a=0}) \  &= \  P_{\mathcal{M}}(C^{a=0} \mid B^{a=0})P_{\mathcal{M}}(B^{a=0}) P_{\mathcal{M}}(A)  \label{EQ9.1}
\end{align}

\begin{figure}[h!]
    \centering
    \includegraphics[width=0.75\textwidth]{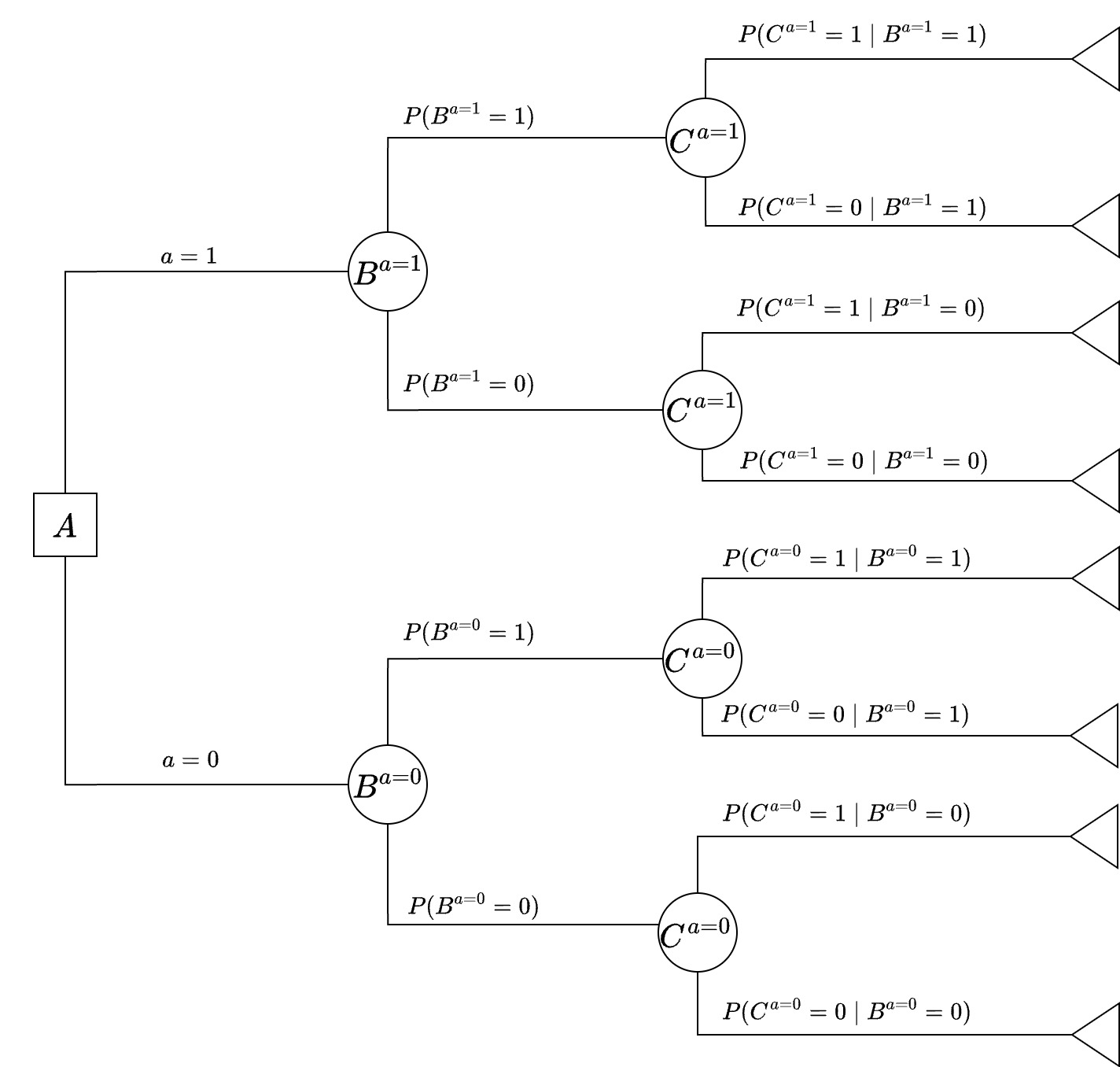}
    \caption{Target decision-analytical model, decision tree with causal probability input parameter estimands, with the square representing a decision node ($A$), circles representing chance nodes ($B$ and $C$), and triangles denoting terminal nodes.}
    \label{fig:C_decisionT}
\end{figure}

\noindent
Accordingly, based on the preceding factorization, we define the collection of all target probability parameters corresponding to the decision-analytical model as $\mathcal{P}_{\mathcal{M}}=\{P(B^{a})=b, P(C^a=c \mid B^a=b) : a,b,c\in\{0,1\}\}$. This set can be directly mapped onto the structure of decision-analytical model $\mathcal{M}$, as illustrated in Figure (\ref{fig:C_decisionT}), thereby defining our target decision-analytical model. This highlights that each chance node requires the identification of a corresponding causal probability parameter.

\begin{figure}[h!]
    \centering
    \includegraphics[width=0.35\textwidth]{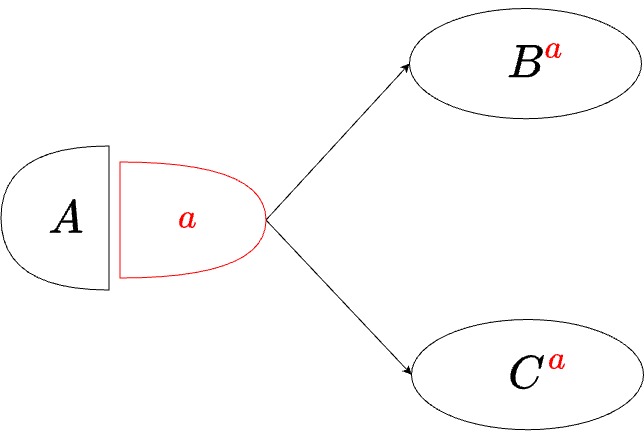}
    \caption{Alternative causal structure that may be consistent with Figure (\ref{fig1})}
    \label{fig:swig_voorbeeld}
\end{figure}

Although factorizing the joint distributions of the potential health outcomes is, in principle, merely a re-expression of the same distribution into components, it becomes particularly valuable when linking data to the target decision-analytical model, that is, within the context of the empirical decision-analytical model. Because the joint distribution of potential health outcomes rarely has directly empirically observed analogs, factorization allows us to work with lower-dimensional components that are more likely to have observed counterparts. Importantly, the target estimand together with the assumed underlying causal structure dictates the causal factorization and thus shapes the decision-analytical model, not the other way around. In fact, a decision tree model presented without accompanying causal context can be consistent with different underlying causal structures. For example, Figure (\ref{fig1}) could also be consistent with a causal structure in which there is no direct causal pathway between $B$ and $C$, as in Figure (\ref{fig:swig_voorbeeld}).

\vspace{0.1cm}

 This highlights an important point: the structure and corresponding target parameter estimands of a decision-analytical model are inherently dependent on the assumed causal structure. It is therefore encouraged to supplement the decision-analytical model with a causal graph, as it helps clarify relationships among variables, particularly in more intricate settings \cite{dijk2025directed}. For instance, when considering three additional outcomes such that $\boldsymbol{V}^a=\{B^a,C^a,D^a,E^a,F^a\}$, it becomes challenging, without an accompanying causal graph, to clearly define the appropriate target parameter estimands and, consequently, to understand precisely what is being estimated.

Thus far, we have employed causal graphs to structure the target decision-analytical model by defining the relevant target parameter estimands. With the target model in place, as depicted in Figure (\ref{fig:C_decisionT}), the next step is to construct the empirical decision-analytical model by identifying these target parameter estimands. For this task, we again turn to causal graphs; however, they serve a different purpose. Rather than defining the target parameter estimands, causal graphs are now used for identification which requires considering both causal and non-causal paths, with the latter defined as any path between the intervention and the outcome of interest that is not a directed path from intervention to outcome (e.g. backdoor paths, collider paths). The following section highlights this important distinction and illustrates how causal graphs support different objectives within the decision-analytical modelling process. The section thereafter turns to the actual identification of the target parameters.

\subsubsection{Causal graphs for different aims}

Defining the target causal parameter estimands, as previously discussed, focuses exclusively on causal paths \cite{dijk2025directed}. In contrast, identification connects data to the target causal parameter estimands, necessitating causal assumptions and, therefore, the inclusion of non-causal paths. Separate causal graphs are needed for distinct target parameters when these are estimated from different data sources, because each source may involve a different data-generating mechanism affecting identification.

\begin{figure}[h!]
    \centering
    \includegraphics[width=0.6\textwidth]{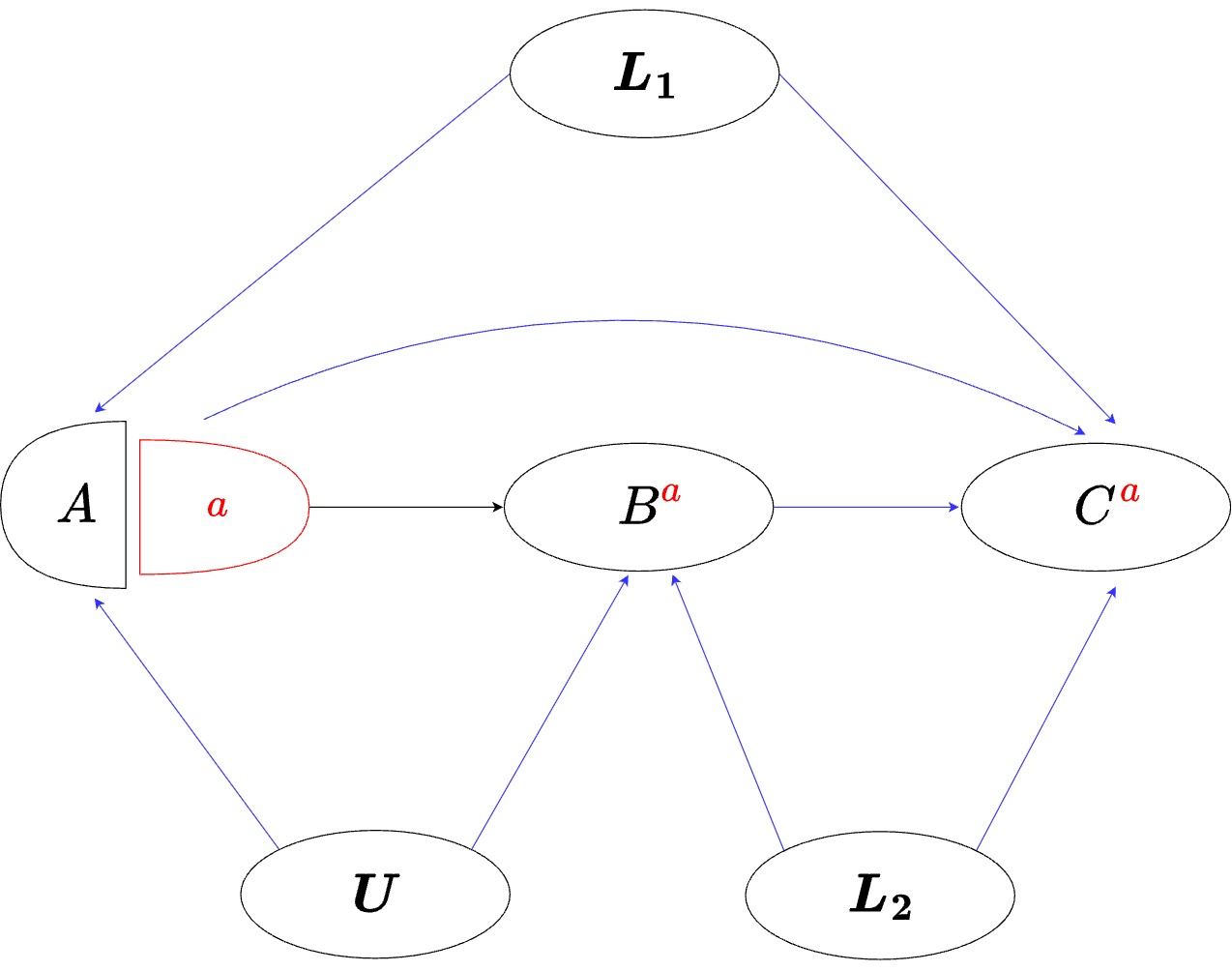}
    \caption{An incorrectly constructed SWIT in which different data generating mechanisms are combined into a single causal graph. The data used to estimate the impact of $A$ on $B$ comes from an ideal randomized controlled trial (relevant edge denoted in black), whereas the impact of $A$ on $C$ is estimated using distinct observational data (additionally requiring the inclusion of all blue edges). Here, $\boldsymbol{U}$ represents unmeasured confounders, and $\boldsymbol{L_1}$ and $\boldsymbol{L_2}$ denote measured ones.}
    \label{fig:aim_swig}
\end{figure}

Conceptually, consider Figure (\ref{fig:aim_swig}), which resembles the SWIGs in Figure (\ref{fig:swigs_combined}) but incorporates non-causal paths, illustrating that representing different data-generating mechanisms within a single SWIT is invalid. For instance, suppose that the impact of $A$ on $B$ is estimated from an ideal randomized controlled trial, with $\boldsymbol{U}$ representing unmeasured confounders, while the impact of $A$ on $C$ is estimated using different observational data that include measured confounders $\boldsymbol{L_1}$ and  $\boldsymbol{L_2}$. The fundamental issue with depicting these different data-generating mechanisms in a single causal graph is that we either randomize $A$ or we do not, leading to an apparent contradiction. When $A$ is randomized, incoming edges into $A$ (e.g. from $\boldsymbol{U}$) are no longer operative and are therefore omitted, such that only the black edge ($A\mid a \longrightarrow B^a$) remains relevant for the randomized controlled trial source. In contrast, estimating the impact of $A$ on $C$ using observational data additionally requires the inclusion of all blue edges. Importantly, this distinction directly governs the conditions under which the target parameters can be unbiasedly identified. More generally, this example highlights the need for separate causal graphs when different data sources are used to estimate distinct target parameters, in order to appropriately represent the data-generating mechanism relevant to each parameter.

\subsection{The internal and external validity components}

Having specified the target decision-analytical model, the empirical decision-analytical model is constructed by identifying and estimating the target parameters from available data. For a general decision-analytical model $\mathcal{M}$ with $N$ target input probability parameters, such that $\mathcal{P_M}=\{P_k : k=1,\dots,N\}$, we define $\boldsymbol{P}_{\mathcal{M}} \subseteq \mathcal{P_M}$ as the minimal subset of target parameters that require direct estimation, where $|\boldsymbol{P}_{\mathcal{M}}|=n$ and $n\leq N$. While the notation introduced so far accommodates settings in which multiple sources can be integrated to estimate a single parameter, as in causal meta-analytic contexts, such settings involve additional methodological considerations that must be taken into account (see, e.g., \cite{dahabreh2020toward,dahabreh2023efficient}). Here, we restrict attention to the case in which each parameter $P_k$ is estimated from a single source, so that $\mathcal{S}_k$ contains exactly one element for all $k$.

In this setting, let $\boldsymbol{P}_{\mathcal{M} \mid \boldsymbol{S}}=\{P_{k \mid\mathcal{S}_k}: k=1,\dots,n\}$ denote the sample-specific target parameter estimands, which correspond to those in $\boldsymbol{P}_{\mathcal{M}}$ but are defined with respect to the sample-specific context $\mathcal{S}_k$ (e.g. source-specific population). The corresponding estimates, $\hat{\boldsymbol{P}}_{\mathcal{M} \mid \boldsymbol{S}}=\{\hat{P}_{k \mid\mathcal{S}_k}: k=1,\dots,n\}$, allow us to formalize internal validity as the agreement between the sample-specific targets and the expected values of their estimates. That is, internal validity holds when $\displaystyle\mathop{\mathbb{E}}[\hat{P}_{k \mid \mathcal{S}_k}]=P_{k \mid \mathcal{S}_k}$ for all $k$. External validity requires that each sample-specific estimate is an unbiased estimate of its corresponding target quantity in $\boldsymbol{P}_{\mathcal{M}}$. Sample-specific estimates may require some form of adjustment to account for differences between the source and target context, which, as previously noted, refers to any external validity dimension (e.g. population, intervention, time \cite{findley2021external}) deemed relevant that could meaningfully change the estimate. Accordingly, we define $\hat{\boldsymbol{P}}_{\mathcal{M}}=\{\hat{P}_{k}: k=1,\dots,n\}$ as the set of sample-specific estimates that have been adjusted to the target context. Dropping the subscript $\mathcal{S}_k$ emphasizes that each $\hat{P}_k$ now corresponds to the target context rather than the original source-specific context $\mathcal{S}_k$, with adjustments made for the external validity dimensions relevant to each parameter $k$ given the data. Crucially, the sample-specific estimates $\hat{P}_{k \mid \mathcal{S}_k}$ must be adjusted so that each resulting estimate $\hat{P}_k$ provides an unbiased estimate of its corresponding target parameter $P_k$, that is, external validity holds if $\displaystyle\mathop{\mathbb{E}}[\hat{P}_{k}]=P_{k}$ for all $k$. For example, inverse probability of selection weighting (IPSW) has been proposed to adjust for differences in distribution of covariates between the study sample and target population, so that the estimated causal parameter is representative of the target population \cite{degtiar2023review}.

\subsubsection{Causal synthesis}

We may view the task of ensuring that each parameter is target valid as a form of \textit{causal synthesis}. This reflects the objective of combining parameters estimated from different source contexts such that, when integrated within $\mathcal{M}$, they approximate a single data-generating process that is representative of the target context. This can be seen more clearly in Figure (7), which schematically summarizes the estimation process for all relevant probability parameters and highlights the stages at which assumptions are invoked.

\begin{center}
    \begin{figure}[h!]
    \centering
    \includegraphics[width=0.65\textwidth]{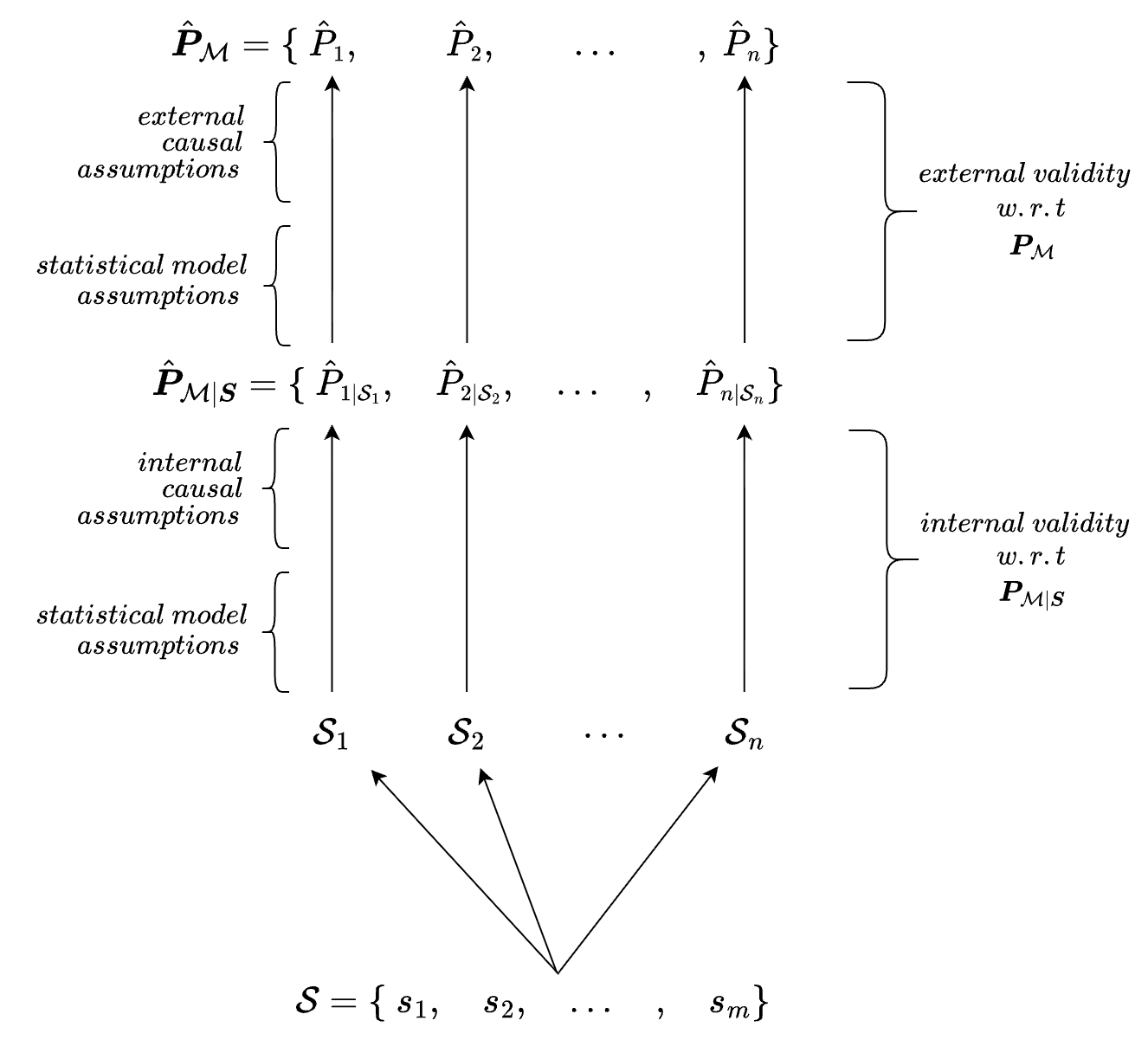}
    \caption{From data to target parameter estimates}
    \label{fig:P}
\end{figure}
\end{center}

The availability of data relevant to each target parameter, $\mathcal{S}_k$, constitutes the first and foremost consideration in the identification process. Both the nature of the available data and the processes by which it was generated constrain the set of feasible identification strategies. The identification strategy employed, in turn, determines the corresponding set of causal assumptions (i.e. identification conditions), which vary across strategies. For example, $\hat{P}_{1 \mid\mathcal{S}_1}$ may be estimated using an instrumental variables design \cite{hernan2020causal}, $\hat{P}_{2 \mid \mathcal{S}_2}$ using a target trial emulation–based design \cite{hernan2022target}, and $\hat{P}_{3 \mid \mathcal{S}_3}$ estimated using a difference-in-differences approach \cite{quasiexperimental}. In all cases, the plausibility of the required assumptions must be carefully assessed and explicitly justified. In addition to the causal assumptions, we further rely on the correct specification of the statistical models used to estimate the parameters, which we refer to as \textit{statistical model assumptions}, as well as the absence of measurement bias in the relevant variables in each data source consulted \cite{hernan2020causal}.

Even if each parameter is internally valid, combining them into a single model $\mathcal{M}$ may be ill-defined when the source contexts underlying the parameters differ from one another or from the intended target context in which the decision-analytical model is applied, a challenge central to decision-analytical modelling. Linking sample-specific estimates to the target therefore requires causal external validity assumptions, shaped both by differences between source and target contexts and by the identification strategies employed for ensuring internal validity \cite{bareinboim2013causal, bertanha2020external, degtiar2023review}. Most work on external validity, however, has focused on population differences, implicitly assuming equivalence along all other dimensions, with methodological developments reflecting this emphasis \cite{degtiar2023review}. An important consideration in this context is whether the sample is a subset of the target population: if so, the task concerns generalization, addressing bias from non-representative sampling; if not, it concerns transportability, a more challenging task as it requires accounting for differences in both distributions and causal mechanisms \cite{pearl2022external, correa2019adjustment}. In either case, beyond the causal assumptions, it is assumed that the statistical models used to adjust the sample estimates are well-specified (i.e. again referred to as statistical model assumptions), and the absence of measurement bias in the target population data when such data are used. The target population data are not explicitly represented in the notation, as they serve only to inform the weights. Their contribution is instead implicitly reflected in the adjustment of the sample-specific estimates, with the additional assumption that these data are free from measurement bias.

While we have emphasized the importance of the causal assumptions for both internal and external validity, upon which the identification of each parameter depends, we have not yet provided a detailed substantive discussion of those assumptions. The reason is that the assumptions required for each parameter will be context specific, particularly with respect to the structure and availability of data, which makes an exhaustive treatment of all possible scenarios infeasible. Instead, Figure (7) provides a general outline, emphasizing that establishing target validity requires an approach tailored to each individual parameter. Nonetheless, to provide a concrete illustration, we revisit the earlier decision tree example and its accompanying target decision-analytical model depicted in Figure (4), highlighting the substantive role of causal assumptions and illustrating how even under simplified conditions, the availability of data can critically shape these assumptions.

\subsubsection{Identification: revisiting the decision tree example}

The ability to reconstruct the target decision-analytical model in Figure (\ref{fig:C_decisionT}), primarily depends on the available data. To demonstrate this, we examine three scenarios of data availability, distinguishing them by the combination of data sources involved: (i) a reference scenario in which a single data source contains all necessary information; (ii) a setting where the intervention contrast of interest is not available from a single source; and (iii) a setting where outcome data along the intervention are not jointly observed within a single source. In particular, we consider the following cases:

\begin{enumerate}
    \item[(i)] Data on $A$, $B$, and $C$ from a single source, $\mathcal{S}=\{s_1\}$;
    \item[(ii)] Data from source $s_2$ on $A=1$, $B$, and $C$, and from source $s_3$ on $A=0$, $B$, and $C$, $\mathcal{S}=\{s_2,s_3\}$;
    \item[(iii)] Data on $A$ and $B$ from source $s_4$ and on $B$ and $C$ from source $s_5$, $\mathcal{S}=\{s_4,s_5\}$.
\end{enumerate}
\vspace{0.1cm}

\noindent
Irrespective of the available data, the goal is to identify and estimate the following minimal subset of target parameters, which suffices to recover the full set of estimands $\mathcal{P_M}$ corresponding to Figure (4):

\[
\boldsymbol{P}_{\mathcal{M}} = \left\{ P(B^a = 1),\ P(C^a = 1 \mid B^a = b) \;  : \;  a, b \in \{0,1\} \right\}
\]

\noindent
For the purpose of this illustration, we consider each data source, whether experimental or observational, to be a non-random sample drawn from the target population. We assume that each target parameter can be identified via the backdoor adjustment strategy to ensure internal validity, and that external validity can be addressed by recalibrating for differences in the distributions of observed covariates between the sample and target populations \cite{bareinboim2016causal}.

Specifically, the baseline covariates necessary for confounding adjustment and for external validity, those whose distributions differ between the sample and target population and which influence the potential outcomes of interest, are assumed to be discrete and serve both roles simultaneously, denoted $\boldsymbol{L} \in \mathbb{Z}^q$. It is further assumed that there is complete adherence and no missing data. Under these simplifications, the challenges under consideration arise primarily from the structure and availability of the data; in more intrincate scenarios, however, the same reasoning continues to apply, although additional complexities must be addressed.

\vspace{0.5cm}
\noindent
\textbf{Scenario (i)} \\
\noindent
Since all relevant information is available from a single source, $s_1$, we define the corresponding sample-specific target parameters as follows:
\[
\boldsymbol{P}_{\mathcal{M} \mid \boldsymbol{S}} = 
\left\{ 
P(B^a = 1 \mid S_1=1),\ 
P(C^a = 1 \mid B^a = b, S_1=1) \; : \; a, b \in \{0,1\} 
\right\}
\]

\noindent
These estimands are then evaluated using the observed data in $s_1$, resulting in the following estimates:
\[
\hat{\boldsymbol{P}}_{\mathcal{M} \mid \boldsymbol{S}} = \left\{ \hat{P}(B = 1 \mid A=a, \boldsymbol{L}=\boldsymbol{l}, S_1=1),\ \hat{P}(C = 1 \mid A=a,B = b, \boldsymbol{L}=\boldsymbol{l}, S_1=1) \;  : \;  a, b \in \{0,1\}, \boldsymbol{l} \right\}
\]

\noindent
To unbiasedly link each estimate in $\hat{\boldsymbol{P}}_{\mathcal{M} \mid \boldsymbol{S}}$ to its corresponding sample-specific target in $\boldsymbol{P}_{\mathcal{M} \mid \boldsymbol{S}}$, we rely on causal assumptions \textbf{(A.1)} - \textbf{(A.3)}, along with the conditions that the data in $s_1$ are not subject to measurement bias and that the statistical models used for estimation are well-specified. For notational convenience, we collectively refer to these two conditions as assumption \textbf{(A.4)}. Note that correct model specification entails both selecting an appropriate model class (e.g. the correct link function) and correctly modelling the relationships between variables (e.g. including all relevant statistical interactions and non-linearities), which can be both verified using the data.

Concerning the causal assumptions, conditional exchangeability \textbf{(A.1)} requires that, within source $s_1$, and conditional on $\boldsymbol{L}$, the outcomes observed among individuals who received $A=a$ are representative of the potential outcomes that would be observed if all individuals in $s_1$ had been assigned $A=a$. Intuitively, we assume that the only systematic difference between individuals receiving different intervention levels is the intervention itself, after adjusting for baseline covariates $\boldsymbol{L}$. As $\boldsymbol{L}$ is defined over observed variables, conditional exchangeability inherently requires that there are no unmeasured confounders \cite{hernan2020causal}. Moreover, if covariate profile $\boldsymbol{L}=\boldsymbol{l}$ is observable within $s_1$, we assume that there is a probability higher than zero of receiving each intervention level. Note that this concerns structural positivity, not stochastic positivity \cite{zivich2022positivity}. Even if \textbf{(A.2)} holds, random violations may occur and must be appropriately addressed by restricting to the region of overlap (i.e. altering the estimand), or by relying on additional extrapolation assumptions \cite{petersen2012diagnosing}. The Stable Unit Treatment Value Assumption (SUTVA) \textbf{(A.3)} entails three conditions: consistency, meaning that for each individual in $s_1$ the observed outcomes equal their potential outcomes; no interference, implying one individual’s treatment does not affect another’s outcome; and no multiple versions of treatment, ensuring each intervention level is well-defined \cite{vanderweele2009concerning}.

\begin{align}
\textbf{(A.1)}    \quad & \text{Conditional Exchangeability} \quad &&  (B^a,C^a) \independent A \mid \boldsymbol{L}=\boldsymbol{l}, S_1=1 \quad \forall a
 \nonumber \\
 & \text{of Treatment Assignment} && \nonumber \\[0.2cm]
\textbf{(A.2)}   \quad & \text{Positivity of Treatment} \quad && P(\boldsymbol{L}=\boldsymbol{l} \mid S_1=1)>0 \implies \forall a \  P(A=a \mid \boldsymbol{L}=\boldsymbol{l}, S_1=1) >0  \nonumber \\
& \text{Assignment} &&  \text{with probability 1} \nonumber \\[0.2cm]
\textbf{(A.3)}   \quad & \text{Stable Unit Treatment} \quad && \text{if } A=a \text{ and } S_1=1, \text{ then } (B^a,C^a)=(B,C)  \nonumber \\
& \text{Value Assumption} && \nonumber \\
& \text{for Treatment Assignment} && \nonumber
\end{align}

\noindent
To make explicit how assumptions \textbf{(A.1)}–\textbf{(A.3)} permit identification, we consider the target parameter $P(B^{a=1}=1 \mid S_1=1)$ as an illustrative example and apply these assumptions successively, as shown in Equation (\ref{EQ:identification1}). This demonstrates that the potential-outcome probability can be expressed in terms of observable quantities. In particular, standardization is applied to marginalize over the confounders in $\boldsymbol{L}$, resulting in the corresponding marginal target quantity. In practice, the observable quantities in the final expression of Equation (\ref{EQ:identification1}) must be estimated from data, thereby moving from identification to estimation and invoking assumption $\boldsymbol{(A.4)}$.

\begin{align} \label{EQ:identification1}
  P(B^{a=1}=1 \mid S_1=1) 
  =&\sum_{\boldsymbol{l}} P(B^{a=1}=1 \mid \boldsymbol{L}=\boldsymbol{l}, S_1=1) P(\boldsymbol{L}=\boldsymbol{l} \mid S_1=1)  \\[0.1cm] 
  =& \sum_{\boldsymbol{l}} P(B^{a=1}=1 \mid A=1, \boldsymbol{L}=\boldsymbol{l}, S_1=1) P(\boldsymbol{L}=\boldsymbol{l} \mid S_1=1) 
  \quad \text{by} \quad \textbf{(A.1, A.2)} \nonumber \\[0.1cm]
  =& \sum_{\boldsymbol{l}} P(B=1 \mid A=1, \boldsymbol{L}=\boldsymbol{l}, S_1=1) P(\boldsymbol{L}=\boldsymbol{l} \mid S_1=1) 
  \qquad \; \  \text{by} \quad \textbf{(A.3)} \nonumber
\end{align}

Beyond internal validity considerations, the generalizability of each sample-specific quantity to the target context must be assessed, recalling that $S_1=1$ represents a subset of the target population. Under the simplifying assumption introduced above, that the only difference between the source and target contexts lies in the distribution of $\boldsymbol{L}$, sample-specific marginal probabilities may be externally biased relative to the corresponding target estimands. For example, if the proportion of males is higher in the sample than in the target population and outcomes depend on sex, then $P(B^{a=1} \mid S_1=1)$ differs from target $P(B^{a=1}=1)$. To address this bias, each sample-specific conditional probability can be standardized (or equivalently reweighted) to the distribution of $\boldsymbol{L}$ in the target population, denoted by $P(\boldsymbol{L=\boldsymbol{l}})$. However, the validity of this procedure depends not only on internally valid quantities, but also on the credibility of the causal external validity assumptions \textbf{(A.5)}–\textbf{(A.7)} \cite{hernan2020causal, degtiar2023review}.

Specifically, conditional exchangeability of study selection \textbf{(A.5)}, posits that, given the same intervention level $a$ and covariate profile $\boldsymbol{L}=\boldsymbol{l}$, the outcomes in the source and target populations are comparable in distribution. Since $\boldsymbol{L}$ represents observed covariates, this entails the absence of unmeasured covariates that both influence the outcomes of interest and differ in distribution between the sample and target populations. Therefore, this assumption holds if all relevant covariates are observed in $\boldsymbol{L}$, or, alternatively, if $s_1$ is a random sample from the target population. Positivity of selection \textbf{(A.6)} requires that within every stratum of $\boldsymbol{L}$, each unit has a positive probability of being included in the source or being represented by the study participants in the source. This ensures that all subgroups in the target population are represented in the source, though selection may be preferential. Such preferential selection constitutes an issue of generalizability rather than transportability, as transportability would violate positivity of selection \cite{lesko2017generalizing}. SUTVA for selection \textbf{(A.7)} requires treatment variation irrelevance and the absence of interference between individuals in the source and those in the target population. It further assumes that outcomes are measured equivalently across settings, the intervention is implemented consistently, and the underlying data-generating mechanism is common to both source and target populations \cite{degtiar2023review, westreich2019target}.

\begin{align}
\textbf{(A.5)}    \quad & \text{Conditional Exchangeability} \quad &&  (B^a,C^a) \independent S_1 \mid \boldsymbol{L}=\boldsymbol{l} \quad \forall a
 \nonumber \\
 & \text{for Selection} && \nonumber \\[0.2cm]
 \textbf{(A.6)}   \quad & \text{Positivity of Selection} \quad && P(\boldsymbol{L}=\boldsymbol{l})>0 \implies P(S_1=1 \mid \boldsymbol{L}=\boldsymbol{l}) >0  \nonumber \\
&  &&  \text{with probability 1} \nonumber \\[0.2cm]
\textbf{(A.7)}   \quad & \text{Stable Unit Treatment} \quad && \text{if } A=a \text{ and } S_1=s \ \text{ then } (B^a,C^a)=(B,C)  \nonumber \\
& \text{Value Assumption} && \nonumber \\
& \text{for Selection} && \nonumber
\end{align}

\noindent
Extending the previous illustrative example on internal validity, Equation (\ref{EQ:identification2}) demonstrates the role of external validity assumptions \textbf{(A.5)}–\textbf{(A.7)} in identifying the target parameter $P(B^{a=1}=1)$, based on the internally valid sample-specific conditional probability. Consistent with the previous example, we apply standardization; however, each sample-specific probability is now weighted according to the target distribution of $\boldsymbol{L}$, with the target population quantity indicated by the absence of conditioning on $S_1=1$ \cite{westreich2019target}.

\begin{align} \label{EQ:identification2}
P(B^{a=1}=1) =&\sum_{\boldsymbol{l}} P(B^{a=1}=1 \mid \boldsymbol{L}=\boldsymbol{l}) P(\boldsymbol{L}=\boldsymbol{l}) \quad & \\[0.1cm] 
    =&\sum_{\boldsymbol{l}} P(B^{a=1}=1 \mid \boldsymbol{L}=\boldsymbol{l}, S_1=1) P(\boldsymbol{L}=\boldsymbol{l}) \qquad \qquad \text{by} \quad \textbf{(A.4-A.6)} \nonumber \\[0.1cm]
    =&\sum_{\boldsymbol{l}} P(B=1 \mid A=1,\boldsymbol{L}=\boldsymbol{l}, S_1=1) P(\boldsymbol{L}=\boldsymbol{l})\qquad \; \, \text{by} \quad \textbf{(A.1-A.3)}\nonumber
\end{align}

\noindent
The subsequent step involves estimating these observable quantities. As noted previously, the sample-specific probabilities are estimated from data $s_1$ under assumptions \textbf{(A.1)}–\textbf{(A.4)}, whereas the target probabilities, $P(\boldsymbol{L=l})$, are typically informed by an alternative source, if available. Accordingly, we must also assume that the data used to estimate the target weights are not subject to measurement bias and that the corresponding statistical model used for estimating those weights is well-specified \cite{hernan2020causal}. For convenience, these conditions are referred to as assumption \textbf{(A.8)}. The resulting estimate of the final expression in Equation (16) is denoted as $\hat{P}(B^a=1)$, representing the marginal, sample-specific adjusted quantity, which, together with all other relevant estimates, is captured in $\hat{\boldsymbol{P}}_{\mathcal{M}}$:

\[
\hat{\boldsymbol{P}}_{\mathcal{M}} = \left\{ \hat{P}(B^a = 1),\ \hat{P}(C^a = 1 \mid B^a = b) \;  : \;  a, b \in \{0,1\} \right\}
\]

\noindent
Altogether, under assumptions \textbf{(A.1)}–\textbf{(A.8)}, the estimates in $\hat{\boldsymbol{P}}_{\mathcal{M}}$ are target-valid, meaning that the expected value of each estimate equals its corresponding target estimand in $\boldsymbol{P}_{\mathcal{M}}$, for example, $\displaystyle\mathop{\mathbb{E}}[\hat{P}(B^a=1)]=P(B^{a=1}=1)$. It is worth noting, however, that weaker versions of the causal assumptions may be applicable. For instance, assumption \textbf{(A.1)} can be replaced by mean conditional exchangeability \cite{dahabreh2019generalizing}, and assumption \textbf{(A.7)} may be replaced by the assumption that the distribution of treatment versions is equivalent between the sample and target population \cite{lesko2017generalizing}.

Thus far, we have considered a scenario in which all required data are available from a single source,
which, in practice, is rarely the case. The analyst must instead rely on multiple data sources, each providing only partial information on the quantities of interest. While such settings generally require careful consideration, particularly when the objective is to recover the joint distribution $P(B^a,C^a)$ under each $a$, as will be illustrated in scenarios (ii) and (iii), it may sometimes suffice to identify the marginal distributions $P(B^a)$ and $P(C^a)$. For example, if we believe that:

\[f(B^a,C^a)= f_1(B^a) + f_2(C^a) \quad \text{and} \quad g(B^a,C^a)= g_1(B^a) + g_2(C^a), \]

\noindent
so that 

\[\displaystyle\mathbb{E}[f(B^a,C^a)]= \displaystyle\mathbb{E}[f_1(B^a)] + \displaystyle\mathbb{E}[f_2(C^a)], \quad \displaystyle\mathbb{E}[g(B^a,C^a)]= \displaystyle\mathbb{E}[g_1(B^a)] + \displaystyle\mathbb{E}[g_2(C^a)],\]

\noindent
we could identify the potential outcomes under appropriately adapted versions of \textbf{(A.1)}–\textbf{(A.8)}, defined separately for each outcome and with respect to its corresponding data source. Specifially, $B^a$ could be identified from $P(A,B,\boldsymbol{L}_0\mid S_0=1)$ and $C^a$ from $P(A,C,\boldsymbol{L}_1\mid S_1=1)$, where $S_0$ and $S_1$ denote the respective sources and $\boldsymbol{L}_0$ and $\boldsymbol{L}_1$ the relevant covariates. For example, \textit{conditional exchangeability of treatment assignment} with respect to $B^a$ would be expressed as: $B^a \independent A \mid \boldsymbol{L}_0, S_0=1 \quad \forall a$, with analogous adaptations for the remaining assumptions and for the outcome $C^a$ relative to source $S_1$.

\vspace{0.5cm}
\noindent
\textbf{Scenario (ii)} \\
\noindent
In this setting, each sample-specific estimand is defined relative to the data source corresponding to the intervention level of interest: outcomes under intervention $A=1$ are informed by source $s_2$, whereas outcomes under $A=0$ are informed by source $s_3$. We suppose that, for the intervention levels of interest, each level is observed only in its corresponding source (i.e. $A=1$ in $s_2$ and $A=0$ in $s_3$, although other intervention levels may be available across sources. Consequently, the set of sample-specific estimands is defined as follows:

\begin{align*}
\boldsymbol{P}_{\mathcal{M} \mid \boldsymbol{S}} = \Big\{ \; &
P(B^{a=1} = 1 \mid S_2=1), P(C^{a=1} = 1 \mid B^{a=1} = b, S_2=1), \\
& P(B^{a=0} = 1 \mid S_3=1), P(C^{a=0} = 1 \mid B^{a=0} = b, S_3=1) \; : \; b \in \{0,1\} \; \Big\}
\end{align*}

\noindent
Each estimand is subsequently estimated from the observed data in their respective source, $s_2$ or $s_3$. The baseline covariates, previously denoted by $\boldsymbol{L}$, are now defined separately for each source as $\boldsymbol{L}_2$ and $\boldsymbol{L}_3$, which are not necessarily the same. It follows that the set of sample-specific estimates can be expressed as:

\begin{align*}
\hat{\boldsymbol{P}}_{\mathcal{M} \mid \boldsymbol{S}} = \Big\{ \; &
\hat{P}(B = 1 \mid A=1, \boldsymbol{L}_2=\boldsymbol{l}_2, S_2=1), \hat{P}(C = 1 \mid B = b, A=1, \boldsymbol{L}_2=\boldsymbol{l}_2, S_2=1), \\
& \hat{P}(B = 1 \mid A=0, \boldsymbol{L}_3=\boldsymbol{l}_3, S_3=1),\hat{P}(C = 1 \mid B = b, A=0, \boldsymbol{L}_3=\boldsymbol{l}_3, S_3=1) \; : \; 
b \in \{0,1\}, \boldsymbol{l}_2,\boldsymbol{l}_3 \; \Big\}
\end{align*}

\noindent
To ensure target validity, each sample-specific estimate is linked to its corresponding sample-specific target under internal and external validity assumptions analogous to those discussed previously, now defined relative to each source. However, as we are interested in a single intervention level within each source, rather than the source-specific contrast, it suffices that the causal assumptions hold only for the intervention level of interest.

Specifically, with respect to internal validity, it is sufficient, though not in general necessary, to invoke partial conditional exchangeability assumptions \textbf{(A.1a)}–\textbf{(A.1b)} within each source, rather than full exchangeability \cite{sarvet2020graphical}. In particular, when restricting to a single-arm (i.e. $A=a$), partial exchangeability holds trivially, yet it is nonetheless formally invoked to justify replacing the potential outcome with its corresponding observed outcome. Furthermore, because $A$ is binary, referring to $A$ in the independence statement is equivalent to explicitly specifying the treatment level $A = a$; while this equivalence no longer holds when $A$ is multi-level, in which case the relevant level $a$ must be explicitly specified. Analogously, partial positivity of treatment assignment, \textbf{(A.2a)}–\textbf{(A.2b)}, is trivial in this case because $\boldsymbol{L}_2$ and $\boldsymbol{L}_3$ are empty. In contrast, the Stable Unit Treatment Value Assumptions, \textbf{(A.3a)}–\textbf{(A.3b)}, are necessary and reformulated accordingly.

\begin{align}
\textbf{(A.1a)}    \quad & \text{Partial Conditional Exchangeability} \quad &&  (B^{a=1},C^{a=1}) \independent A \mid \boldsymbol{L}_2=\boldsymbol{l}_2, S_2=1,
 \nonumber \\
\textbf{(A.1b)}    \quad & \text{of Treatment Assignment} && (B^{a=0},C^{a=0}) \independent A \mid \boldsymbol{L}_3=\boldsymbol{l}_3, S_3=1 \nonumber \\[0.2cm]
\textbf{(A.2a)}   \quad & \text{Partial Positivity of Treatment} \quad && P(\boldsymbol{L}_2=\boldsymbol{l}_2 \mid S_2=1)>0 \implies \  P(A=1 \mid \boldsymbol{L}_2=\boldsymbol{l}_2, S_2=1) >0  \nonumber \\
& \text{Assignment} &&  \text{with probability 1} \nonumber \\
\textbf{(A.2b)}   \quad & && P(\boldsymbol{L}_3=\boldsymbol{l}_3 \mid S_3=1)>0 \implies \  P(A=0 \mid \boldsymbol{L}_3=\boldsymbol{l}_3, S_3=1) >0  \nonumber \\
& && \text{with probability 1} \nonumber \\[0.2cm]
\textbf{(A.3a)}   \quad & \text{Stable Unit Treatment} \quad && \text{if } A=1 \text{ and } S_2=1, \text{ then } (B^{a=1},C^{a=1})=(B,C)  \nonumber \\
\textbf{(A.3b)}   \quad  & \text{Value Assumption} && \text{if } A=0 \text{ and } S_3=1, \text{ then } (B^{a=0},C^{a=0})=(B,C) \nonumber \\
& \text{for Treatment Assignment} && \nonumber
\end{align}

\noindent
Following the same reasoning as demonstrated in Equation (\ref{EQ:identification1}), each potential outcome probability can be expressed in terms of observable quantities under assumptions \textbf{(A.1a)}–\textbf{(A.3b)}. These quantities are then estimated from the data available within their respective sources to obtain the corresponding sample-specific estimates. For each source, it is assumed that measurement bias is absent and that the employed statistical models are correctly specified, collectively referred to as assumption  \textbf{(A.4)}.

Akin to how the internal causal validity assumptions have been adapted, we adapt the external causal validity assumptions, as detailed in \textbf{(A.5a)–(A.7b)}. Note that generalizability must be assessed for each source separately with respect to the target, requiring that the assumptions hold across both sources for valid synthesis. The qualitative interpretation remains, in principle, unchanged, except that each assumption explicitly specifies the source–target relation to which it applies.

\begin{align}
\textbf{(A.5a)}    \quad & \text{Partial Conditional Exchangeability} \quad &&  (B^{a=1},C^{a=1}) \independent S_2 \mid \boldsymbol{L}_2=\boldsymbol{l}_2, 
 \nonumber \\
 \textbf{(A.5b)} \quad & \text{for Selection} && (B^{a=0},C^{a=0}) \independent S_3 \mid \boldsymbol{L}_3=\boldsymbol{l}_3 \nonumber \\[0.2cm]
 \textbf{(A.6a)}   \quad & \text{Partial Positivity of Selection} \quad && P(\boldsymbol{L}_2=\boldsymbol{l}_2)>0 \implies P(S_2=1 \mid \boldsymbol{L}_2=\boldsymbol{l}_2) >0  \nonumber \\
&  &&  \text{with probability 1} \nonumber \\
\textbf{(A.6b)}   \quad & && P(\boldsymbol{L}_3=\boldsymbol{l}_3)>0 \implies P(S_3=1 \mid \boldsymbol{L}_3=\boldsymbol{l}_3) >0 \nonumber \\
& && \text{with probability 1} \nonumber \\[0.2cm]
\textbf{(A.7a)}   \quad & \text{Stable Unit Treatment} \quad && \text{if } A=1 \text{ and } S_2=s \ \text{ then } (B^{a=1},C^{a=1})=(B,C)  \nonumber \\
\textbf{(A.7b)}   \quad & \text{Value Assumption} && \text{if } A=0 \text{ and } S_3=s \ \text{ then } (B^{a=0},C^{a=0})=(B,C)  \nonumber \\
& \text{for Selection} && \nonumber
\end{align}

\noindent
Provided that the sample-specific conditional probability estimates are internally valid and that the causal external validity assumptions are satisfied, attention turns to the estimation of the corresponding target weights. For simplicity, we assume that a single target data source is available whose covariate set $\boldsymbol{L}$ encompasses all covariates from both sources, $s_2$ and $s_3$, so that $\boldsymbol{L} \supseteq \boldsymbol{L}_2 \cup \boldsymbol{L}_3$. In case where $\boldsymbol{L}_2=\boldsymbol{L_3}$, the target weights need only be estimated once, as the same weights can be used for reweighting each source. In contrast, when $\boldsymbol{L}_2 \neq \boldsymbol{L}_3$, the target weights must be estimated separately, based on the covariates relevant to each specific source. In the former case, this requires one additional modelling step, whereas in the latter case, two additional modelling steps are required, and thus relies on both models being well-specified. For ease of exposition, we group the requirement of no measurement bias in the target data together with the corresponding assumptions of having well-specified statistical models under assumption \textbf{(A.8)}. Consequently, having estimated the sample-specific quantities and adjusted them by their respective estimated target weights, we again obtain the marginal sample-specific adjusted estimates:

\[
\hat{\boldsymbol{P}}_{\mathcal{M}} = \left\{ \hat{P}(B^a = 1),\ \hat{P}(C^a = 1 \mid B^a = b) \;  : \;  a, b \in \{0,1\} \right\}
\]

\noindent
In summary, assumptions \textbf{(A.1a)}-\textbf{(A.8)} ensure equality in distribution such that;

\begin{align*}
P(B,C \mid \boldsymbol{L}_2&=\boldsymbol{l}_2, S_2=1,A=1) = P(B^{a=1}, C^{a=1} \mid \boldsymbol{L}_2 = \boldsymbol{l}_2, S_2=1, A=1) = P(B^{a=1}, C^{a=1} \mid \boldsymbol{L}_2=\boldsymbol{l}_2), \\[0.1cm]
P(B,C \mid \boldsymbol{L}_3&=\boldsymbol{l}_3, S_3=1,A=0) = P(B^{a=0}, C^{a=0} \mid \boldsymbol{L}_3 = \boldsymbol{l}_3, S_3=1, A=0) = P(B^{a=0}, C^{a=0} \mid \boldsymbol{L}_3=\boldsymbol{l}_3)
\end{align*}

\noindent
That is, the observed single-arm outcomes in each respective source correspond to the potential outcomes under the same treatment and are representative of the target population as if all individuals had received that same level of intervention. Consequently, the expected value of the estimates in $\boldsymbol{\hat{P}}_{\mathcal{M}}$ equal their corresponding target parameters in $\boldsymbol{P}_{\mathcal{M}}$, thereby ensuring target validity.

\vspace{0.5cm}
\noindent
\textbf{Scenario (iii)} \\
\noindent
In this setting, source $s_4$ provides data on $(A,B)$, whereas source $s_5$ on $(B, C)$. The goal is to identify and estimate $P(B^a)$ from $s_4$ and $P(C^a \mid B^a)$ from $s_5$, and thus define the sample-specific estimands conditional on each source as follows:

\[
\boldsymbol{P}_{\mathcal{M} \mid \boldsymbol{S}} = 
\left\{ 
P(B^a = 1 \mid S_4=1),\ 
P(C^a = 1 \mid B^a = b, S_5=1) \; : \; a, b \in \{0,1\} 
\right\}
\]

\noindent
These estimands are subsequently evaluated using the observed data from $S_4$ or $S_5$, with each source providing access to its own set of observed baseline covariates, denoted by $\boldsymbol{L}_4$ and $\boldsymbol{L}_5$. Here, $\boldsymbol{L}_4$ must capture those relevant for the impact of $A$ on $B^a$, whereas $\boldsymbol{L}_5$ must collectively capture those affecting the effect of $A$ on $C^a$ as well as those relevant for the relation between $B^a$ and $C^a$. Notably, in this scenario, the outcomes $B$ and $C$ have not been observed jointly along intervention $A$ within any single data source. Consequently, the quantities that can be estimated from the available data are given by:

\[
\hat{\boldsymbol{P}}_{\mathcal{M} \mid \boldsymbol{S}} = \left\{ \hat{P}(B = 1 \mid A=a, \boldsymbol{L}_4=\boldsymbol{l}_4, S_4=1),\ \hat{P}(C = 1 \mid B = b, \boldsymbol{L}_5=\boldsymbol{l}_5, S_5=1) \;  : \;  a, b \in \{0,1\}, \boldsymbol{l}_4, \boldsymbol{l}_5 \right\}
\]

\noindent
Identification of the target quantities for the outcome $B^a$ proceeds analogously to setting (i), using the observed data in $s_4$ under assumptions \textbf{(A.1)}-\textbf{(A.3)}, together with the absence of measurement bias and correct model specification collectively denoted as \textbf{(A.9)}.

 \begin{align}
\textbf{(A.1)}    \quad & \text{Conditional Exchangeability} \quad &&  B^a \independent A \mid \boldsymbol{L}_4=\boldsymbol{l}_4, S_4=1 \quad \forall a
 \nonumber \\
 & \text{of Treatment Assignment} && \nonumber \\[0.2cm]
\textbf{(A.2)}   \quad & \text{Positivity of Treatment} \quad && P(\boldsymbol{L}_4=\boldsymbol{l}_4 \mid S_4=1)>0 \implies \forall a \  P(A=a \mid \boldsymbol{L}_4=\boldsymbol{l}_4, S_4=1) >0  \nonumber \\
& \text{Assignment} &&  \text{with probability 1} \nonumber \\[0.2cm]
\textbf{(A.3)}   \quad & \text{Stable Unit Treatment} \quad && \text{if } A=a \text{ and } S_4=1, \text{ then } B^a=B  \nonumber \\
& \text{Value Assumption} && \nonumber \\
& \text{for Treatment Assignment} && \nonumber
\end{align}

\noindent
In contrast, the target quantities involving the potential outcome $C^a$ cannot be identified in the current setting without invoking additional strong assumptions. In particular, the challenge lies in relating estimands that depend on $a$, namely $P(C^a \mid B^a, S_5=1)$, to quantities that can be estimated from the available data in which $A$ does not appear, since no information on $A$ is observed in $s_5$. Fundamentally, due to the absence of $A$ in $s_5$, we can only proceed if we believe that the effect of $a$ on $C^a$ operates completely through $B^a$, formalized as in assumption \textbf{(A.4)}. Next, assumption \textbf{(A.5)} expresses that, within source $S_5$, conditional on the potential mediator strata $B^a$ and on the set of potential confounders $\boldsymbol{L}_5$ affecting both $A$ and $C^{B^a}$, the treatment assignment $A$ provides no further information about the potential outcome $C^{B^a}$. In essence, we assume that there are no unobserved confounders between $A$ and $C^{B^a}$.  To ensure that each level of the mediator $B^a$ can occur under every treatment level $a$, within every stratum of $\boldsymbol{L}_5$, we invoke the conditional positivity assumption \textbf{(A.6)}. The stable unit treatment value assumption consists of two parts. Assumption \textbf{(A.7a)} states that we assume the observed $B$ in source $S_5$ represents $B^a$ under intervention $A=a$; that is, we regard observed $B$ as if it were generated under intervention $A=a$, even though actual treatment assignment is unobserved in $S_5$. Assumption \textbf{(A.7b)} expresses that, following from \textbf{(A.7a)}, the observed outcomes $C$ in $S_5$ represent the counterfactual outcomes that would arise if $B$ were set to its natural value under $A=a$. Importantly, assumption \textbf{(A.5)} is defined with respect to counterfactual strata $B^a$ and outcomes $C^{B^a}$, however, this does not imply that the observed intervention $A$ is independent of outcomes $C$ conditional on $B$ \cite{hernan2020causal}. Specifically, if there exists a common cause $U$ of $B$ and $C$, then conditioning on $B$ opens up the collider path $A \rightarrow B \leftarrow U \rightarrow C$, thereby inducing collider bias. Therefore, we must also invoke assumption \textbf{(A.8)}, to ensure that we do not introduce collider bias when conditioning on observed $B$.

 \begin{align}
\textbf{(A.4)}    \quad & \text{Complete Mediation} \quad && \text{if } S_5=1, \text{ then } \forall a \  C^a = C^{B^a} \nonumber \\[0.2cm]
\textbf{(A.5)}   \quad & \text{Conditional Exchangeability of} \quad && C^{B^a} \independent A \mid B^a, \boldsymbol{L}_5=\boldsymbol{l}_5, S_5=1 \quad \forall a \nonumber \\
& \text{Treatment Assignment in $B^a$} &&  \nonumber \\[0.2cm]
\textbf{(A.6)}   \quad & \text{Conditional Positivity} \quad && P(\boldsymbol{L}_5=\boldsymbol{l}_5 \mid S_5=1)>0 \implies \forall a,b \  P(B^a=b \mid \boldsymbol{L}_5=\boldsymbol{l}_5, S_5=1) >0 \nonumber \\
& \text{of Counterfactual Strata $B^a$} && \text{with probability 1}  \nonumber \\[0.2cm]
\textbf{(A.7a)}   \quad & \text{Stable Unit Treatment} \quad && \text{if } A=a \text{ and } S_5=1, \text{ then } B^a=B  \nonumber \\
\textbf{(A.7b)}   \quad & \text{Value Assumption} && \text{if } B^a=B \text{ and } S_5=1, \text{ then } C^{B^a}=C \nonumber \\
& \text{for Treatment Assignment} && \nonumber \\[0.2cm]
\textbf{(A.8)}   \quad & \text{Conditional Independence Observables} \quad && C \independent A \mid B, \boldsymbol{L}_5=\boldsymbol{l}_5, S_5=1 \nonumber \\
&  &&  \nonumber
\end{align}

\noindent
To illustrate how assumptions \textbf{(A.4)}-\textbf{(A.8)} permit identification of quantities of the form $P(C^a \mid B^a,S_5=1)$ from observable quantities $P(C \mid B, S_5=1)$, we apply these assumptions step-by-step in Equation (17). This results in an expression that contains only quantities that are observed in source $s_5$, as presented in the final line of Equation (\ref{EQ:identification3}). This resulting expression must then be estimated from the data in $s_5$, invoking again assumption \textbf{(A.9)}, which presumes the absence of measurement bias and correct specification of the statistical model.

\begin{align} \label{EQ:identification3}
  &P(C^{a}=c \mid B^a=b, S_5=1)  \\[0.2cm]
  =& \sum_{\boldsymbol{l}_5} P(C^{B^{a}}=c \mid B^a=b, \boldsymbol{L}_5=\boldsymbol{l}_5, S_5=1) P(\boldsymbol{L}_5=\boldsymbol{l}_5 \mid B^a=b, S_5=1)  \quad &&\text{by} \quad \textbf{(A.4)}  \nonumber \\[0.2cm] 
  =& \sum_{\boldsymbol{l}_5} P(C^{B^{a}}=c \mid B^a=b, \boldsymbol{L}_5=\boldsymbol{l}_5, A=a, S_5=1) P(\boldsymbol{L}_5 = \boldsymbol{l}_5 \mid B^a=b, S_5=1) \quad &&\text{by} \quad \textbf{(A.5-A.6)} \nonumber \\[0.2cm]
  =& \sum_{\boldsymbol{l}_5} P(C=c \mid B=b, \boldsymbol{L}_5=\boldsymbol{l}_5, A=a, S_5=1) P(\boldsymbol{L}_5 = \boldsymbol{l}_5 \mid B=b, S_5=1) \quad &&\text{by} \quad \textbf{(A.7a,b)} \nonumber \\[0.2cm]
  =& \sum_{\boldsymbol{l}_5} P(C=c \mid B=b, \boldsymbol{L}_5=\boldsymbol{l}_5, S_5=1) P(\boldsymbol{L}_5 = \boldsymbol{l}_5 \mid B=b, S_5=1) \quad &&\text{by} \quad \textbf{(A.8)} \nonumber
\end{align}

In addition to considering internal validity, the generalizability of each sample-specific quantity with respect to the target context must be evaluated. For quantities of the form $P(B^a \mid S_4=1)$, the assumptions are conceptually analogous to that in setting (i), here represented by \textbf{(A.10)}-\textbf{(A.12)}. However, identifying the target quantities involving $C^a$ from source $S_5$ necessitated strong additional assumptions to ensure internal validity, which must naturally also hold in the target context. In particular, the assumption of a complete mediation structure in $S_5$ must likewise be valid in the target, as formalized in assumption \textbf{(A.13)}. This states that, regardless of whether an individual is included in sample $S_5=1$ or not $S_5=0$, the outcome $C$ under intervention $A=a$ can be expressed as outcome $C$ that would occur if $B$ took the value it naturally would under $A=a$. The conditional exchangeability for selection in $B^a$ assumption \textbf{(A.14)} expresses that, knowing whether a unit is selected into $S_5$ or not provides no additional information about the potential outcome $C^{B^a}$. That is, we assume that there are no unobserved covariates that differ in distribution between $S_5$ and the target that affect the relation $B^a \rightarrow C^{B^a}$. Next, as we are concerned with a generalizability setting, we assume that within every stratum of $\boldsymbol{L}_5=\boldsymbol{l}_5$ in the target population, each unit has a positive probability of being included in $S_5$ or being represented by those units that are included, formalized as positivity of selection \textbf{(A.15)}. Note that excluding the condition $S_5=1$ refers to the target population distribution, for example, $P(\boldsymbol{L}_5 = \boldsymbol{l}_5)$ refers to the distribution in the target population with the source index merely describing that we are restricted to those covariates available in the source. Even if the external data used to construct the target weights contain additional relevant covariates beyond those observed in $s_5$, we are nonetheless restricted to the covariates available in $s_5$. The SUTVA for selection, \textbf{(A.16)}, postulates that outcomes are measured consistently across individuals in the target and source, with no interference between them, and that producing the same mediator value by different means yields the same outcome.

\begin{align}
\textbf{(A.10)}    \quad & \text{Conditional Exchangeability} \quad &&  B^a \independent S_4 \mid \boldsymbol{L}_4=\boldsymbol{l}_4 \quad \forall a
 \nonumber \\
 & \text{for Selection} && \nonumber \\[0.2cm]
 \textbf{(A.11)}   \quad & \text{Positivity of Selection} \quad && P(\boldsymbol{L}_4=\boldsymbol{l}_4)>0 \implies P(S_4=1 \mid \boldsymbol{L}_4=\boldsymbol{l}_4) >0  \nonumber \\
&  &&  \text{with probability 1} \nonumber \\[0.2cm]
\textbf{(A.12)}   \quad & \text{Stable Unit Treatment} \quad && \text{if } A=a \text{ and } S_4=s \ \text{ then } B^a=B  \nonumber \\
& \text{Value Assumption} && \nonumber \\
& \text{for Selection} && \nonumber \\[0.2cm]
\textbf{(A.13)}    \quad & \text{Complete Mediation Target Setting} \quad && \text{if } S_5=s, \text{ then } \forall a \  C^a = C^{B^a} \nonumber \\[0.2cm]
\textbf{(A.14)}   \quad & \text{Conditional Exchangeability} \quad && C^{B^a} \independent S_5 \mid B^a, \boldsymbol{L}_5=\boldsymbol{l}_5 \quad \forall a \nonumber \\
& \text{for Selection in $B^a$} &&  \nonumber \\[0.2cm]
 \textbf{(A.15)}   \quad & \text{Positivity of Selection} \quad && P(\boldsymbol{L}_5=\boldsymbol{l}_5)>0 \implies P(S_5=1 \mid \boldsymbol{L}_5=\boldsymbol{l}_5) >0  \nonumber \\
&  &&  \text{with probability 1} \nonumber \\[0.2cm]
\textbf{(A.16)}   \quad & \text{Stable Unit Treatment} \quad && \text{if } B^a=B \text{ and } S_5=s \ \text{ then } C^{B^a}=C  \nonumber \\
& \text{Value Assumption} && \nonumber \\
& \text{for Selection} && \nonumber
\end{align}

Provided the sample-specific quantities under its identifying assumptions, Equation (\ref{EQ:identification4}) demonstrates how the target quantity $P(C^a \mid B^a)$ together with external valiidty assumptions \textbf{(A.13)}-\textbf{(A.16)} becomes identifiable.

\begin{align} \label{EQ:identification4}
  &P(C^{a}=c \mid B^a=b)  \\[0.2cm]
  =& \sum_{\boldsymbol{l}} P(C^{B^{a}}=c \mid B^a=b, \boldsymbol{L}=\boldsymbol{l}) P(\boldsymbol{L}=\boldsymbol{l} \mid B^a=b)   \nonumber \\[0.2cm] 
  =& \sum_{\boldsymbol{l}_5} P(C^{B^{a}}=c \mid B^a=b, \boldsymbol{L}_5=\boldsymbol{l}_5, S_5=1) P(\boldsymbol{L}_5 = \boldsymbol{l}_5 \mid B^a=b) \quad &&\text{by} \quad \textbf{(A.13-A.16)} \nonumber \\[0.2cm]
  =& \sum_{\boldsymbol{l}_5} P(C=c \mid B=b, \boldsymbol{L}_5=\boldsymbol{l}_5, S_5=1) P(\boldsymbol{L}_5 = \boldsymbol{l}_5 \mid B=b) \quad &&\text{by} \quad \textbf{(A.4-A.8)} \nonumber
\end{align}

\noindent
For estimation, adjusting each sample-specific estimate from its respective source to the corresponding target distribution involves an additional modelling step within the context of that source. This adjustment presumes, for each source, the absence of measurement bias in the data used to estimate the respective target weights and correct specification of the statistical models, collectively represented by assumption \textbf{(A.17)}. Consequently, after estimating the sample-specific quantities and adjusting them using the corresponding target weights, the marginal sample-specific adjusted estimates are obtained, denoted by:

\[
\hat{\boldsymbol{P}}_{\mathcal{M}} = \left\{ \hat{P}(B^a = 1),\ \hat{P}(C^a = 1 \mid B^a = b) \;  : \;  a, b \in \{0,1\} \right\}
\]

\noindent
Under causal assumptions \textbf{(A.1)}-\textbf{(A.8)} and \textbf{(A.10)}-\textbf{(A.16)} together with the assumptions of no measurement bias and correctly specified statistical models, \textbf{(A.9)} and \textbf{(A.17)}, the expected values of the estimates in $\boldsymbol{\hat{P}}_{\mathcal{M}}$ equal their corresponding target parameters in $\boldsymbol{P}_{\mathcal{M}}$, thereby ensuring target validity.

Taken together, scenarios (i)–(iii) show that, even under simplified conditions, identification of the target parameters in the decision-analytical model fundamentally relies on a set of causal assumptions whose nature depends on various factors, particularly the available data. As complexity increases, whether due to the number of outcomes considered, the use of multiple identification strategies, or the joint presence of generalizability and transportability considerations, both the number and specificity of these assumptions grow, increasing the potential for one or more assumptions to be violated. This is a concern of particular importance, as violations of assumptions for any individual parameter induce target bias that propagates through the decision-analytical model and may ultimately lead to sub-optimal decisions.

\subsection{Propagation of bias in decision-analytical models}

To clarify how bias in a single target parameter can propagate through a decision-analytical model, we expand the counterfactual ICER from Equation (\ref{EQ:ICER_example_M}) by adopting the rollback decision tree representation for $\mathcal{M}$. Under this algorithm, expected costs and effectiveness are estimated sequentially, beginning at the terminal nodes and progressing back to the root node, a process known as \textit{averaging out and folding back} \cite{hunink2014decision,TreeAge}.
Mathematically, this is equivalent to summing over the expected branch-specific costs and QALYs under each intervention; however, the rollback representation intuitively illustrates how bias in a single parameter may propagate throughout the model. As an example, consider the expected cost target term from the counterfactual ICER in Equation (\ref{EQ:ICER_example_M}), shown on the first line of Equation (\ref{EQ:target_functions}).

    \begin{align} \label{EQ:target_functions}
    \mathop{\mathbb{E}_{\mathcal{P_M}}[\lambda(a) + f(\boldsymbol{V}^a)]} 
                                &= \lambda(a) + \sum_{b,c} \bigl( f(B^{a}=b) + f(C^{a}=c) \bigr) \  P_{\mathcal{M}}(B^{a}=b,C^{a}=c)   \\
                                &= \lambda(a) + \sum_{b,c} \bigl( f(B^{a}=b) + f(C^{a}=c) \bigr) \  P_{\mathcal{M}}(B^a=b) \  P_{\mathcal{M}}(C^a = c \mid B^a = b) \nonumber  \\
                                &= \lambda(a) + \sum_{b} f(B^{a}=b) \  P_{\mathcal{M}}(B^a=b) \  \sum_{c} P_{\mathcal{M}}(C^a = c \mid B^a = b) + \sum_{b,c} f(C^{a}=c) \   \nonumber \\
                                & \quad \,  \hspace{1.7cm} P_{\mathcal{M}}(B^a=b) \  P_{\mathcal{M}}(C^a = c \mid B^a = b) \nonumber \\
                                &= \lambda(a) + \sum_{b} \biggl(f(B^a=b) \  P_{\mathcal{M}}(B^a=b) \ + \ \biggl( \sum_{c} f(C^a=c) \  P_{\mathcal{M}}(C^a = c \mid B^a = b) \biggr)\biggr) \nonumber
     \end{align}

For simplicity, we assume that costs are additive and independent, and factorize the joint distribution of potential outcomes under intervention $a$ following Equation (\ref{EQ9.1}). By reordering terms, the final expression corresponds exactly to the expected cost calculation performed by the rollback algorithm under intervention $a$, given the target input parameters. It is this final expression that makes explicit how target bias may propagate: if one or more assumptions are violated, for example, we cannot identify $P(C^{a=1}=1\mid B^{a=1}=1)$ but proceed to plug in the estimated value, the resulting bias does not merely affect the branch-specific costs but can propagate through the decision-analytical model and impact the overall expected costs under the specific level of intervention. A concern that may be especially consequential for certain target estimands, such as the counterfactual ICER, being a ratio, for which small deviations in expected costs or effectiveness under a specific level of the intervention can substantially impact the decision criterion and thus potentially alter the resulting decision. We demonstrate this numerically in Appendix A.3, showing that even a violation of a single underlying causal assumption for one parameter can compromise optimal decision making.

This underscores the importance of explicitly stating and justifying the causal assumptions, as doing so allows potential violations, and the resulting sources of bias, to be identified and formally incorporated into causal bias analyses. While sensitivity analyses are typically conducted to quantify uncertainty arising from random error, systematic error (i.e. bias) has generally received limited consideration. Yet, systematic error can be investigated using causal bias analyses \cite{fox2021applying}, which provide a principled framework to quantify uncertainty arising from such sources of bias and complement sensitivity analyses of random error, allowing for a more comprehensive assessment of robustness of findings.

\newpage
\section{Discussion}

Health economic evaluations are fundamentally concerned with answering causal questions by targeting estimands that contrast the costs and effectiveness that would be observed under at least two different competing interventions. This involves the joint distribution of potential outcomes under a given level of the intervention, which, under appropriate causal assumptions, can be identified from the corresponding observed joint distribution. In practice, however, such observable analogues are rarely, if ever, available from a single source. This limitation has motivated the use of decision-analytical models to approximate these intervention-specific joint distributions directly, informed by causal parameters drawn from and synthesized across multiple sources. Under this approach, decision-analytical models provide a means for approximating the potential outcomes of interest from which the target estimand that would otherwise remain unattainable can be estimated. In this work, we have made this task explicit as one of causal inference, thereby formalizing decision-analytical models as causal models within the potential outcomes framework.

To underscore that the validity of this approach, and consequently the resulting decision, fundamentally relies on the extent to which the underlying (causal) assumptions are deemed plausible, we formally defined and decomposed total decision-analytical model bias. Specifically, it was defined as a discrepancy between the target estimand and the expected estimate produced under the employed decision-analytical model (i.e. empirical decision-analytical model), and was decomposed into two broad sources: \textit{model bias} and \textit{target bias}. Model bias refers to discrepancies that arise from the structural assumptions made when using a model to approximate the true data-generating process, which may not accurately reflect that process. Target bias, in contrast, arises from the causal nature of each probability parameter within the decision-analytical model, as these must be identified under appropriate causal assumptions. Violations of these assumptions induce target bias in the corresponding parameter, which could itself be further decomposed into contributions arising from internal or external validity violations.

With respect to the model component, the relationship between the target estimand, the decision-analytical model, and the assumed underlying causal structure was made explicit. To this end, the \textit{target decision-analytical model} was introduced to represent the decision-analytical model in its idealized form, in which all causal parameter estimands are explicitly specified and governed by the assumed underlying causal structure. Explicit specification of the target decision-analytical model not only facilitates a structured analysis, but also reveals that unconventional target parameters, such as potential outcome quantities conditional on other potential outcome quantities under a given intervention level, are not uncommon, even in otherwise simple settings. Provided the target model, each causal parameter estimand must be identified, that is, to ensure each is internally and externally valid with respect to the target context. The \textit{empirical decision-analytical model} was then introduced to represent the decision-analytical model that can, in practice, be estimated from the available data. Juxtaposing the target and empirical decision-analytical models provides a structured framework for assessing how each causal parameter is informed by data, thereby making explicit where violations of internal or external validity may arise. Fundamentally, this rests on evaluating the causal assumptions. When any of these assumptions are violated, target bias arises at the parameter level, which may propagate through the model and affect the higher-level target quantities, ultimately risking suboptimal decision making.

Although the principles discussed in this work were illustrated in the context of a decision tree model, they hold more generally, irrespective of the specific class of decision-analytical model, though additional considerations may arise. For instance, in practice, many applications employ Markov models (e.g. microsimulation), which introduce an explicit time dimension. While the presented rationale remains to hold, the notation and formal framework must be accordingly extended to account for time, moving into the domains of causal survival analysis and multi-state modelling. More generally, this work presented the fundamental rationale in a relatively straightforward setting, which already introduced nontrivial complexities, whereas most practical applications often employ more sophisticated models to address more intrincate settings, entailing additional (causal) assumptions.

To conclude, given the inherently causal nature of the estimands targeted in health economic evaluations, we recommend explicitly formalizing the target estimand as a causal quantity. Doing so makes clear that the analysis is grounded in causal inference, directly highlighting its reliance on underlying causal assumptions. To structurally reason about these assumptions and their implications, we recommend specifying the target decision-analytical model together with the assumed causal structure, ideally represented using a causal graph. This approach not only aids the reader but, importantly, assists the analyst in identifying potential violations of causal assumptions based on the available data. In this way, the analyst can systematically scrutinize the assumptions and incorporate them into causal bias analyses, complementing broader sensitivity analyses to assess the robustness of the findings.

\vspace{0.5cm}
\noindent
\textbf{Data availability statement:} All code required to reproduce the simulation and results is openly available on GitHub at: \url{https://github.com/mauricekorf/Simulation_DAM}

\vspace{0.5cm}
\noindent
\textbf{Funding information:} J.A.L is supported by a Veni grant (09150162010213) from the Netherlands Organisation for Scientific Research (NWO) / Netherlands Organisation for Health Research and Development (ZonMW).

\newpage
\printbibliography

\newpage
\appendix
\renewcommand{\thefigure}{\thesection\arabic{figure}}
\renewcommand{\thetable}{\thesection\arabic{table}}
\setcounter{figure}{0}
\setcounter{table}{0}

\section{Numerical example}

Suppose we are interested in the incremental cost-effectiveness ratio (ICER) of intervention $a=1$ relative to $a=0$, where the relevant set of dichotomous health outcomes influenced by each intervention level is denoted by $\boldsymbol{V}^a=\{V_1^a,V_2^a,V_3^a\}$. Each health outcome influences both the costs and effectiveness, with effectiveness measured in terms of quality-adjusted life years (QALYs). The corresponding target estimand is defined in Equation (\ref{EQ:A1}).

\begin{equation} \label{EQ:A1}
    \frac{\displaystyle\mathop{\mathbb{E}_{\mathcal{P}}}  \Bigl[ \lambda_1 + f(V_1^{a=1},V_2^{a=1},V_3^{a=1}) \Bigr] \  - \  \displaystyle\mathop{\mathbb{E}_{\mathcal{P}}} \Bigl[ \lambda_0 + f(V_1^{a=0},V_2^{a=0},V_3^{a=0}) \Bigr] }{\displaystyle\mathop{\mathbb{E}_{\mathcal{P}}} \Bigl[ g(V_1^{a=1},V_2^{a=1},V_3^{a=1}) \bigr] \  - \ \displaystyle\mathop{\mathbb{E}_{\mathcal{P}}} \bigl[ g(V_1^{a=0},V_2^{a=0},V_3^{a=0}) \Bigr]}
\end{equation}
\vspace{0.1cm}

\noindent
The assumed data-generating mechanism is depicted in Figure (A1) using a Single World Intervention Template (SWIT), which illustrates the underlying causal structure and encodes the assumptions about how the intervention, covariates, and health outcomes are related. Specifically, define $\boldsymbol{L}=\{L_1,L_2,L_3\}$ as the set of observed baseline confounders, and $\boldsymbol{E}=\{E_1,E_2,E_3\}$ denote the set of observed baseline effect modifiers that also act as confounders. While, in practice, it is often simplified by considering a single set in which each element may simultaneously act as a confounder and an effect modifier \cite{degtiar2023review}, we explicitly distinguish $\boldsymbol{E}$ for conceptual clarity, a distinction that is particularly useful when discussing generalizability.

To reduce visual clutter in the SWIT, we represent the sets rather than each as a separate node. However, we explicitly define their roles as follows: $L_1$ confounds the relationship between $A$ and $V_1^a$, $L_2$ between $A$ and $V_2^a$, $L_3$ between $A$ and $V_3^a$; whereas $E_1$ both confounds and modifies the effect of $A$ on $V_1^a$, $E_2$ of $A$ on $V_2^a$, and $E_3$ of $A$ on $V_3^a$. Importantly, $S$ denotes a binary selection node with $S=1$ representing selection into the sample. The edge from $\boldsymbol{E}$ into node $S$ encodes a distributional shift, indicating that the distribution of $\boldsymbol{E}$ in the study sample differs from that in the target population, denoted by $\boldsymbol{E}^{*}$. This reflects an external validity concern, specifically, a generalizability issue arising from sample selection bias \cite{correa2019adjustment}.

\begin{center} \label{Simulation_causaldiagram}
    \begin{figure}[h!]
    \centering
    \includegraphics[width=0.7\textwidth]{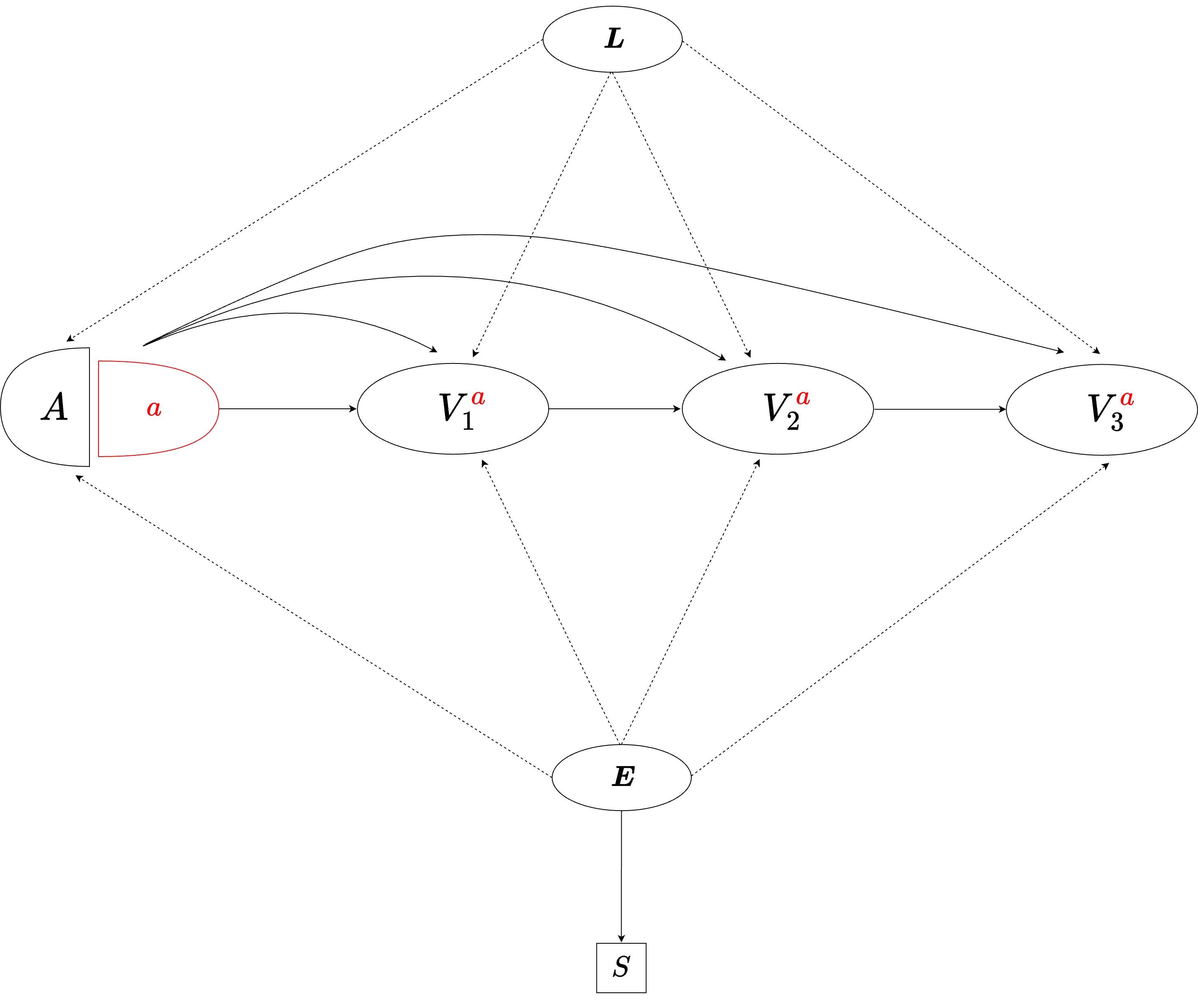}
    \caption{Single-world intervention template representing the hypothetical data-generating mechanism. Here, $\boldsymbol{L}=\{L_1,L_2,L_3\}$ represents the set of measured baseline confounders, whereas $\boldsymbol{E}=\{E_1,E_2,E_3\}$ denotes the set of measured baseline effect modifiers, which also act as confounders, but differ in distribution between the study sample ($S=1$) and target population.}
    \label{fig:SWIT_numerical_example}
\end{figure}
\end{center}

\noindent
Having specified the causal structure, we suppose that a decision tree model (i.e. $\mathcal{M}$) is employed to target the estimand in Equation (\ref{EQ:A1}), expressed under the model as:

\begin{equation}
    \frac{\displaystyle\mathop{\mathbb{E}_{\mathcal{P_M}}}  \Bigl[ \lambda_1 + f(V_1^{a=1},V_2^{a=1},V_3^{a=1}) \Bigr] \  - \  \displaystyle\mathop{\mathbb{E}_{\mathcal{P_M}}} \Bigl[ \lambda_0 + f(V_1^{a=0},V_2^{a=0},V_3^{a=0}) \Bigr] }{\displaystyle\mathop{\mathbb{E}_{\mathcal{P_M}}} \Bigl[ g(V_1^{a=1},V_2^{a=1},V_3^{a=1}) \bigr] \  - \ \displaystyle\mathop{\mathbb{E}_{\mathcal{P_M}}} \bigl[ g(V_1^{a=0},V_2^{a=0},V_3^{a=0}) \Bigr]}
\end{equation}

\noindent
From this, the estimands of the target parameters (i.e. $\mathcal{P_M}$), obtained by factorizing the joint distribution of potential outcomes under intervention $a$ as implied by the SWIT in Figure (A1), can be directly mapped onto the corresponding decision tree representation, giving rise to the target decision-analytical model in Figure (A2).

\begin{center} \label{Simulation_causaldiagram}
    \begin{figure}[h!]
    \centering
    \includegraphics[width=0.67\textwidth]{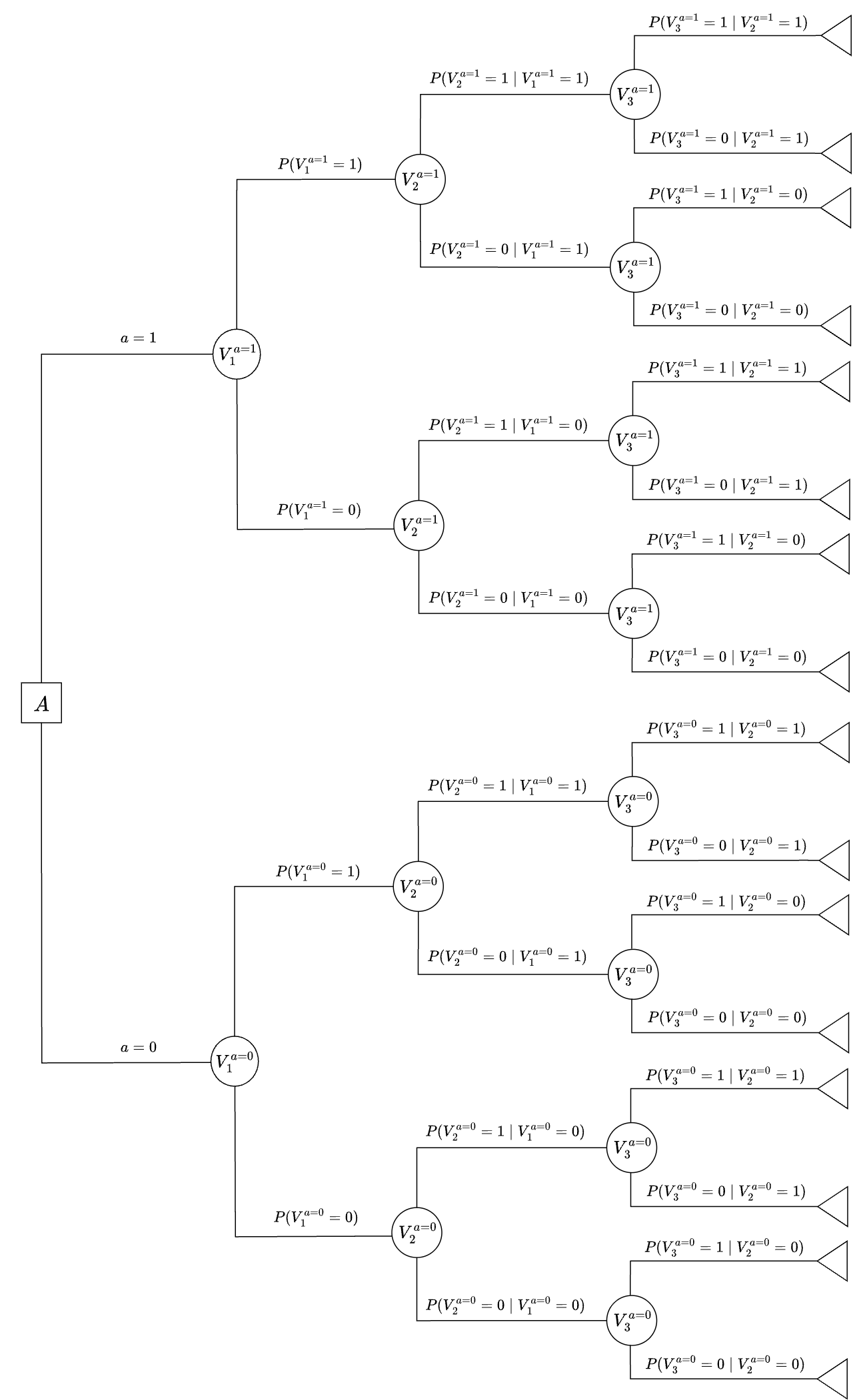}
    \caption{Target decision model corresponding to the causal structure specified in Figure (A1)}
    \label{fig:Counterfactual_DT_numerical_example}
\end{figure}
\end{center}

\noindent
Provided the target decision-analytical model and the accompanying causal graph, we next specify a structural causal model (SCM) to simulate data, which enables us to define a ground truth and subsequently to illustrate the potential impact of confounding bias, sample selection bias, and bias propagation. The parametric specification is constructed to be fully compatible with the decision tree structure, ensuring that no model bias is introduced and allowing us to focus exclusively on target bias. Conceptually, the parametric specification corresponds to having access to the full joint distribution from a single source, in which case a decision-analytical model would not be strictly necessary. Nonetheless, this setup allows us to mimic a setting in which a decision tree model is applied, while retaining full control to isolate and illustrate sources of target bias. The parametric specification of the SCM over the variables $\{A,V_1,V_2,V_3,L_1,L_2,L_3,E_1,E_2,E_3\}$ is given as follows:

\begin{align*}
L_i &\sim \text{Bernoulli}(0.5), \quad \text{for } i = 1, 2, 3 \\
E_j &\sim \text{Bernoulli}(0.6), \quad \text{for } j = 1, 2, 3 \\
S &\sim \text{Bernoulli}\left( \text{expit}(1 - 2E_1 + 1E_2 - 2E_3) \right) \\
A &\sim \text{Bernoulli}\left( \text{expit}(-2 - 2.5L_1 + 3L_2 + 2.5L_3 + 2E_1 - 1E_2 + 1.5E_3) \right) \\
V_1 &\sim \text{Bernoulli}\left( \text{expit}(-0.1 -1.5A + 3L_1 + 0.5E_1 + 1.5AE_1) \right) \\
V_2 &\sim \text{Bernoulli}\left( \text{expit}(-0.1 -1.2A + 0.5V_1 + 4L_2 + 0.5E_2 + 2AE_2) \right) \\
V_3 &\sim \text{Bernoulli}\left( \text{expit}(-0.1 -1.5A - 0.5V_2 + 3.5L_3 + 0.5E_3 + 1AE_3) \right)
\end{align*}
\vspace{0.1cm}

\noindent
Here, \(\text{expit}(x) = \frac{1}{1 + e^{-x}}\) denotes the inverse function of the logit with \(x = \log\big(\frac{p}{1 - p}\big)\) representing the log-odds for a given probability \(p\), such as \(P(A=1)\). In this setting, an outcome is considered adverse when the outcome variable takes the value 1. Intuitively, intervention $A=1$ tends to reduce, on average, the log-odds of experiencing adverse outcomes as compared to $A=0$, although the magnitude depends on baseline covariates, effect modifiers, and interaction terms. Given the SCM, we assign cost and QALY values as functions of the potential outcomes. The cost and QALY values assigned to each potential outcome are summarised in Table (A1), where fixed intervention-specific costs are denoted by $\lambda_1$ under $a=1$ and $\lambda_0$ under $a=0$. All other reward values are assumed to be invariant between interventions. Note that higher costs and lower QALYs are assigned to adverse outcomes.

\begin{table}[htbp]
\centering
\caption{True cost (\euro) and quality-adjusted life year values}
\label{TAB:causal_probs}
\begin{tabular}{llll}
\toprule
\textbf{Costs} & \textbf{Value} & \textbf{QALYs} & \textbf{Value} \\
\midrule
$V_1^{a}=1$ & 5400 & $V_1^{a}=1$ & 0.4\\
$V_1^{a}=0$ & 2700 & $V_1^{a}=0$ & 0.8 \\
$V_2^{a}=1$ & 2700 & $V_2^{a}=1$ & 0.3 \\
$V_2^{a}=0$ & 900  & $V_2^{a}=0$ & 0.75  \\
$V_3^{a}=1$ & 1800 & $V_3^{a}=1$ & 0.2 \\
$V_3^{a}=0$ & 270  & $V_3^{a}=0$ & 0.65 \\
$\lambda_1$ & 9000 &  &  \\
$\lambda_0$ & 0 &  &  \\
\bottomrule
\end{tabular}
\end{table}

\noindent
Based on the assigned cost and QALY values, we can specify the total cost and QALY function, with the latter defined as follows:

\[
\begin{aligned}
\text{QALY}^a 
&= g(V_1^a, V_2^a, V_3^a) \\[2mm]
&=
\begin{cases}
\text{QALY}^a_{\text{base}} + 0.4, & \ \text{if} \quad n_\text{events} = 0, \\[6pt]
\text{QALY}^a_{\text{base}}
- \min\!\big(0.05\,n_\text{events} + 0.03\,n_\text{events}^2,\; \text{QALY}^a_{\text{base}}\big), & \ \text{if} \quad n_\text{events} > 0
\end{cases}
\end{aligned}
\]

\vspace{0.2cm}
\noindent
In this expression, $n_{\text{events}}=\sum_{i=1}^{3} V^a_i$ denotes the total number of adverse events, whereas QALY$^a_{\text{base}}$ represents the baseline QALYs derived from the potential outcomes and corresponding reward values specified in Table (A1), with its functional structure defined as:

\begin{align}
    \text{QALY}^a_{\text{base}} &=   V_1^a \times 0.4 \ + \  (1 - V_1^a) \times 0.8 
  + V_2^a \times 0.3 \ + \ (1 - V_2^a) \times 0.75\notag \\  
  &\quad \  + \  V_3^a \times 0.2 \ + \  (1 - V_3^a) \times 0.65 \notag
\end{align}

\vspace{0.2cm}
\noindent
The baseline QALYs are reduced in the presence of adverse events ($n_{\text{events}} > 0$), with the reduction limited to the baseline value to ensure that QALYs remain non-negative. In contrast, if no adverse events occur ($n_{\text{events}} = 0$), a bonus of 0.4 QALYs is added. For the cost function, we define:

\[
\begin{aligned}
\text{costs}^a 
&= \lambda_a + f(V_1^a, V_2^a, V_3^a) =
\begin{cases}
\lambda_a \ + \ \text{costs}^a_{\text{base}}, & \ \text{if} \quad n_\text{events} \le 1, \\[6pt]
\lambda_a \ + \ \text{costs}^a_{\text{base}} \ + \ 2000,
 & \ \text{if} \quad n_\text{events} = 2 \\[6pt]
\lambda_a \ + \  \text{costs}^a_{\text{base}} \ + \ 4000,
 & \ \text{if} \quad n_\text{events} = 3
\end{cases}
\end{aligned}
\]

\vspace{0.2cm}
\noindent
where the baseline cost component under intervention $a$ (i.e. costs$^{a}_{\text{base}}$) is defined as;

\begin{align}
\text{cost}^a_{\text{base}} &= 
 V_1^a \times 5400 + (1 - V_1^a) \times 2700 \ + \ V_2^a \times 2700 \ + \ (1 - V_2^a) \times 900 \notag\\
&\quad + \  V_3^a \times 1800 + (1 - V_3^a) \times 270 \notag
\end{align}

\noindent
The baseline costs are increased by additional cost components whenever two or more events occur, represented by an increment of 2,000 when two events occur and 4,000 when three events occur. With the setup in place, we next turn to discussing examples of confounding bias, sample selection bias, and the propagation of bias.

\subsection{Confounding bias}
Since confounding bias concerns internal validity, we focus on the sample-specific setting, represented by conditioning on $S=1$. The true expected potential costs and QALYs for the study sample under each specific intervention level, as determined by the specified data-generating mechanism, are reported in the first row of Table (A2), along with the corresponding expected ICER. Notably, the reported $\displaystyle\mathbb{E}[\text{ICER}]$ represents the mean of the iteration-specific ICERs, whereas the reported true ICER is computed directly from the true marginal quantities. Although confounding bias arises at the probability parameter level, specifically when the expected value of the sample-specific estimate deviates from the true sample-specific value due to incomplete adjustment for $\{\boldsymbol{L,E}\}$, we evaluate its impact on the corresponding target outcome quantities. For this reason, instead of reporting the values of the probability parameters, we focus on comparing the true sample costs, QALYs, and ICER to their estimated counterparts obtained under incomplete adjustment.

In line with recommended practice \cite{korf2025causal}, we report results under different estimators; here, we consider two: G-computation and inverse probability of treatment weighting (IPTW) \cite{hernan2020causal}. As the models are correctly specified in our simulation, we use the comparison of G-computation and IPTW primarily to illustrate the behavior of bias across different estimation approaches and to confirm that, as expected in this setting, both yield bias in the same direction for each target outcome quantity. Note that, consistent with the decision tree framework, individual probability parameters must be estimated. However, since IPTW targets marginal quantities by design, an outcome model is subsequently required to estimate each (conditional) probability. While the individual conditional probabilities at the chance nodes are not guaranteed to be unbiased under this approach, IPTW ensures that any marginal expectation of a function of the potential outcomes under $a$ are unbiased in expectation. Therefore, the marginal expected costs and QALYs under intervention level $a$ remain unbiased even if the conditional probabilities are not. Importantly, this does not generally hold in multi-source settings, underscoring that the current simulation assumes a single, complete data source.

As shown in Table (A2), failing to adjust for $\{\boldsymbol{L},\boldsymbol{E}\}$ leads to biased estimates of costs and QALYs under each intervention level, which in turn inflates the corresponding mean ICER. Across all simulation scenarios presented, we assume a willingness-to-pay (WTP) threshold of 50.000 euros, corresponding to an intermediate disease burden following Dutch guidelines \cite{zin_ziektelast_2018}. Given this WTP, the bias induced by not adjusting for $\{\boldsymbol{L},\boldsymbol{E}\}$ would consistently lead to an incorrect decision, erroneously suggesting that intervention $a=1$ is not cost-effective compared to $a=0$. To better illustrate this, and given that the uncertainty interval around the ICER is not straightforward to interpret, being a ratio constructed from four estimated quantities, it is recommended to visualise uncertainty on a cost-effectiveness (CE) plane \cite{bilcke2022generating}. Therefore, Figure (A3) presents the corresponding CE plane, plotting the incremental cost and QALY coordinates from each sample iteration (10,000 iterations, blue points) alongside the WTP threshold line at 50,000 euros. The baseline intervention, here $a=0$, is represented by a single point at the origin (0,0), corresponding to zero incremental cost and QALYs relative to itself. All points lying on a line through the origin share the same incremental cost-effectiveness ratio relative to the baseline. Hence, the cost-effectiveness threshold appears as a line on the CE plane, representing the boundary at which interventions are considered cost-effective compared to the baseline \cite{bilcke2022generating}. As illustrated in Figure (A3) and consistent with Table (A2), without adjustment for $\{\boldsymbol{L},\boldsymbol{E}\}$, the mean sample estimate (red point) lies to the left of the WTP threshold, reflecting that intervention $a=1$ is not cost-effective relative to $a=0$. In fact, there is no decision uncertainty here, as all sample mean estimates (blue points) lie to the left of the WTP threshold.
\vspace{0.2cm}

\begin{table}[htbp]
\centering
\caption{True sample-specific costs, quality-adjusted life-years (QALYs), and incremental cost-effectiveness ratio (ICER) under interventions $a=1$ and $a=0$. Estimates are reported to quantify the impact of confounding bias resulting from incomplete adjustment of $\{\boldsymbol{L},\boldsymbol{E}\}$. Estimates and corresponding 95\% confidence intervals (CI) were obtained by simulating 10.000 data sets of size 10.000 from the specified data-generating mechanism, under either parametric G-computation or inverse probability of treatment weighting (IPTW). Cost values are rounded to the nearest euro, and QALY values are presented with four decimal places.}
\label{tab:regression-style-results}
\resizebox{\textwidth}{!}{ 
\begin{tabular}{lccccc}
\toprule
 Scenario & $\displaystyle\mathbb{E}[\text{costs}^{a=1} \mid S=1]$ & $\displaystyle\mathbb{E}[\text{costs}^{a=0} \mid S=1]$ & $\displaystyle\mathbb{E}[\text{QALY}^{a=1} \mid S=1]$ & $\displaystyle\mathbb{E}[\text{QALY}^{a=0} \mid S=1]$ & $\displaystyle\mathbb{E}[\text{ICER}]$ \\
\midrule
 True sample values & 18740 & 10910 & 1.1351 & 0.9522 & 42810 \\[0.5cm]
 \textbf{Unadjusted} &  & &  & &  \\[0.3cm]
Mean estimate & 19031 & 10668 & 1.0607  & 1.0129 & 211459 \\
95 \% CI & [19030, 19033] & [10667, 10670] & [1.0605, 1.0609]  & [1.0126, 1.0131] & [199649, 223268] \\[0.5cm]
 \textbf{G-computation} &  & &  & &  \\[0.3cm]
Mean estimate: adjusted $\{\boldsymbol{E}\}$& 18853 & 10710 & 1.0842 & 1.0070  & 111272 \\
95 \% CI & [18851, 18854] & [10709, 10712] & [1.0840, 1.0844]  & [1.0068, 1.0073] & [110663, 111881] \\[0.5cm]
Mean estimate: adjusted $\{L_1,L_3, \boldsymbol{E}\}$& 18946 & 10657 & 1.0956 & 0.9979  & 86999 \\
95 \% CI & [18945, 18947] & [10656, 10659] & [1.0954, 1.0958]  & [0.9977, 0.9981] & [86702, 87296] \\[0.5cm]
\textbf{IPTW + outcome model} &  & &  & &  \\[0.3cm]
Mean estimate: adjusted $\{\boldsymbol{E}\}$& 18878 & 10684 & 1.0778 & 1.0125  & 136006 \\
95 \% CI & [18877, 18880] & [10683, 10686] & [1.0776, 1.0780]  & [1.0123, 1.0127] & [134927, 137086] \\[0.5cm]
Mean estimate: adjusted $\{L_1,L_3,\boldsymbol{E}\}$ & 19008 & 10562 & 1.0801  & 1.0141 & 147733 \\
95 \% CI & [19007, 19010] & [10560, 10564] & [1.0798, 1.0803]  & [1.0138, 1.0144] & [137213, 158255] \\
\bottomrule
\end{tabular}
}
\end{table}
\vspace{0.1cm}

Analogous to the scenario without adjustment for $\{\boldsymbol{L},\boldsymbol{E}\}$, Table (A2) shows that adjusting only for a subset of covariates, such as $\{\boldsymbol{E}\}$ or $\{L_1,L_3,\boldsymbol{E}\}$, likewise results in biased estimates of the target outcomes, which in turn inflate the ICER and mislead one to infer that $a=1$ is not cost-effective relative to $a=0$. Note, however, that the magnitude of the bias in the estimated target outcome quantities differs slightly across estimators, provided that a sample size of 10,000 individuals is considered, although both estimators agree on the direction of the bias, as expected in this simulation. As a complementary visual illustration, Figure (A4) presents the CE plane corresponding to G-computation adjusting for $\{L_1,L_3,\boldsymbol{E}\}$. This figure clearly demonstrates that, when $L_2$ is not adjusted for, all sample estimates (blue points) lie to the left of the WTP threshold, reflecting no decision uncertainty, and would therefore consistently, yet incorrectly, lead to the conclusion that intervention $a=1$ is not cost-effective compared to $a=0$. For clarity, we do not present plots for all scenarios and estimators, as doing so would not provide additional insight beyond what is already illustrated.

\begin{figure}[ht]
    \centering
    \begin{subfigure}[t]{0.48\textwidth}
        \centering
        \includegraphics[width=\linewidth]{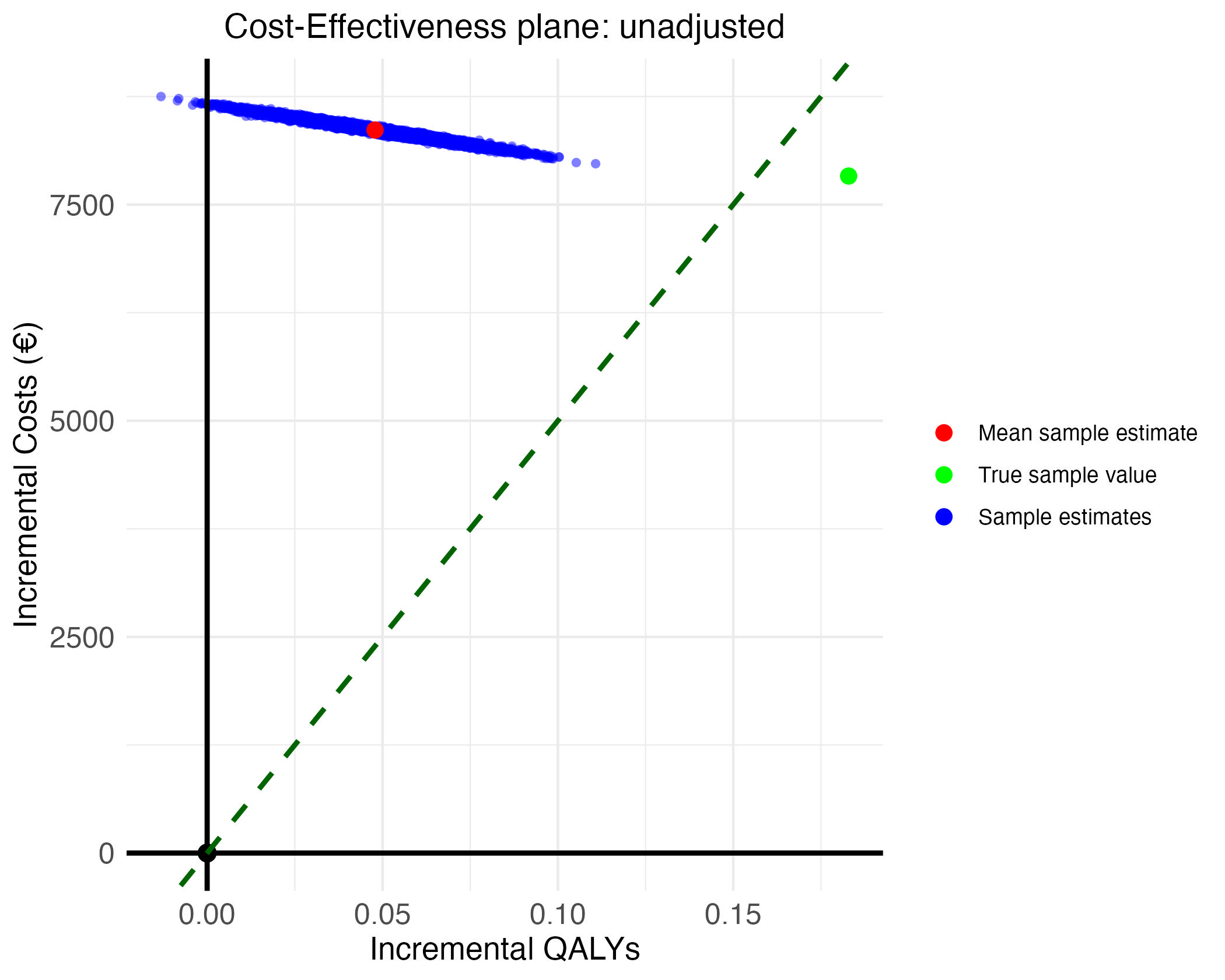}
        \caption{Zoomed out}
        \label{fig:adjusted_subset1}
    \end{subfigure}
    \hfill
    \begin{subfigure}[t]{0.48\textwidth}
        \centering
        \includegraphics[width=\linewidth]{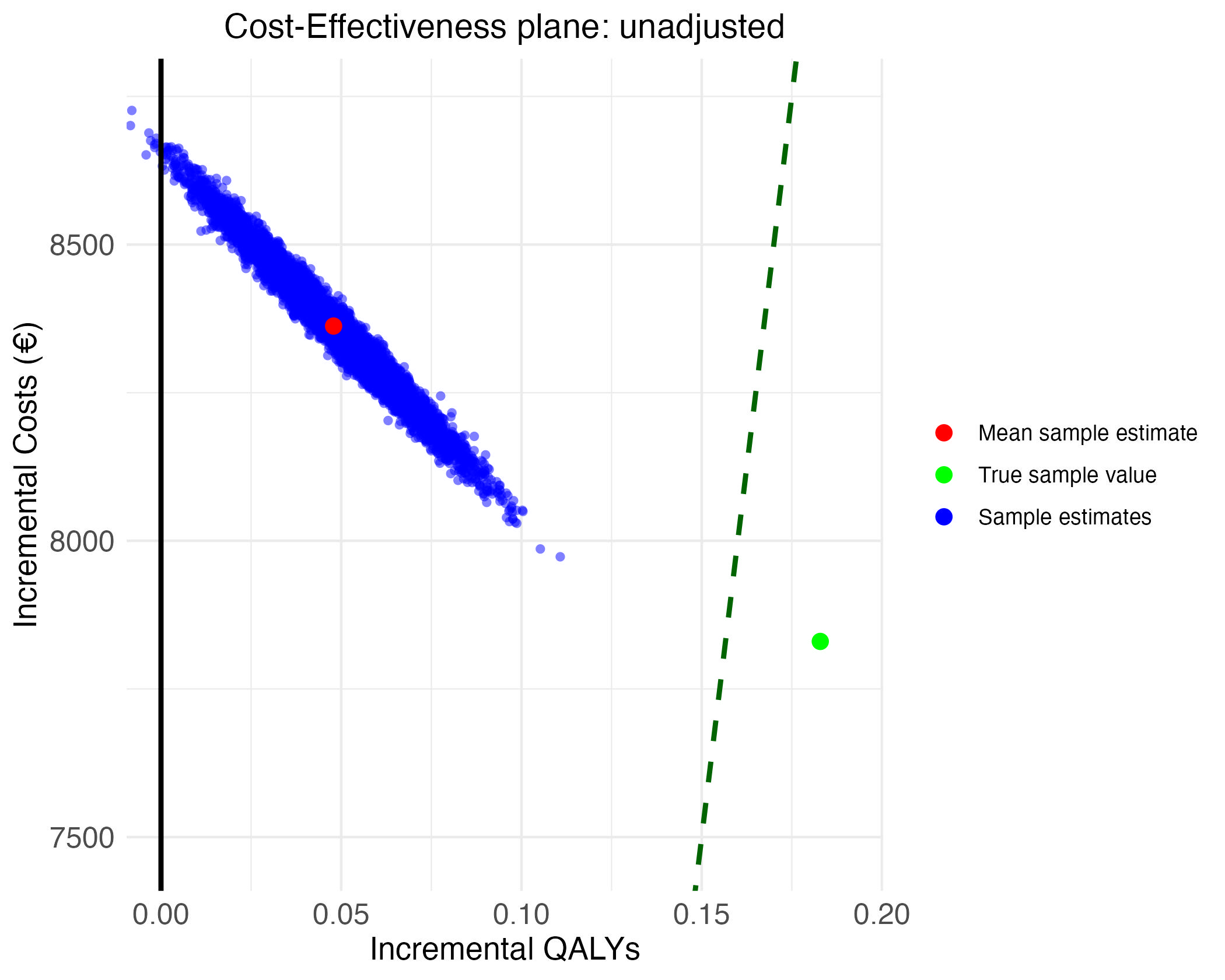}
        \caption{Zoomed in}
        \label{fig:adjusted_subset_zoom1}
    \end{subfigure}
    \caption{Cost-effectiveness plane when not adjusting. Blue points represent mean estimates from each simulated sample (10,000 samples, size 10,000), the red point shows the mean of these sample means, and the green point indicates the true sample value derived from the data-generating mechanism. The dashed line represents the willingness to pay threshold of 50,000 euro}
    \label{fig:adjusted_combined}
\end{figure}

\begin{figure}[ht]
    \centering
    \begin{subfigure}[t]{0.48\textwidth}
        \centering
        \includegraphics[width=\linewidth]{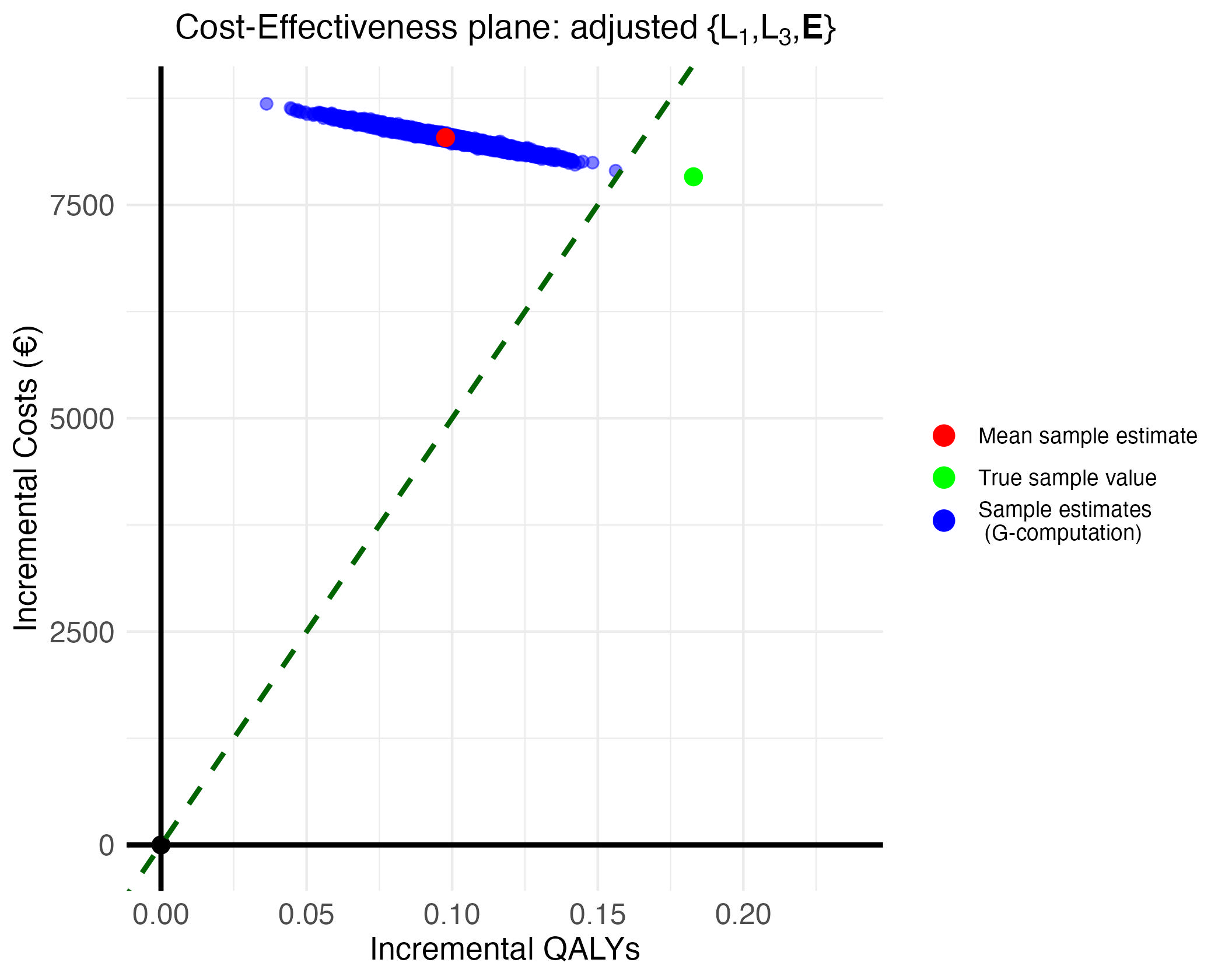}
        \caption{Zoomed out}
        \label{fig:adjusted_subset2}
    \end{subfigure}
    \hfill
    \begin{subfigure}[t]{0.48\textwidth}
        \centering
        \includegraphics[width=\linewidth]{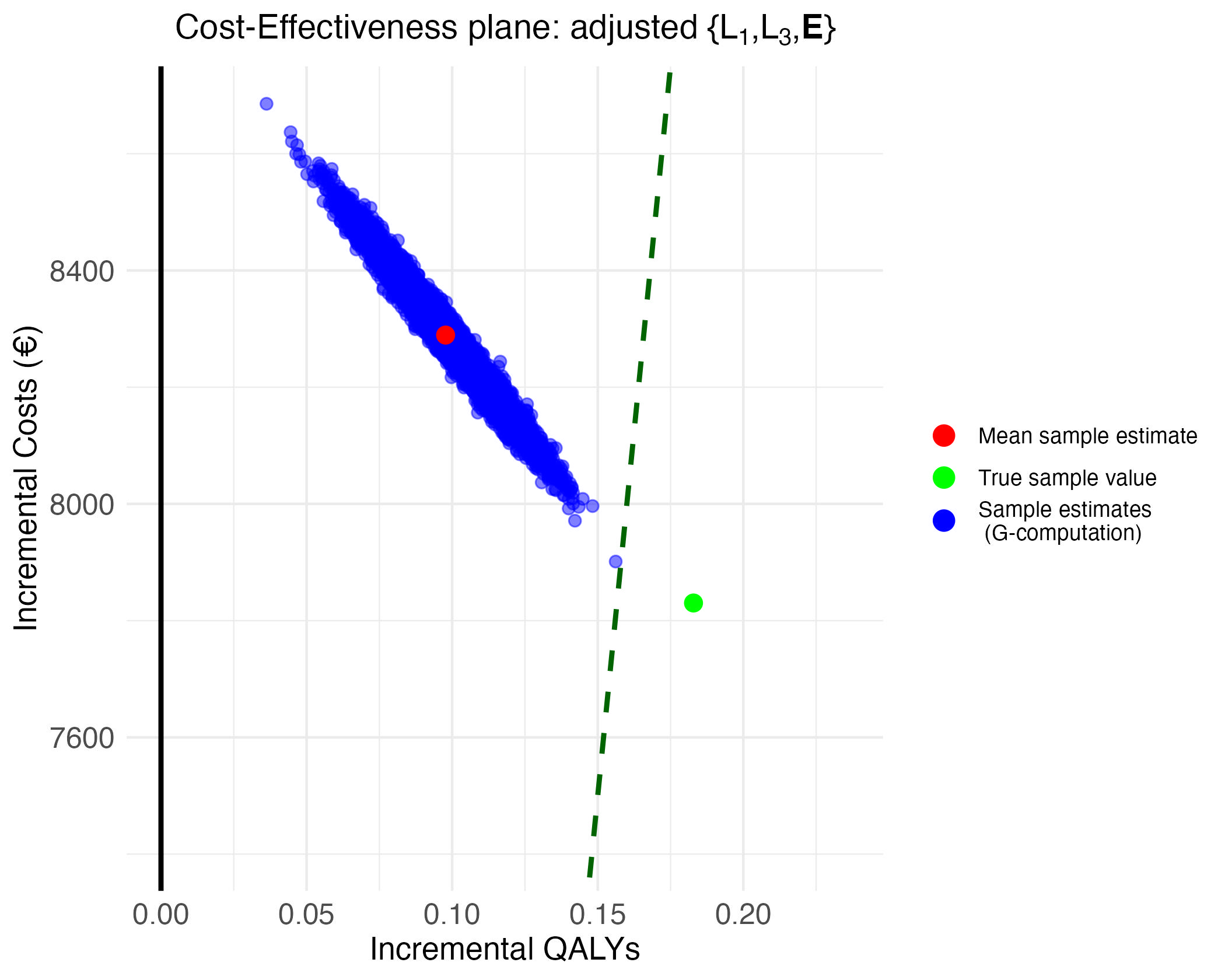}
        \caption{Zoomed in}
        \label{fig:adjusted_subset_zoom2}
    \end{subfigure}
    \caption{Cost-effectiveness plane adjusting for $\{L_1,L_3,\boldsymbol{E}\}$ using G-computation. Blue points represent mean estimates from each simulated sample (10,000 samples, size 10,000), the red point shows the mean of these sample means, and the green point indicates the true sample value derived from the data-generating mechanism. The dashed line represents the willingness to pay threshold of 50,000 euro}
    \label{fig:partly_adjusted_combined}
\end{figure}

\clearpage
\subsection{Sample selection bias (generalizability)}

In general, inferring a decision based on a model that draws on multiple data sources, even if each is assumed to result in internally valid estimates, remains inherently challenging, as the credibility of the resulting decision depends on the alignment of the underlying contexts (i.e. external validity). This concern persists even in the idealized case considered here, where all parameters originate from a single source; even then, the inferred decision is not guaranteed to hold beyond the study sample and may lead to misguided decision-making. In particular, a comparison of the true sample values with the true population values, presented in the first two lines of Table (A3), illustrates that the decision inferred from the study sample does not generalize to the target population, which is the relevant population for decision-making. In fact, based on the study sample reported values, one would conclude that intervention $a=1$ is cost-effective relative to $a=0$, whereas based on the true population values one would conclude the opposite, since the ICER exceeds the WTP of 50.000 euros.

This discrepancy arises from sample selection bias due to a shift in the distribution of $\boldsymbol{E}$ between the study sample and target population, $\boldsymbol{E}^{*}$. To account for this, we adjust the $\{\boldsymbol{L},\boldsymbol{E}\}$ stratum-specific sample-based estimates according to the target distribution $\boldsymbol{E}^{*}$ using G-computation (henceforth weighted G-computation), or alternatively, by reweighting the sample population using selection weights, a procedure referred to as inverse probability of selection weighting (IPSW) \cite{lesko2017generalizing}. As noted previously, an additional outcome model is required within the weighting approach to recover the full set of (conditional) probability parameters.

\begin{table}[htbp]
\centering
\caption{True sample-specific and target population-level costs, quality-adjusted life-years (QALYs), and incremental cost-effectiveness ratios (ICERs) under interventions $a=1$ and $a=0$. Adjusted estimates are reported to quantify the impact of sample selection bias resulting from incomplete adjustment and reweighting. Estimates and corresponding 95\% confidence intervals (CI) were obtained by simulating 10.000 data sets of size 10.000 from the specified data-generating mechanism, under either parametric weighted G-computation or combined inverse probability of treatment and selection weighting (IPTW x IPSW). Cost values are rounded to the nearest euro, and QALY values are presented with four decimal places.}

\label{tab:regression-style-results}
\resizebox{\textwidth}{!}{ 
\begin{tabular}{lccccc}
\toprule
 Scenario & $\displaystyle\mathbb{E}[\text{costs}^{a=1}]$ & $\displaystyle\mathbb{E}[\text{costs}^{a=0}]$ & $\displaystyle\mathbb{E}[\text{QALY}^{a=1}]$ & $\displaystyle\mathbb{E}[\text{QALY}^{a=0}]$ & $\displaystyle\mathbb{E}[\text{ICER}]$ \\
\midrule
 True sample values & 18740 & 10910 & 1.1351 & 0.9522 & 42810 \\[0.5cm]
 True population values & 19126 & 11027 & 1.08 & 0.9347 & 55740  \\[0.5cm]
 \textbf{Weighted G-computation} &  & &  & &  \\[0.3cm]
 Mean estimate: adjusted $\{\boldsymbol{L},\boldsymbol{E}\}$ \& & 19126 & 11026 & 1.0801  & 0.9348 & 56524 \\
 \hspace{2.4cm} weighted $\{\boldsymbol{E}^{*}\}$ &  & &  & & \\
95 \% CI & [19124, 19127] & [11024, 11028] & [1.0798, 1.0803]  & [0.9346, 0.9351] & [56385, 56664] \\[0.5cm]
Mean estimate: adjusted $\{\boldsymbol{L},E_2\}$ \& & 18821 & 10836 & 1.1274  & 0.9642 & 49381 \\
 \hspace{2.4cm} weighted $\{E^{*}_2\}$ &  & &  & & \\
95 \% CI & [18820, 18822] & [10834, 10837] & [1.1272, 1.1276]  & [0.9640, 0.9645] & [49282, 49479] \\[0.5cm]
Mean estimate: adjusted $\{L_2,\boldsymbol{E}\}$ \& & 19371 & 10762 & 1.0353  & 0.9821 & 198451 \\
 \hspace{2.4cm} weighted $\{\boldsymbol{E}^{*}\}$ &  & &  & & \\
95 \% CI & [19370, 19372] & [10760, 10763] & [1.0351, 1.0356]  & [0.9819, 0.9824] & [179717, 217184] \\[0.5cm]
\textbf{IPTW + IPSW + outcome model} &  & &  & &  \\[0.3cm]
Mean estimate: adjusted $\{\boldsymbol{L},\boldsymbol{E}\}$ \& & 19128 & 10969 & 1.08  & 0.9485 & 65505 \\
 \hspace{2.4cm} weighted $\{\boldsymbol{E}^{*}\}$ &  & &  & & \\
95 \% CI & [19121, 19136] & [10954, 10983] & [1.0787, 1.0813]  & [0.9462, 0.9508] & [56590, 74421] \\[0.5cm]
Mean estimate: adjusted $\{\boldsymbol{L}\}$ \& & 18623.906 & 10879.023 & 1.157  & 0.9579 & 41003.232 \\
 \hspace{2.4cm} weighted $\{E_2\}$ &  & &  & & \\
95 \% CI & [18620.64, 18627.171] & [10875.151, 10882.896] & [1.1564, 1.1576]  & [0.9573, 0.9585] & [40778.9142, 41227.55] \\[0.5cm]
\bottomrule
\end{tabular}
}
\end{table}

\noindent
Under both estimation procedures, the fully adjusted and weighted estimates are reported in Table (A3), demonstrating that the true values can be closely recovered when both the model and causal assumptions are satisfied. However, note that in this specific simulation setting, the weighting approach exhibits considerably greater uncertainty than the weighted G-computation approach. For the weighted G-computation approach, the corresponding cost-effectiveness plane is shown in Figure (A5), while the analogous plot for the weighting approach is omitted, as it provides no additional insight.  Because the expected sample-adjusted estimate closely matches the true population value (orange point), it is depicted with a hollow red point for visibility.

Note that decision uncertainty is present as the cloud of sample-adjusted estimates (blue points) crosses the WTP threshold line. This uncertainty arises primarily from the additional variability introduced when adjusting the sample-specific estimates to the target distribution or, under the weighting approach, from estimating and combining selection weights with treatment weights, while the sample size remains at 10,000. The increased variability is largely driven by rare strata in the sample receiving large weights, which inflates the uncertainty of the weighted estimates even after procedures such as stabilizing and truncating extreme weights. As shown in Figure (A5) plots (c) and (d), which are analogous to Figure (A5) plots (a) and (b), increasing the sample size to 1 million effectively removes this decision uncertainty. Nonetheless, for the remainder, we retain a sample size of 10,000 to maintain consistency in reporting and because it is more representative of a realistic study setting.

\begin{figure}[ht]
    \centering
    \begin{subfigure}[t]{0.48\textwidth}
        \centering
        \includegraphics[width=\linewidth]{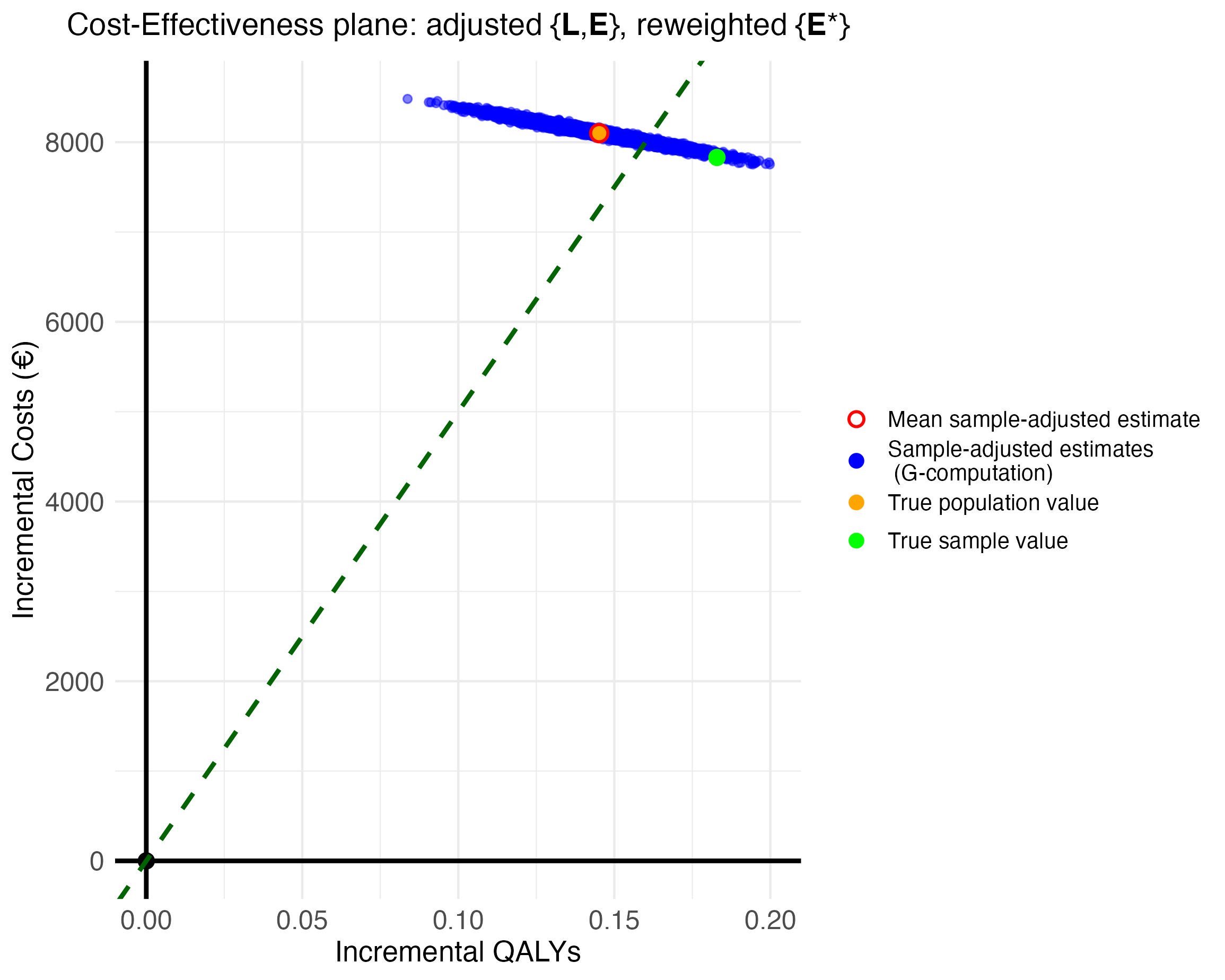}
        \caption{Zoomed out (sample size 10,000)}
        \label{fig:ce_adj_10k_out}
    \end{subfigure}
    \hfill
    \begin{subfigure}[t]{0.48\textwidth}
        \centering
        \includegraphics[width=\linewidth]{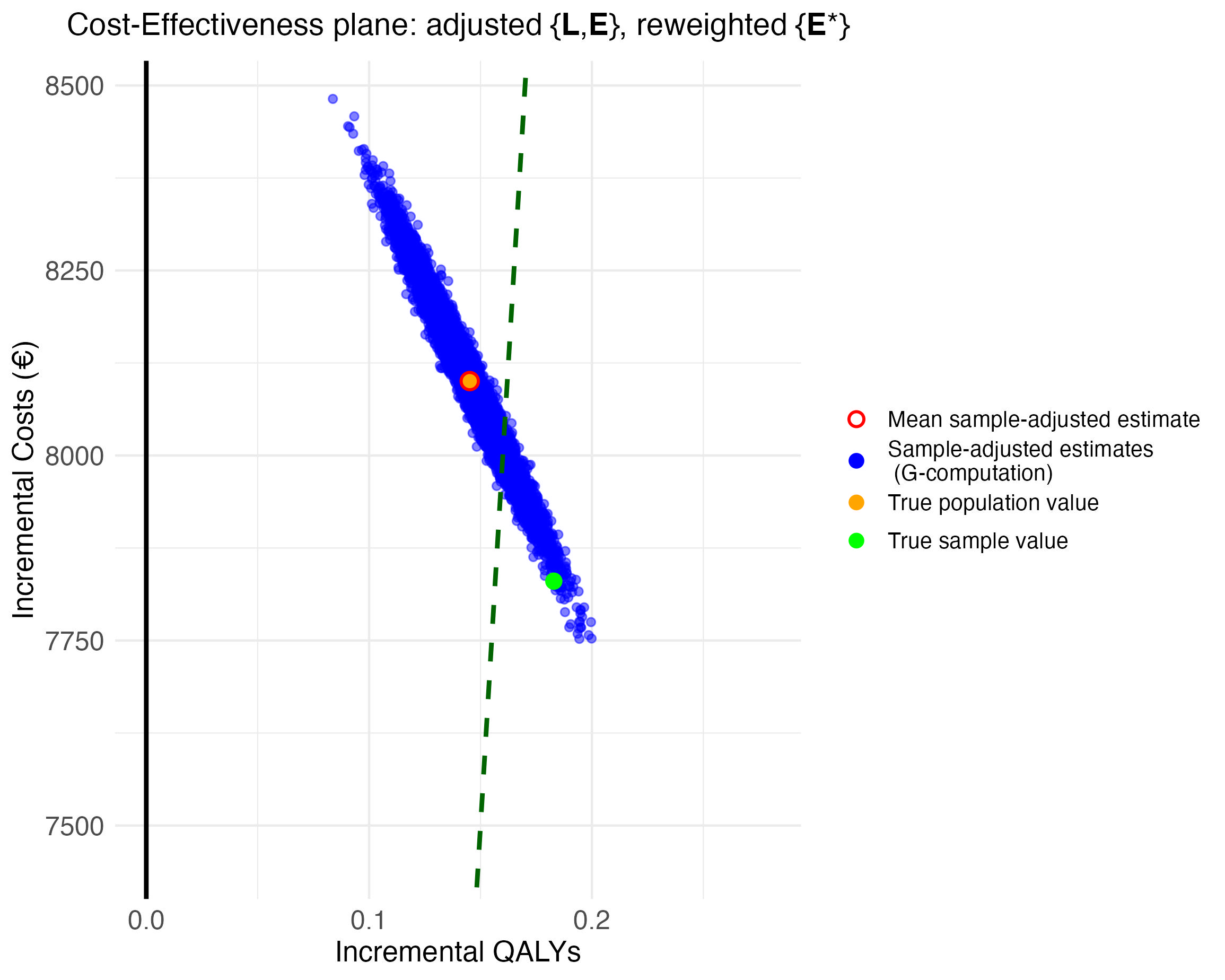}
        \caption{Zoomed in (sample size 10,000)}
        \label{fig:ce_adj_10k_in}
    \end{subfigure}
    \vspace{1cm}  

    \begin{subfigure}[t]{0.48\textwidth}
        \centering
        \includegraphics[width=\linewidth]{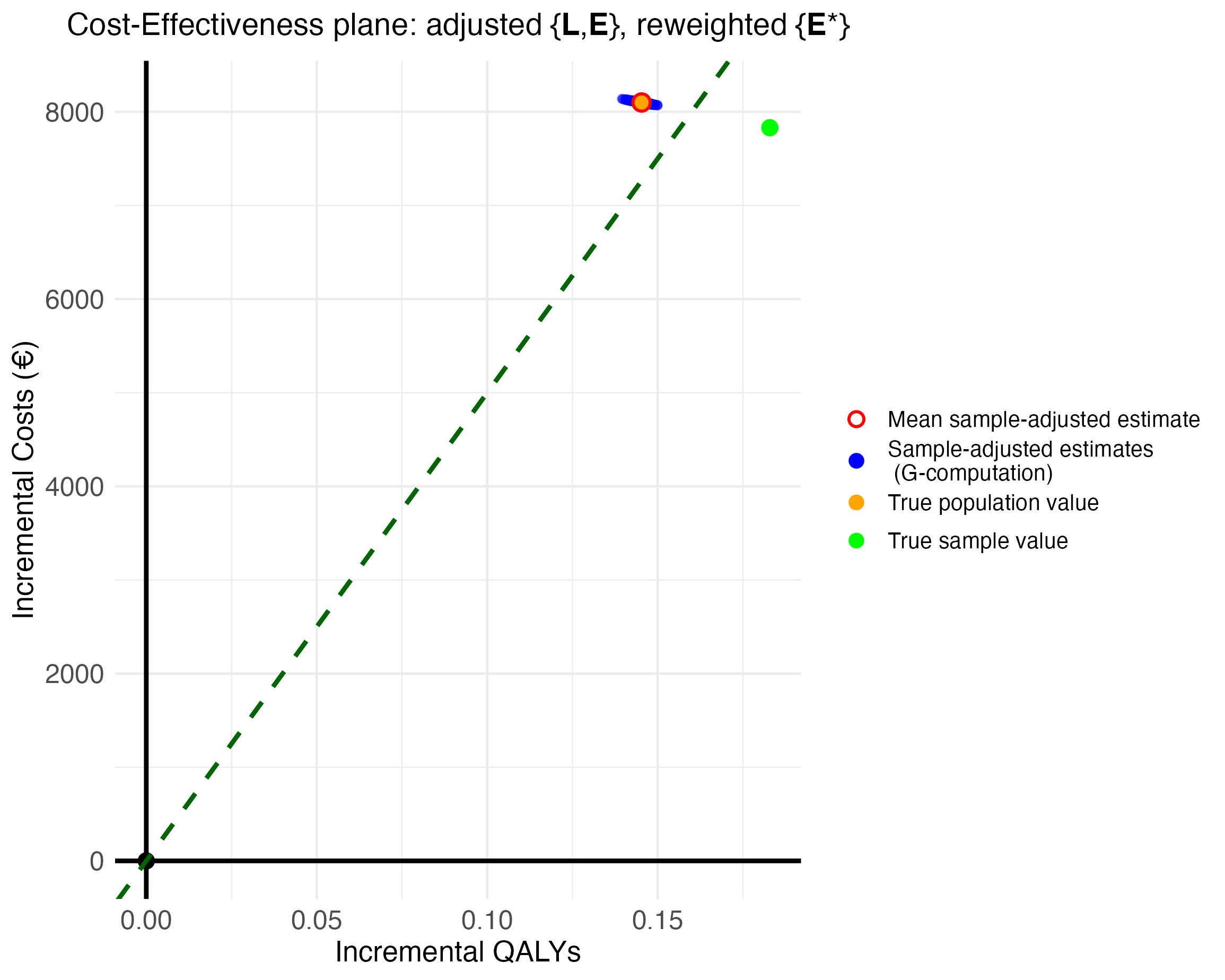}
        \caption{Zoomed out (sample size 1,000,000)}
        \label{fig:ce_adj_1m_out}
    \end{subfigure}
    \hfill
    \begin{subfigure}[t]{0.48\textwidth}
        \centering
        \includegraphics[width=\linewidth]{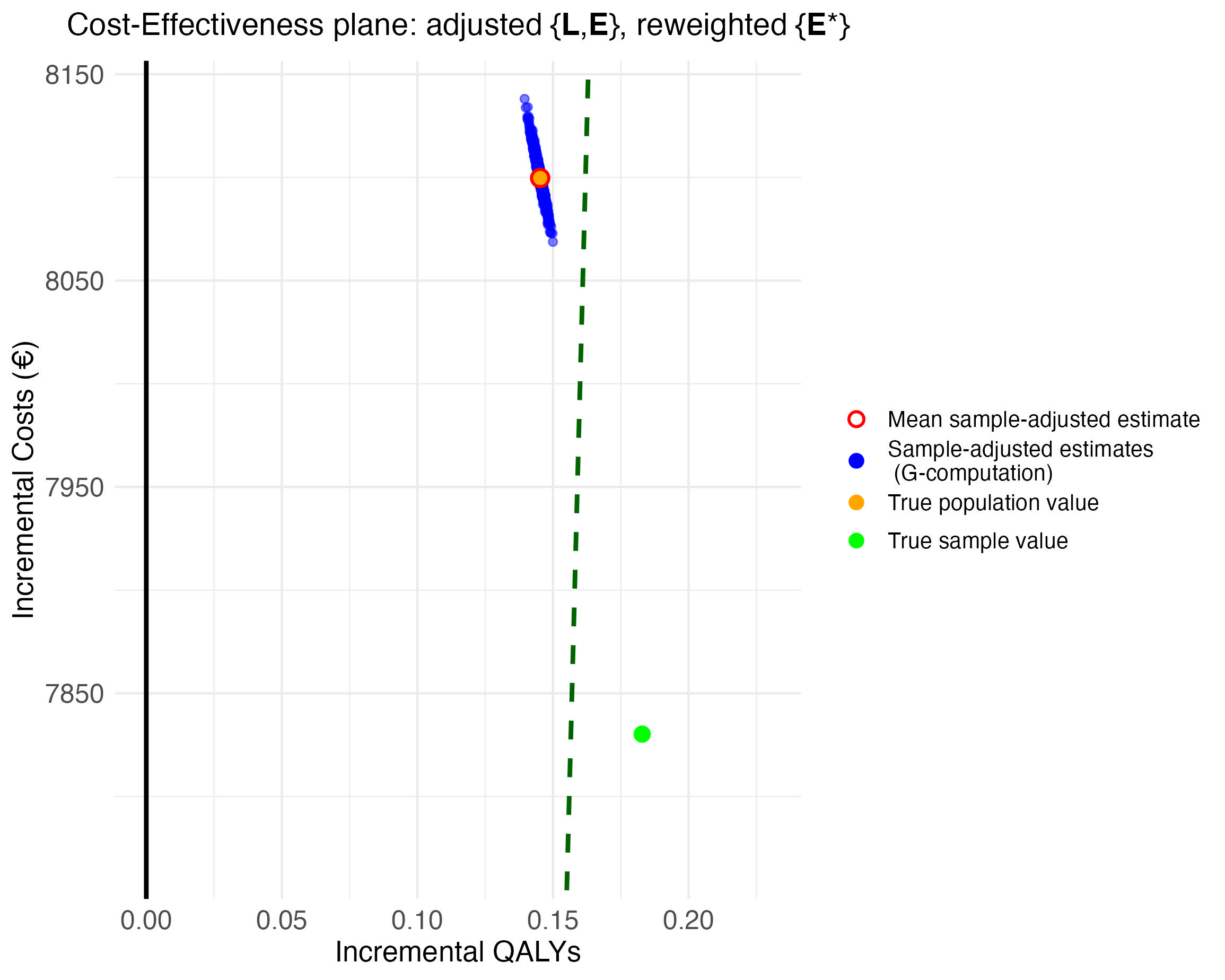}
        \caption{Zoomed in (sample size 1,000,000)}
        \label{fig:ce_adj_1m_in}
    \end{subfigure}

    \caption{
        Cost-effectiveness planes when adjusting for $\{\boldsymbol{L},\boldsymbol{E}\}$ and reweighting to the target distribution $\{\boldsymbol{E}^{*}\}$ using G-computation. Blue points represent mean adjusted estimates from each simulated sample, the red point shows the mean of these sample-adjusted means, whereas the orange and green point, respectively, indicate the true population and true sample value derived from the data-generating mechanism. Plots (a) and (b) were generated using 10,000 iterations and size of 10,000, while (c) and (d) using 1000 iterations with a size of 1,000,000. The dashed line represents the willingness to pay threshold of 50,000 euro
    }
    \label{fig:ce_adjusted_fourpanel}
\end{figure}

\clearpage
\noindent
When $\boldsymbol{E}$ is not fully observed, the study sample–specific estimates become subject to confounding bias, and the absence of complete information precludes the proper weighting required to adjust for discrepancies between the sample and target populations (i.e. unmeasured confounding and unmeasured effect modification). For instance, if both $E_1$ and $E_3$ are unobserved, reweighting solely based on $E_2$ produces biased estimates of the target outcomes which consequently deflates the population-level ICER, as illustrated on line 5 in Table (A3). Consequently, this would, on average, lead to the incorrect conclusion that $a=1$ is cost-effective relative to $a=0$ in the target population. A corresponding visual depiction is presented in Figure (A6), which shows that the mean sample-adjusted estimate lies to the right of the threshold line, in contrast to the true population value. 
\vspace{0.1cm}

\begin{figure}[H]
    \centering
    \begin{subfigure}[t]{0.48\textwidth}
        \centering
        \includegraphics[width=\linewidth]{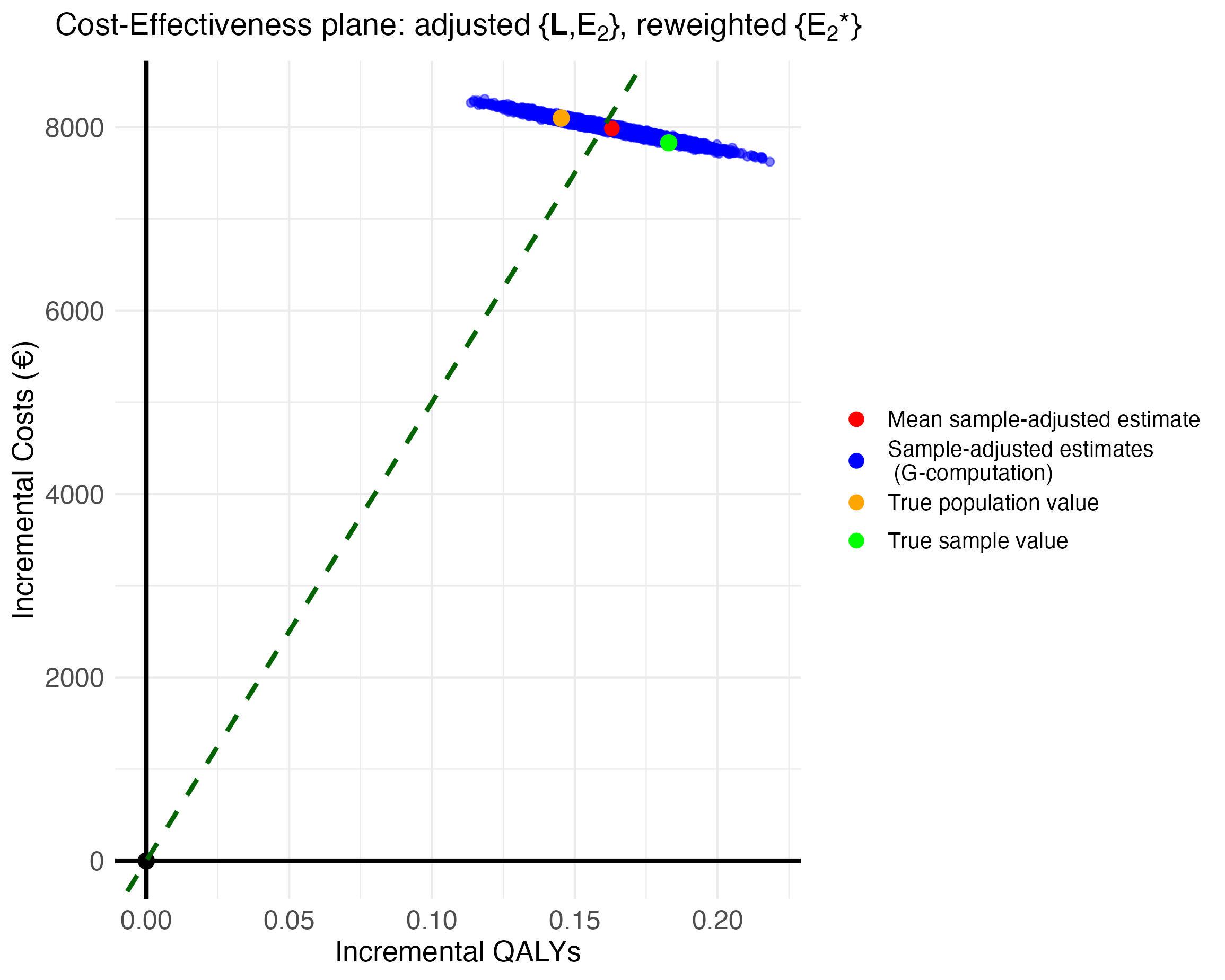}
        \caption{Zoomed out}
        \label{fig:adjusted_modifiers_subset}
    \end{subfigure}
    \hfill
    \begin{subfigure}[t]{0.48\textwidth}
        \centering
        \includegraphics[width=\linewidth]{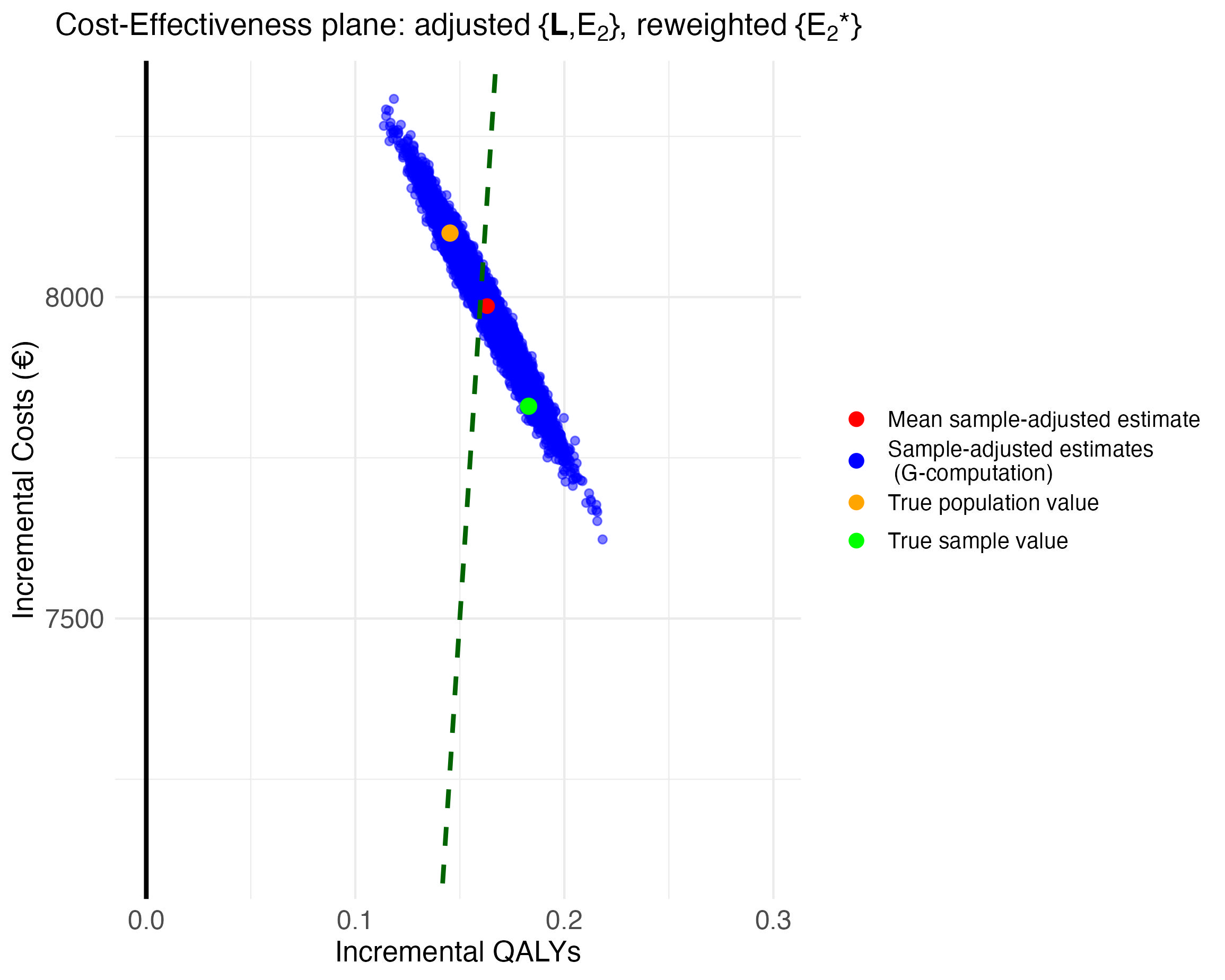}
        \caption{Zoomed in}
        \label{fig:adjusted_modifiers_subset_zoom}
    \end{subfigure}
    \caption{Cost-effectiveness plane when adjusting for $\{\boldsymbol{L},E_2\}$ and reweighting by $\{E^{*}\}$ from the target population. using G-computation. Blue points represent mean adjusted estimates from each simulated sample (10,000 samples), the red point shows the mean of these sample-adjusted means, whereas the orange and green point, respectively, indicate the true population and true sample value derived from the data-generating mechanism. The dashed line represents the willingness to pay threshold at 50.000 euro}
    \label{fig:adjusted_modifiers_combined}
\end{figure}
\vspace{0.1cm}

\newpage
\subsubsection{Propagation of bias}

Having demonstrated the implications of confounding and sample selection bias, we wish to underscore that decision-analytical models are inherently prone to bias propagation. Bias in even a single component, irrespective of its source, can propagate throughout the model and ultimately compromise decision-making. As an illustrative example, consider a setting with unmeasured confounding in which the confounder $L_2$ is supposed to be unobserved. In this case, the target probability parameters involving $V^{a}_2$ as the outcome, as well as any parameters conditional on $V^{a}_2$ are subject to bias when estimated from the data. More specifically, this bias propagates through all downstream parameters that depend on the unidentifiable term and is carried forward accordingly, thereby affecting the estimated expected costs, QALYs and ICER.

This is demonstrated numerically in Table (A2) under the G-computation approach (line 6) and is correspondingly visualised in the cost-effectiveness plane in Figure (A7). The induced confounding bias inflates the ICER in the sample-specific setting, thereby incorrectly suggesting that intervention $a=1$ is not cost-effective relative to $a=0$ for this study sample population.
\vspace{0.5cm}

\begin{figure}[ht]
    \centering
    \begin{subfigure}[t]{0.48\textwidth}
        \centering
        \includegraphics[width=\linewidth]{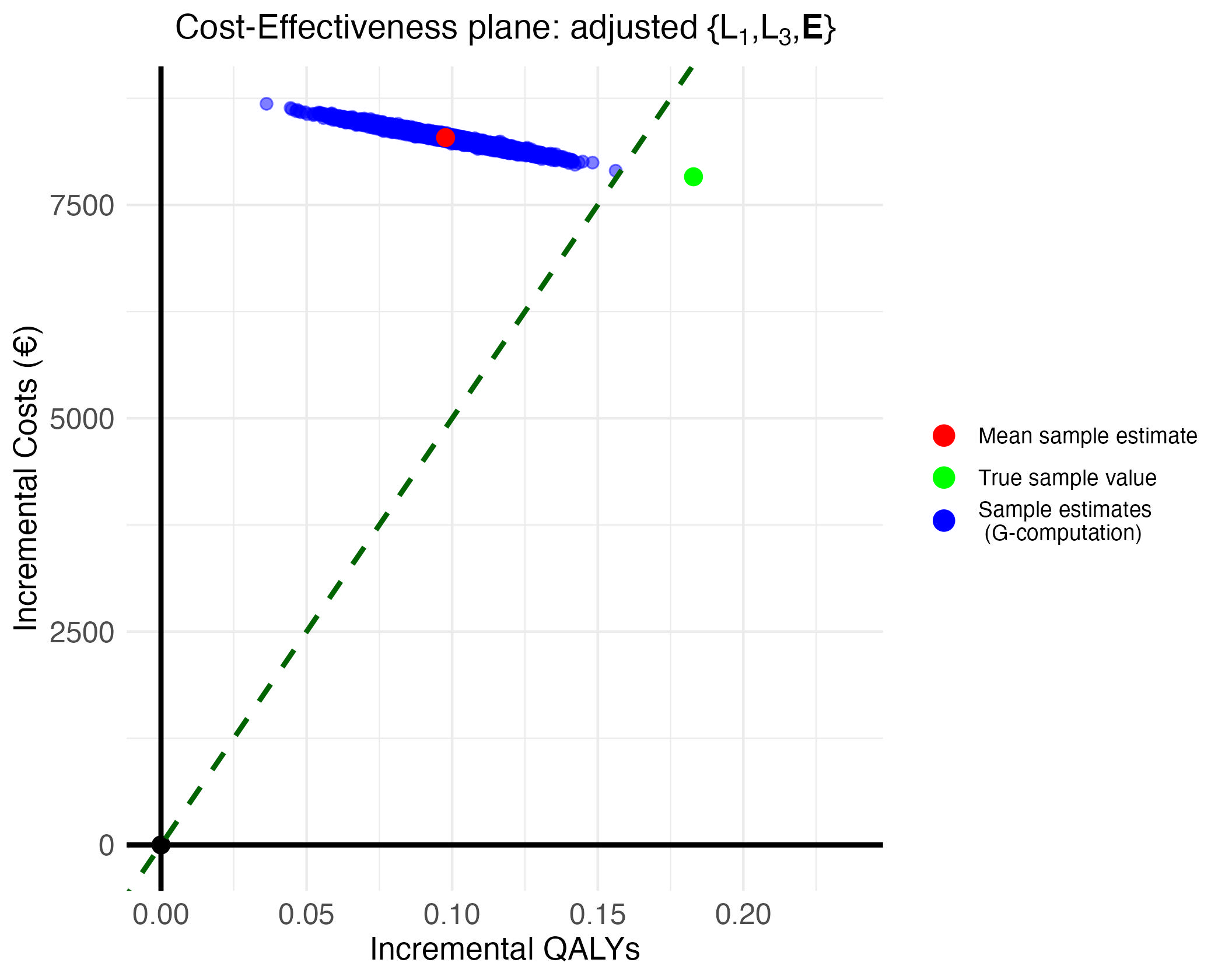}
        \caption{Zoomed out}
        \label{fig:adjusted_subset_propagation}
    \end{subfigure}
    \hfill
    \begin{subfigure}[t]{0.48\textwidth}
        \centering
        \includegraphics[width=\linewidth]{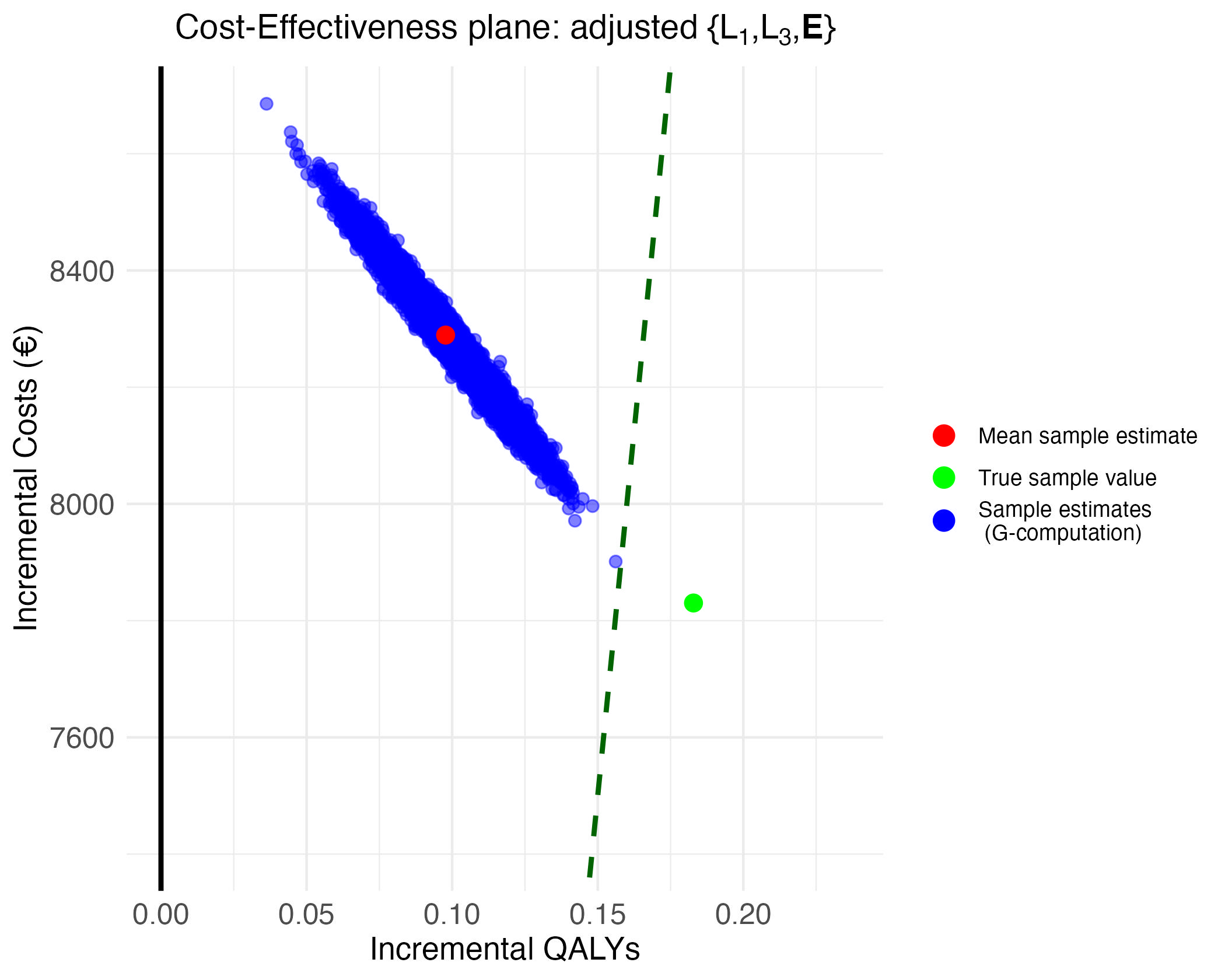}
        \caption{Zoomed in}
        \label{fig:adjusted_subset_zoom_propagation}
    \end{subfigure}
    \caption{Cost-effectiveness plane when not adjusting for $\{L_2\}$. Blue points represent mean estimates from each simulated sample (10,000 samples, size 10,000), the red point shows the mean of these sample means, and the green point indicates the true sample value derived from the data-generating mechanism. The dashed line represents the willingness to pay threshold of 50.000 euro}
    \label{fig:adjusted_combined_propagation}
\end{figure}

\noindent
Although $L_2$ is distributed equally across the sample and target populations, the induced confounding bias propagates to affect external validity, as shown numerically in Table (A3) on line 7 and graphically in the corresponding cost-effectiveness plane in Figure (A8). Despite the presence of this bias, the resulting conclusion would, on average, remain correct; reflecting that intervention $a=1$ is not cost-effective compared to $a=0$ for the target population. While the decision for the target population remains unchanged in the present example, it is important to recognize that the bias inflating the ICER is substantial, even with only a single source of bias. In more intricate settings, the propagation of bias may be more complex and could potentially shift the decision. This underscores the importance of assessing each parameter individually, complemented by causal bias analyses, since even when nearly all parameters are target-valid, a single biased parameter can meaningfully affect decision-making.

\begin{figure}[H]
    \centering
    \begin{subfigure}[t]{0.48\textwidth}
        \centering
        \includegraphics[width=\linewidth]{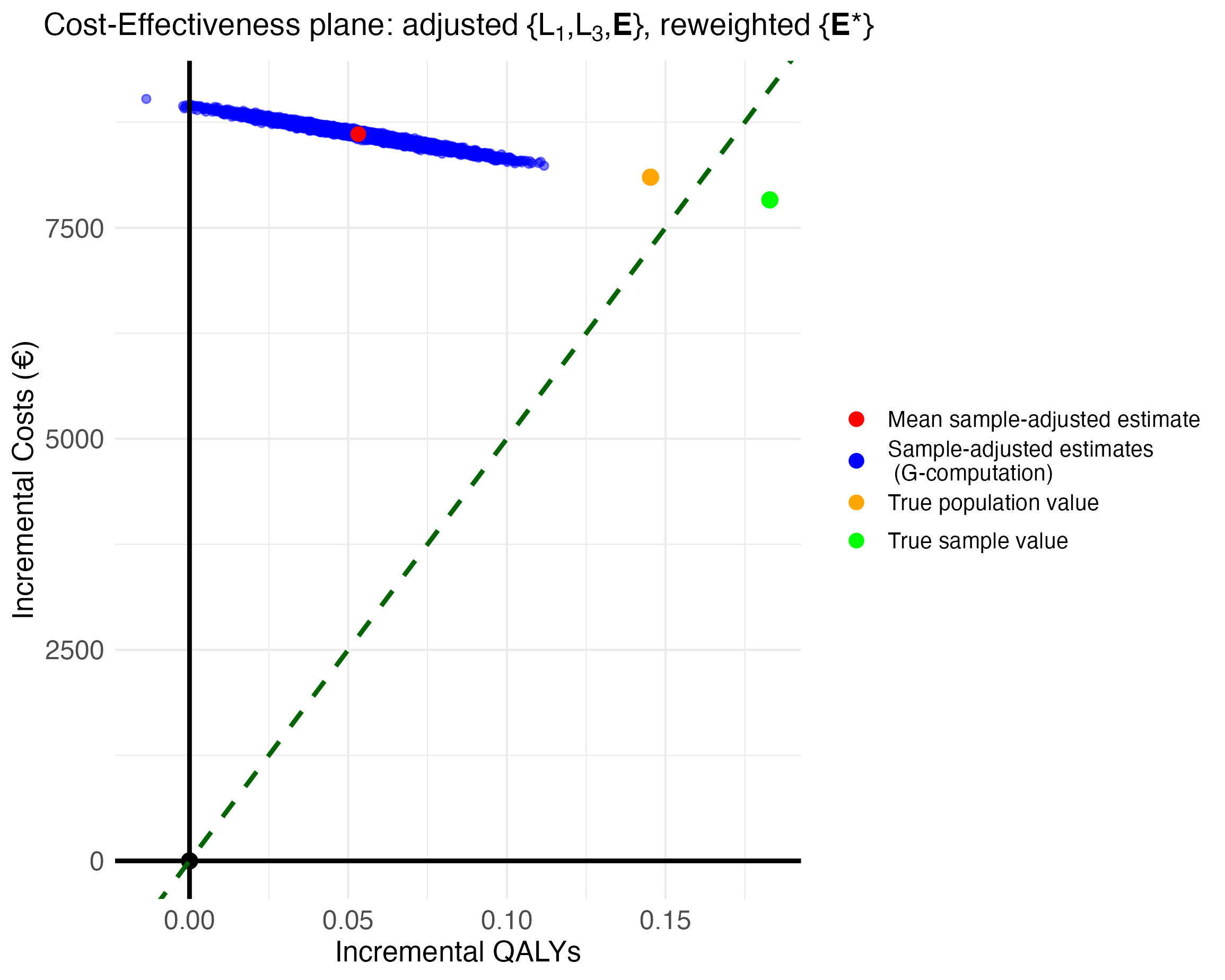}
        \caption{Zoomed out}
        \label{fig:adjusted_subset_propagation_pop}
    \end{subfigure}
    \hfill
    \begin{subfigure}[t]{0.48\textwidth}
        \centering
        \includegraphics[width=\linewidth]{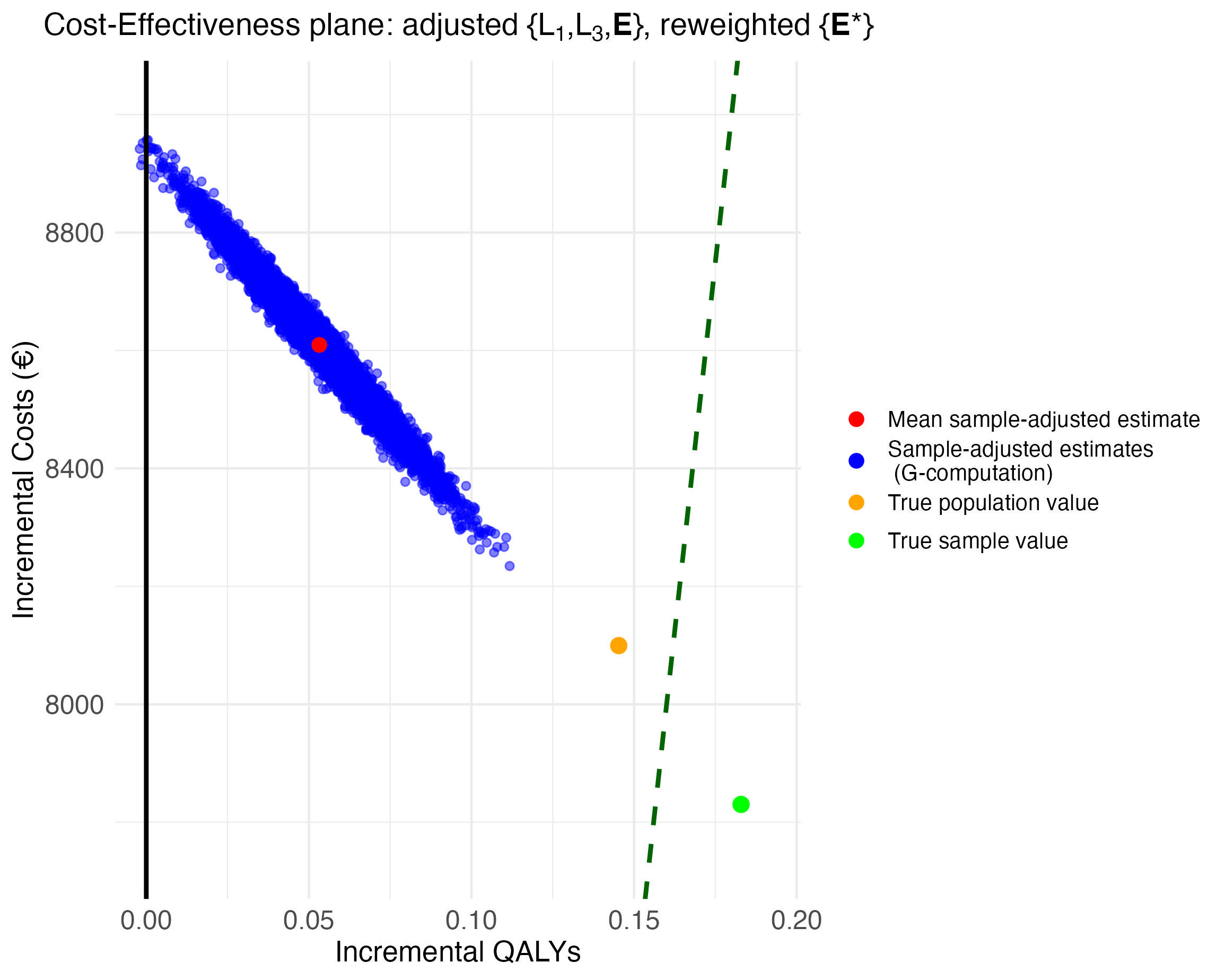}
        \caption{Zoomed in}
        \label{fig:adjusted_subset_zoom_propagation_pop}
    \end{subfigure}
    \caption{Cost-effectiveness plane when not adjusting for $L_2$ using weighted G-computation, with the weights defined by distribution of $\boldsymbol{E}$ in the target population. Cost-effectiveness plane when not adjusting for $\{L_2\}$ using weighted G-computation with weighs $\{ \boldsymbol{E}^{*}\}$. Blue points represent mean adjusted estimates from each simulated sample (10,000 samples), the red point shows the mean of these sample-adjusted means, whereas the orange and green point, respectively, indicate the true population and true sample value derived from the data-generating mechanism. The dashed line represents the willingness to pay threshold at 50.000 euro}
    \label{fig:adjusted_combined_propagation_pop}
\end{figure}

\clearpage
\section{Bias decomposition of the incremental cost-effectiveness ratio}

We consider the incremental cost-effectiveness ratio (ICER) and demonstrate how total bias can be decomposed into components attributable to model bias, internal validity bias, and external validity bias (i.e. collectively target bias), separately for both costs and effectiveness under each intervention. This decomposition highlights that accurate estimation of the ICER necessitates valid identification and estimation of all four underlying components. Moreover, because the ICER is a ratio, the decomposition reveals non-linear dependencies between biases in the costs and effectiveness.

\begin{small}
\begin{align}
\text{Bias}_{\text{DM}} &=  \mathcal{T} - \hat{\mathcal{T}}_{\mathcal{M}} \notag \\[0.3cm]
&= \frac{
    \displaystyle\mathbb{E}_{\mathcal{P}} \Bigl[ \lambda_1 + f(\mathbf{V}^{a=1}) \Bigr] - \displaystyle\mathbb{E}_{\mathcal{P}} \Bigl[\lambda_0 + f(\mathbf{V}^{a=0}) \Bigr]
}{
    \displaystyle\mathbb{E}_{\mathcal{P}} \Bigl[ g(\mathbf{V}^{a=1}) \Bigr] - \displaystyle\mathbb{E}_{\mathcal{P}} \Bigl[g(\mathbf{V}^{a=0}) \Bigr]
} \notag \\
&\quad - 
\frac{
    \displaystyle\mathbb{E}\biggl[\mathbb{E}_{\hat{\mathcal{P}}_{\mathcal{M} \mid \boldsymbol{S}}} \bigl[ \lambda_1 + f(\mathbf{V} \mid A=1, \boldsymbol{S}) \bigr]\biggr] - \displaystyle\mathbb{E}\biggl[\mathbb{E}_{\hat{\mathcal{P}}_{\mathcal{M} \mid \boldsymbol{S}}} \bigl[ \lambda_0 + f(\mathbf{V} \mid A=0, \boldsymbol{S}) \bigr]\biggr]
}{
    \displaystyle\mathbb{E}\biggl[\mathbb{E}_{\hat{\mathcal{P}}_{\mathcal{M}\mid \boldsymbol{S}}} \bigl[ g(\mathbf{V} \mid A=1, \boldsymbol{S}) \bigr]\biggr] - \displaystyle\mathbb{E}\biggl[\mathbb{E}_{\hat{\mathcal{P}}_{\mathcal{M}\mid \boldsymbol{S}}} \bigl[ g(\mathbf{V} \mid A=0, \boldsymbol{S}) \bigr]\biggr]
}  \\[0.3cm]
                        &= \frac{\Delta_{f\vphantom{\hat{f}}}}{\Delta_{g\vphantom{\hat{g}}}} - \frac{\Delta_{\hat{f}}}{\Delta_{\hat{g}}} \notag \\[0.3cm]
                        &= \frac{\big(\Delta_{f} - \Delta_{\hat{f}} \big) \Delta_{\hat{g}} \ + \  \Delta_{\hat{f}} \big(\Delta_{\hat{g}} - \Delta_{g} \big)}{\Delta_{g} \Delta_{\hat{g}}} \notag \\[0.3cm] 
                        &= \frac{\Delta_{f} - \Delta_{\hat{f}}}{\Delta_{g}} \ + \ \Delta_{\hat{f}} \ \frac{\big( \Delta_{g}  - \Delta_{\hat{g}} \big)} {\Delta_{g} \Delta_{\hat{g}}}\notag \\[0.3cm]
                        &= \frac{1}{\Delta_{g}}\Big( \big(\mathbb{E}_{\mathcal{P}}[f(\boldsymbol{V}^{a=1})] - \mathbb{E}\bigl[\mathbb{E}_{\hat{\mathcal{P}}_{\mathcal{M} \mid \boldsymbol{S}}}[f(\boldsymbol{V} \mid A=1, \boldsymbol{S})]\bigr] \big) - \big(\mathbb{E}_{\mathcal{P}}[f(\boldsymbol{V}^{a=0})] - \mathbb{E}\bigl[\mathbb{E}_{\hat{\mathcal{P}}_{\mathcal{M} \mid \boldsymbol{S}}}[f(\boldsymbol{V} \mid A=0, \boldsymbol{S})]\bigr] \big)\Big) \ + \notag \\
                        &\quad \frac{\Delta_{\hat{f}}}{\Delta_{g} \Delta_{\hat{g}}} \Big( \big(\mathbb{E}_{\mathcal{P}}[g(\boldsymbol{V}^{a=1})] - \mathbb{E}\bigl[\mathbb{E}_{\hat{\mathcal{P}}_{\mathcal{M} \mid \boldsymbol{S}}}[g(\boldsymbol{V} \mid A=1, \boldsymbol{S})]\bigr] \big) - \big(\mathbb{E}_{\mathcal{P}}[g(\boldsymbol{V}^{a=0})] - \mathbb{E}\bigl[\mathbb{E}_{\hat{\mathcal{P}}_{\mathcal{M} \mid \boldsymbol{S}}}[g(\boldsymbol{V} \mid A=1, \boldsymbol{S})]\bigr] \big)\Big)  \notag \\[0.3cm]
                        &= \frac{1}{\Delta_{g}} \Bigg(\Big(\underbrace{\mathop{\mathbb{E}_{\mathcal{P}}}[f(\boldsymbol{V}^{a=1})] - \mathop{\mathbb{E}_{\mathcal{P_M}}[f(\boldsymbol{V}^{a=1})]}}_{\text{model bias } (f,a=1)} +  \underbrace{\mathop{\mathbb{E}_{\mathcal{P_M}}[f(\boldsymbol{V}^{a=1})]} - \mathop{\mathbb{E}_{\mathcal{P}_{\mathcal{M} \mid \boldsymbol{S}}}[f(\boldsymbol{V}^{a=1} \mid \boldsymbol{S})]}}_{\text{external validity }(f,a=1)} + \notag \\
                        &\quad \underbrace{\mathop{\mathbb{E}_{\mathcal{P}_{\mathcal{M} \mid \boldsymbol{S}}}[f(\boldsymbol{V}^{a=1} \mid \boldsymbol{S})]} - \mathbb{E}\bigl[\mathop{\mathbb{E}_{\hat{\mathcal{P}}_{\mathcal{M} \mid \boldsymbol{S}}}[f(\boldsymbol{V} \mid A=1,\boldsymbol{S})]}\bigr]}_{\text{internal validity }(f,a=1)} \Bigr) + \Big(\underbrace{\mathop{\mathbb{E}_{\mathcal{P}}}[f(\boldsymbol{V}^{a=0})] - \mathop{\mathbb{E}_{\mathcal{P_M}}[f(\boldsymbol{V}^{a=0})]}}_{\text{model bias } (f,a=0)} + \notag \\
                        &\quad \underbrace{\big( \mathop{\mathbb{E}_{\mathcal{P_M}}[f(\boldsymbol{V}^{a=0})]} - \mathop{\mathbb{E}_{\mathcal{P}_{\mathcal{M} \mid \boldsymbol{S}}}[f(\boldsymbol{V}^{a=0} \mid \boldsymbol{S})]} \big)}_{\text{external validity }(f,a=0)} + \underbrace{\mathop{\mathbb{E}_{\mathcal{P}_{\mathcal{M} \mid \boldsymbol{S}}}[f(\boldsymbol{V}^{a=0} \mid \boldsymbol{S})]} - \mathbb{E}\bigl[\mathop{\mathbb{E}_{\hat{\mathcal{P}}_{\mathcal{M} \mid \boldsymbol{S}}}[f(\boldsymbol{V} \mid A=0,\boldsymbol{S})]}\bigr]}_{\text{internal validity }(f,a=0)} \Bigr)\Bigg) \ + \notag \\
                        &\quad \frac{\Delta_{\hat{f}}}{\Delta_{g} \Delta_{\hat{g}}} \Bigg(\Big(\underbrace{\mathop{\mathbb{E}_{\mathcal{P}}}[g(\boldsymbol{V}^{a=1})] - \mathop{\mathbb{E}_{\mathcal{P_M}}[g(\boldsymbol{V}^{a=1})]}}_{\text{model bias } (g,a=1)} +  \underbrace{\mathop{\mathbb{E}_{\mathcal{P_M}}[g(\boldsymbol{V}^{a=1})]} - \mathop{\mathbb{E}_{\mathcal{P}_{\mathcal{M} \mid \boldsymbol{S}}}[g(\boldsymbol{V}^{a=1} \mid \boldsymbol{S})]}}_{\text{external validity }(g,a=1)} + \notag \\
                        &\quad \underbrace{\mathop{\mathbb{E}_{\mathcal{P}_{\mathcal{M} \mid \boldsymbol{S}}}[g(\boldsymbol{V}^{a=1} \mid \boldsymbol{S})]} - \mathbb{E}\bigl[\mathop{\mathbb{E}_{\hat{\mathcal{P}}_{\mathcal{M} \mid \boldsymbol{S}}}[g(\boldsymbol{V} \mid A=1,\boldsymbol{S})]}\bigr]}_{\text{internal validity }(g,a=1)} \Bigr) + \Big(\underbrace{\mathop{\mathbb{E}_{\mathcal{P}}}[g(\boldsymbol{V}^{a=0})] - \mathop{\mathbb{E}_{\mathcal{P_M}}[g(\boldsymbol{V}^{a=0})]}}_{\text{model bias } (g,a=0)} + \notag \\
                        &\quad \underbrace{\big( \mathop{\mathbb{E}_{\mathcal{P_M}}[g(\boldsymbol{V}^{a=0})]} - \mathop{\mathbb{E}_{\mathcal{P}_{\mathcal{M} \mid \boldsymbol{S}}}[g(\boldsymbol{V}^{a=0} \mid \boldsymbol{S})]} \big)}_{\text{external validity }(g,a=0)} + \underbrace{\mathop{\mathbb{E}_{\mathcal{P}_{\mathcal{M} \mid \boldsymbol{S}}}[g(\boldsymbol{V}^{a=0} \mid \boldsymbol{S})]} - \mathbb{E}\bigl[\mathop{\mathbb{E}_{\hat{\mathcal{P}}_{\mathcal{M} \mid \boldsymbol{S}}}[g(\boldsymbol{V} \mid A=0,\boldsymbol{S})]}\bigr]}_{\text{internal validity }(g,a=0)} \Bigr)\Bigg) \notag \\[0.3cm]
                        &\text{with } \Delta_{g}, \Delta_{\hat{g}} \neq 0 \notag
\end{align}
\end{small}

\noindent
We restrict to $\Delta_{g} \neq 0$ and $\Delta_{\hat{g}} \neq 0$, as the bias expression is otherwise undefined. Intuitively, $\Delta_{g} \neq 0$ reflects the assumption (or expectation) that true QALYs under $a=1$ and $a=0$ are not exactly equal \textemdash that is, a decision trade-off exists. If this were not the case, such that $\Delta_{g}=0$, there would be no need for a decision-analytical model, as the choice between $a=1$ and $a=0$ could be made solely on the basis of cost differences. While one might expect that $\Delta_{\hat{g}} \neq 0$ follows from $\Delta_{g} \neq 0$, in practice, this may not be guaranteed. We therefore impose both restrictions explicitly. The main focus here is on ensuring that $(\Delta_{f} = \Delta_{\hat{f}})$ and $(\Delta_{g} =\Delta_{\hat{g}})$, as this would reduce total bias to zero. However, worth noting is that the condition $\Delta_{\hat{f}}=0$ would make the bias due to $(\Delta_{g} \neq \Delta_{\hat{g}})$ irrelevant. In other words, impact of QALY-related bias depends on whether there is an estimated cost-difference.

\end{document}